\documentclass[useAMS,usenatbib,usegraphicx]{mn2e}
\usepackage{wasysym}
\title[Type III migration; Inward migration]{Numerical simulations of type 
III planetary migration:\\ II. Inward migration of massive planets}
\author[A. Pepli\'nski et al.]{A. Pepli\'nski,$^1$\thanks{E-mail: 
adam@astro.su.se} P. Artymowicz$^2$ and G. Mellema$^1$\\
$^1$Stockholm University, AlbaNova University Centre, SE-106 91 Stockholm, Sweden\\
$^2$University of Toronto at Scarborough, 1265 Military Trail, Toronto, 
Ontario M1C 1A4, Canada}
\begin{document}
\voffset=-0.8in
\date{Accepted 0000 . Received 0000 ; in original form 2007 September}
\pagerange{\pageref{firstpage}--\pageref{lastpage}} \pubyear{0000}

\maketitle

\label{firstpage}
%
\begin{abstract}

We present a numerical study of rapid, so called type III migration
for Jupiter-sized planets embedded in a protoplanetary disc. We limit
ourselves to the case of inward migration, and study in detail its
evolution and physics, concentrating on the structure of the
corotation and circumplanetary regions, and processes for stopping
migration. We also consider the dependence of the migration behaviour
on several key parameters. We perform this study using the results of
global, two-dimensional hydrodynamical simulations with adaptive mesh
refinement. The initial conditions are chosen to satisfy the condition
for rapid inward migration. We find that type III migration can be
divided into two regimes, fast and slow. The structure of the
co-orbital region, mass accumulation rate, and migration behaviour
differ between these two regimes. All our simulations show a
transition from the fast to the slow regime, ending type III migration
well before reaching the star. The stopping radius is found to be
larger for more massive planets and less massive discs. A sharp
density drop is also found to be an efficient stopping mechanism. In
the fast migration limit the migration rate and induced eccentricity
are lower for less massive discs, but almost do not depend on planet
mass. Eccentricity is damped on the migration time scale.
\end{abstract}

%
\section{Introduction}

The theory of disk-planet interaction aims at finding a physical
description of the gravitational interaction between gas-dust
protoplanetary disks and protoplanets. A major advance, in comparison
with the theories of solar system formation from 20 years ago, was the
realization that protoplanets can globally migrate and consequently
that planets can be found on orbits different from the formation sites
of planetary cores (or gaseous protoplanets, in an alternative
scenario). This conclusion was first reached by
\citet{1979ApJ...233..857G,1980ApJ...241..425G} and worked out in a
number of investigations summarised by \citet{1993prpl.conf..749L} and
\citet{2000prpl.conf.1111L}. Theories of type I migration (planet
embedded in gas) and type II (in a disk gap opened by the planet's
gravity) for a standard stationary accretion disk both predict an
inward direction of migration, towards the star (Ward 1997). Type II
migration, which carries a gap-opening planet with the viscous flow of
material in the disk, is capable of moving it outward provided the
disk has local density gradients supporting outward viscous spreading
(\citealt{1993prpl.conf..749L}).

The discovery of extra-solar planets (\citealt{1995Natur.378..355M},
\citealt{2000prpl.conf.1285M}, \citealt{2002ApJ...568..352V})
triggered a renewed interest in the migration process. For the
so-called hot Jupiters, giant planets residing at a distance of
typically just 0.05-0.1 AU from the host star, an in situ formation
appears impossible in either the core accretion or the disk
instability scenario of planet formation. Therefore, in the current
consensus, hot Jupiters are understood as bodies that started
forming at orbital radii about 100 times larger, where availability of ice
enhances the rate of growth of planetary cores, and later migrated
toward the star by interacting with the disk (for an extensive recent
review of migration, see \citealt{2007prpl.conf..655P}).  Another
important issue in the theory of planet formation is to understand the
intermediate ($e=0.2-0.5$) orbital eccentricities of planets.  While
in this paper we concentrate primarily on migration, our numerical
investigations also produce results for the eccentricity evolution of
a planet migrating in a disk.

Until recently, disk-planet interaction was understood entirely in
terms of just one type of coupling mechanism, involving
resonances. Disks are perturbed by a planet most strongly at Lindblad
resonances, which are the gas-dynamical equivalents of mean motion
resonances in celestial mechanics. Disks respond by launching density
and bending waves. These waves take the angular momentum and energy
flux deposited by the perturber and transport it away from their
launching point, to later deposit these quantities in the disk. The
non-axisymmetric wave pattern acts back the planet, providing the
torque and energy flux\footnote{We shall neglect the mention of energy
flow and only describe the torque or angular momentum flow. Only the
latter drives migration of bodies whose eccentricity does not grow (a
typical result of numerical simulations such as ours).}  to drive its
orbital evolution (e.g., \citealt{1979ApJ...233..857G,
1980ApJ...241..425G}). The action of Lindblad torques is not only
responsible, for orbital migration of types I and II, but also for the
shepherding effect in the disk-satellite interaction as well as the
gap opening phenomenon.

However, in addition to the Lindblad resonances, effects often
described as corotational resonances can play a significant and even
dominant role, as found independently by \citet{2003ApJ...588..494M}
and \citet{2004ASPC..324...39A}. Those effects are not truly resonant,
in the sense of depending on strict mean-motion commensurabilities,
beyond a long-term 1:1 commensurability of the planet with gas orbits
librating with it. We will therefore refer to them as corotational
torques or corotational flows. The corotational region corresponds to
the horseshoe and tadpole orbit region in a frame corotating with the
planet. It occupies an annulus a few Roche lobe or Hill sphere radii
away from a circular planetary orbit. Gas in this region alternates
between being closer and further from the star than the planet. In
doing so, the gas may flow at small density but significant velocity
through the gap region, if a gap exists.

Misunderstood for a long time as unimportant (partly because of an
assumed smooth density distribution in a disk, and partly due to
previous estimations done for a fixed-orbit, i.e., non-migrating,
planet, which automatically eliminates the effect), the effects of the
flow of gas across the corotational region can provide a far larger
torque on the planet than the competing, wave-launching Lindblad
interactions. This leads to a new type of very rapid migration (called
type III), which apart from speed, also distinguishes itself by having
a far smaller predilection for inward migration than the standard type
I or type II varieties (\citealt{2003ApJ...588..494M};
\citealt{2004ASPC..324...39A}; \citealt{2005CeMDA..91...33P};
\citealt{2006AIPC..843....3A}). This type of migration was studied
numerically by \citet{2003ApJ...588..494M}, who performed
two-dimensional simulations of a freely moving planet in steady-state
migration for a whole range of imposed fixed migration rates.
Global, high resolution two- and three-dimensional simulations of a
freely migrating planet were performed by \citet{2005MNRAS.358..316D}.
\citet{2005CeMDA..91...33P} considered local shearing box simulations
in 2-D, and \citet{Pawel_Miguel1} in 3-D.

This paper is the second one in a series on numerical simulations of
planet migration. In the first paper (\citealt{PaperI}, henceforth
Paper~I) we showed how two new, physically motivated, ingredients in
our model, namely the use of a modified local-isothermal approximation
and a correction for self-gravity effects, are necessary to achieve
consistent and convergent numerical results. In the current paper,
using these methods, we study the evolution of type III migration and
the physical dependencies between the model parameters and the outcome
(mostly, migration rate and range, and the flow of gas around and onto
the planet). In other words, we focus on the physics rather than
numerics of type III migration.  We limit ourselves in this paper to
models describing the inward migration of a freely evolving
planet. Paper III \citep{PaperIII} will describe the fast outward
migration, which is not a simple mirror case of the inward migration.

The layout of the paper is as follows. In Section~\ref{type3teory} we
outline the mechanism of type III migration.
Sections~\ref{disc-model} and~\ref{sect_num_mth} briefly summarise the
basic model equations and the numerical method. In
Section~\ref{inward_mig} we describe and analyse in detail our
`standard case', treating its orbital evolution, evolution of torques,
and the flow pattern near the planet.  Section~\ref{sect_dep_on_par}
deals with the dependence of the migration process on various
parameters.  In Sections~\ref{stop_mig} and~\ref{sec_ecc_ev} we
analyse the stopping of type III migration and the eccentricity
evolution. Finally, in Section \ref{conclusions} we summarise our
findings.

\section{Type III migration regime}

\label{type3teory}

As outlined above, type III migration is mostly driven by the flow of
material through the co-orbital region. Although the corotation torque
dominates, the differential Lindblad torque still retains some
importance by modifying the space of allowed solutions for migration
rates and generally biasing the migration to become inwardly directed,
an in type I migration. The source of the corotation torque is the
interaction between the planet and orbit-crossing fluid elements. The
fluid exchanges angular momentum with the planet when it executes a
U-turn at the end of each horseshoe-like streamline. In the absence of
planet migration, dissipation or a time-dependent effect such as an
initial transient, the fluid elements in the co-orbital region
circulate on closed orbits (in the corotating frame), resulting in a
zero net corotation torque in a smooth disc. This cancellation is
often called the saturation of the corotational torque. In this
stationary case only an actively maintained density or more generally,
a specific vorticity gradient in the planet's vicinity can produce a
non-zero corotation torque.  The density variation can be due to the
internal disc structure (e.g., an edge of a dead zone, where viscosity
changes abruptly, cf.~\citealt{2006AIPC..843....3A}), or to disc
heating by the planet \citep{2006A&A...459L..17P} or due to MHD
turbulence \citep{2004MNRAS.350..849N}.

For a migrating planet, the situation is clearly more susceptible to a
non-cancellation of torques. The horseshoe region becomes asymmetric
and some streamlines are no longer closed (those belonging to the
disk). On the other hand, a region of librating, closed orbits
forms. It is carried along together with the migrating planet as
`ballast' if it is denser than the disk, or as a `balloon' if it is
less dense. In the latter case it so to say, propels the planet along
its orbit by the imbalance of gravitational pull of gas in front and
in the back of the planet. In fact, the initial direction of type III
migration being toward the denser side of the disk (to compensate for
the flow of mass and angular momentum of gas from that dense to the
rarefied side, and conserve the angular momentum of planet-disk system
globally), migration always results in the librating orbit region
being under-dense, i.e.\ the second case just mentioned. We can thus
describe the disk with a planet in type III migration as having at
least a partial gap, but one which is front-back asymmetric.

This configuration provides a positive feedback for the migration.
\citet{2003ApJ...588..494M} showed that the corotation torque is
proportional to the migration rate $\dot a$ for slow migration
(defined quantitatively below).  At large imposed migration rates,
they found that the torque diminishes slightly after reaching a
maximum, and obtained the migration speed estimate of
\begin{equation}
\label{a_MP03}
\dot a = {2 \Gamma_\rmn{LR} \over {\Omega a (M_\rmn{p}' - \delta m)}},
\end{equation}
where $\Gamma_\rmn{LR}$ is the differential Lindblad torque,
$M_\rmn{p}' = M_\rmn{P} + M_\rmn{R}$ is all of the mass content within
the Roche lobe (sum of the planet and gas mass), and $\delta m$ is the
co-orbital mass deficit: the difference between the mass
that the horseshoe region would have if it had a uniform surface
density equal to the upstream surface density $\Sigma_\rmn{s}$ and the
actual horseshoe region mass $M_\rmn{HS}$,
\begin{equation}
\label{eqn_m_delta_small}
\delta m = 4\pi a x_\rmn{s} \Sigma_\rmn{s} - M_\rmn{HS}= 4\pi a x_\rmn{s} 
(\Sigma_\rmn{s} -\Sigma_\rmn{g}),
\end{equation}
where $\Sigma_\rmn{g}$ is an average density in the horseshoe region
(gap region)\footnote{A note of caution: despite its name and physical units,
$\delta m$ provides a measure of surface density deficit rather than
the actual mass difference of the librating region against the
background disk, owing to the variable shape of that region ($\delta
m$ becomes the actual mass deficit only if migration is
infinitesimally slow and the librating region is an annulus of a
constant width $2x_\rmn{s}$).}.  Equation (\ref{a_MP03}) clearly cannot
be applied to the most interesting case, where the denominator of
Eq.~(\ref{a_MP03}) tends to zero, the mass deficit becomes comparable
with or larger than $M_\rmn{p}' $, and we expect a transition to fast
migration.

\citet{PaperIV} considered a simple model in which the angular
momentum of a fluid element undergoes a sudden change at orbital
conjunction with the planet. They found a critical migration speed
$\dot a = \dot{a_\rmn{f}}$ at which one of the ends of the
horseshoe-like orbit cannot reach conjunction and dis-attaches from the
planet, becoming a tadpole orbit.  This speed is the ratio of the
half-width of the corotational horseshoe region $x_\rmn{s}$ (the subscript
`s' stands for separatrix distance) and one-half of the libration time
(synodic period) on the critical horseshoe/tadpole orbit. The
$\dot a_\rmn{f}$ corresponds to the maximum corotation torque in this
model and provides a convenient fiducial speed unit for type III
migration.  For a Keplerian disc with angular speed at planet's
semi-major axis $\Omega$, $\dot a_\rmn{f}$ has the form
\begin{equation}
\label{eq2_a_f}
\dot a_\rmn{f} = {{3 x_\rmn{s}^2 \Omega} \over {8 \pi a}},
\end{equation}
In terms of this fiducial speed, we call migration 
with $|\dot a|/\dot a_\rmn{f} < 1$ slow migration, and  migration 
with $|\dot a|/\dot a_\rmn{f} > 1$ fast migration. 

To simplify the notation, we define two useful, non-dimensional 
quantities. The first is the speed of migration $Z$ that we
will extensively use to describe a momentary state of our numerical models,
\begin{equation}
\label{defin_Z}
Z  = \frac{|\dot a|}{\dot a_\rmn{f}}\,.
\end{equation}
$Z$ expresses the ratio of the migration timescale $T_\rmn{migr}$
(defined as $x_\rmn{s}/{\dot a}$) to the libration timescale
$T_\rmn{lib}$.  We therefore distinguish between the fast ($|Z|>1$)
and slow migration limits ($|Z|<1$).

The second dimensionless quantity is $M_\rmn{\Delta}$, expressing the
co-orbital mass deficit
\begin{equation}
\label{eq_MDelta}
M_\rmn{\Delta} = \delta m /M_\rmn{p}' = \frac{4 \pi a x_\rmn{s}
(\Sigma_\rmn{s} -\Sigma_\rmn{g})}{ M_\rmn{p}'}\,.
\end{equation}

\citet{PaperIV} derived an analytical theory for the whole range of migration
rates based on a guiding centre approximation. They obtained  $x_\rmn{s} = 2.47
\approx 2.5$ Hill sphere radii (Roche lobes) of the planet, and concluded that 
this value provides a good approximation to the numerically determined 
estimators of the corotational region's half-width.
The corotation torque is a linear function of $Z$ for $Z \ll 1$
(slow migration limit) and becomes constant for $Z > 1$ (fast migration limit):
\begin{equation}
\label{eq2_GCR}
\Gamma_\rmn{CR} = {\rmn {sign}}(\dot a){\delta m \over 3} \dot
a_\rmn{f} \left(1-[\max(0,1-|Z|)]^{3/2}\right) \Omega a.
\end{equation}
Unlike \citet{2003ApJ...588..494M}, \citet{PaperIV} do not predict
migration to have the runaway or self-accelerating character of an
instability.  Rather, a steady migration speed $Z$ results from the
torque balance equation (cf. also \citealt{2007prpl.conf..655P}) in
which the following three torques are in balance: the torque required
for migration with speed $Z$ according to orbital mechanics of nearly
circular orbits, the corotation torque $\Gamma_\rmn{CR}$, and the Lindblad
resonant torque $\Gamma_\rmn{LR}$.  This allows one to define $Z$ as a
function of $M_\rmn{\Delta}$ and determine the space of allowed
solutions for a given value of $\Gamma_\rmn{LR}$.  The relation between
$Z$ and $M_\rmn{\Delta}$ is linear in the fast migration limit $|Z| =
2M_\rmn{\Delta}/3$ (for $M_\rmn{\Delta} > 1.5$) and becomes nonlinear
in the slow migration limit ($M_\rmn{\Delta} < 1.5$).

From Eqs.~(\ref{eq2_a_f}) and (\ref{eq2_GCR}) we can see that
$\Gamma_\rmn{CR} \sim M_\rmn{P}$ in the fast migration limit and
increases slower than linearly in the slow migration limit. This means
that the planet's orbital evolution should be weakly dependent on the
planet mass in the fast migration limit. However, in the slow
migration limit any increase of the planet mass should decrease the
migration rate significantly. We verify these predictions below.

To conclude this description of type III migration we want to point
out that it in general depends on the migration history, because the
way the horseshoe region is populated depends on the previous
evolution. In the case of strong variations of $\dot a$ the
streamlines of the horseshoe region are not exactly closed, so the
co-orbital mass deficit can be lost and the fast migration would stop
(\citealt{2007prpl.conf..655P}).

\section{Description of the physical model}

\label{disc-model}

The full description of the disc model and the numerical method was 
given in Paper~I. In this section we give only a short description 
of the governing equations and our methodology. 

\subsection{Disc model}

We adopt in our simulations a two-dimensional, infinitesimally thin
disc model and use vertically averaged quantities. We work in the 
inertial reference frame, in the Cartesian coordinate system $(x,y,z)$. 
The plane of the disc and the star--planet system coincides with the 
$z=0$ plane. The centre of mass of the star--planet system is initially 
placed at the origin of the coordinate system.

The gas in the disc is taken to be inviscid and
non-self-gravitating. The evolution of the disc is given by the
two-dimensional $(x,y)$ continuity equation for $\Sigma$ and the Euler
equations for the velocity components $\bmath{v} \equiv
(v_x,v_y)$
\begin{eqnarray}
{\partial \Sigma \over \partial t} + \nabla(\Sigma \bmath{v}) = 0,\\
{\partial \Sigma \bmath{v} \over \partial t} + \nabla(\Sigma \bmath{v} 
\bmath{v}) + \nabla P = -\Sigma \nabla \Phi,
\end{eqnarray}
where $P$ is two-dimensional (vertically integrated) pressure, and
$\Phi$ is the gravitational potential generated by the protostar and
planet
\begin{equation}
\Phi = \Phi_\rmn{S} +\Phi_\rmn{P} = - {G M_\rmn{S} \over |\bmath{r} - 
\bmath{r}_\rmn{S}|} - {G M^*_\rmn{P} \over |\bmath{r} - \bmath{r}_\rmn{P}|}.
\end{equation}
The positions and masses of the star and the planet we denote by
$\bmath{r}_\rmn{S}$, $M_\rmn{S}$, $\bmath{r}_\rmn{P}$ and
$M^*_\rmn{P}$ respectively. $M^*_\rmn{P}$ is the effective planet
mass, as explained below. The gravitational potential close to the
star and the planet is softened according to Eq.~5 from Paper~I, which
contains a parameter $r_\rmn{soft}$, the smoothing length.

We do not consider the energy equation, but use a local isothermal
approximation which prescribed the temperature as a function of both
distance to the star and to the planet such that the global disc has
a scale height $h_\rmn{s}$ with respect to the star, and the circumplanetary
disc a scale height $h_\rmn{p}$ with respect to the planet. This equation
of state was called EOS2 in Paper~I.

In Paper~I we also showed that it is necessary to include some sort of
self-gravity in the neighbourhood of the planet. For this we use the
extra acceleration given by Eq.~13 from Paper~I for the gas near the
planet. The size of the region to which this correction is applied is
characterised by a parameter $r_\rmn{env}$. For this correction we
introduced the concept of the effective planet mass, $M_\rmn{P}^*$,
which we normally take to be ${\widetilde M}_\rmn{P} = M_\rmn{P} +
M_\rmn{soft}$, the sum of the planet mass and the gas mass within the
softening radius. For experiments where we want to explicitly ignore
the mass accumulation near the planet, we take $M_\rmn{P}^*=
M_\rmn{P}$.

\subsection{Equation of motion for the star and the planet}

The goal of the current study is to investigate the orbital evolution
of the planetary system due to the gravitational action of the disc
material. Since the calculations are done in the inertial reference
frame, the equation of motion for the star and the planet have the
simple form:
\begin{eqnarray}
\ddot {\bmath{r}}_\rmn{S} = - {G M_\rmn{P} (\bmath{r}_\rmn{S} -
\bmath{r}_\rmn{P}) \over {|\bmath{r}_\rmn{P} - \bmath{r}_\rmn{S}|}^3}
- \int\limits_{M_\rmn{D}} {G (\bmath{r}_\rmn{S} - \bmath{r})
dM_\rmn{D}(\bmath{r}) \over {|\bmath{r} - \bmath{r}_\rmn{S}|}^3}, \\
\ddot {\bmath{r}}_\rmn{P} = - {G M_\rmn{S} (\bmath{r}_\rmn{P} -
\bmath{r}_\rmn{S}) \over {|\bmath{r}_\rmn{P} - \bmath{r}_\rmn{S}|}^3}
- \int\limits_{M_\rmn{D}} {G (\bmath{r}_\rmn{P} - \bmath{r})
dM_\rmn{D}(\bmath{r}) \over {|\bmath{r} - \bmath{r}_\rmn{P}|}^3}.
\end{eqnarray}
In both cases the integration is carried out over the disc mass
$M_\rmn{D}$ included inside the radius $r_\rmn{disc}$.

\subsection{Setup description}

\label{sec_setup-desc}

In the simulations we adopt non-dimensional units, where the sum of
star and planet mass $M_\rmn{S} + M_\rmn{P}$ represents the unit of
mass. The time unit and the length unit are chosen to make the
gravitational constant $G=1$. This makes the orbital period of
Keplerian rotation at a radius $a=1$ around a unit mass body equal to $2
\pi$. However, when it is necessary to convert quantities into
physical units, we use a Solar-mass protostar $M_\rmn{S}
=M_{\astrosun}$, a Jupiter-mass protoplanet $M_\rmn{P}=M_{\jupiter}$,
and a length unit of $5.2 AU$. This makes the time unit equal to
$11.8/2\pi$ years.

In all simulations the grid extends from $-4.0$ to $4.0$ in both
directions around the star and planet mass centre. This corresponds to
a disc region with a physical size of $20.8$ AU.

\subsubsection{Initial conditions}

\label{setc_init_cond}

The initial surface density $\Sigma_\rmn{init}$ is given by a modified
power law:
\begin{equation}
\Sigma_\rmn{init} = \psi(r_\rmn{c}) \Sigma_\rmn{0} 
(r_\rmn{c}/r_\rmn{0})^{\alpha_\rmn{\Sigma}},
\end{equation}
where $r_\rmn{c} = |\bmath{r} - \bmath{r}_\rmn{C}|$ is the distance to
the mass centre of the planet-star system, $r_\rmn{0}$ is a unit
distance, and $\psi$ is a function that allows introducing a sharp
edges in the disc (see Fig.~2 in Paper~I).

The disc mass is described by the disc to the primary mass ratio
\begin{equation}
\mu_\rmn{D} = {{\Sigma_\rmn{init}(r_\rmn{0}) \pi r_\rmn{0}^2} \over {M_\rmn{S}}} = 
{{\Sigma_\rmn{0} \pi r_\rmn{0}^2} \over {M_\rmn{S}}}.
\end{equation}
In the simulations $\mu_\rmn{D}$ ranges from $0.0025$ to $0.005$. For
the Minimum Mass Solar Nebula (MMSN) $\mu_\rmn{D}=0.00144$ (for
$\alpha_\rmn{\Sigma}=-3/2$). We investigate different density profiles
by changing $\alpha_\rmn{\Sigma}$ from $-1.5$ to $-0.5$.

To enforce rapid migration the planet is introduced
instantaneously. Since we focus on type III migration only and do not
analyse the problem of orbital stability inside a gap, we do not
introduce the planet smoothly nor keep it on a constant orbit for the
time needed to create a gap. For the simulated cases of inward
migration the initial density gradient given is sufficient to start
rapid migration.

In most simulations the initial planet mass is such that
$M_\rmn{P}/M_\rmn{S} = 0.001$ (i.e. one Jupiter mass, $M_{\jupiter}$,
for a one-solar-mass star). We investigate the migration of the
planets with $M_\rmn{P}/M_\rmn{S}$ equal $0.0007$ and $0.0013$ too.
The planet always starts on a circular orbit of semi-major axis equal
$3.0$.

The aspect ratio for the disc with respect to the star is fixed at
$h_\rmn{s} = 0.05$, whereas the circumplanetary disc aspect ratio
$h_\rmn{p}$ lies in the range $0.4$ to $0.6$.

The smoothing length of the stellar potential, $r_\rmn{soft}$, is
taken to be $0.5$. For the planet this parameter was chosen be a
$r_\rmn{soft} = 0.33 R_\rmn{H}$, where the Hill radius is given by
$R_\rmn{H} = a{[M_\rmn{P}/(3M_\rmn{S})]}^{(1/3)}$.  The
characteristic size of the region where the self-gravity correction is
applied is $r_\rmn{env}=0.5 R_\rmn{H}$.

In our simulations we use an outflow-inflow boundary condition with a
so-called killing wave zone next to the boundaries, where the solution
of the Euler equations are smoothly connected with the initial disc in
sub-Keplerian rotation. For a full description see Paper~I,
Sect.~2.3.2.

To track the flow of material through the corotation region, we use a
tracer fluid that has a value of 1 in the (initial) corotation region
($a_\rmn{init} - 2R_\rmn{H}) < r < (a_\rmn{init} + 2R_\rmn{H})$, and
zero outside it. This allows us to distinguish between gas captured by
the planet in the horseshoe region and gas flowing through the
corotation region.

\section{Numerical method}
\label{sect_num_mth}

For our simulations we use the {\it Flash} hydrodynamics code version
2.3 written by the {\it FLASH Code Group} from the ASC / Alliance
Center for Astrophysical Thermonuclear Flashes at the University of
Chicago\footnote{http://flash.uchicago.edu} in 1997. {\it Flash} is a
modular, adaptive-mesh, parallelised simulation code capable of
handling general compressible flow problems. It is designed to allow
users to configure initial and boundary conditions, change algorithms,
and add new physics modules. It uses the {\it Paramesh} library
\citep{2000CPC.126..330} to manage a block-structured adaptive mesh,
placing resolution elements only where they are needed most.

For our purpose, the code is used in the pure hydrodynamical mode in
two dimensions, and the refinement criteria are set to achieve high
resolution around the planet. Our simulations use a lowest resolution
mesh of $800$ cells in each direction, and a square region around the
planet is refined. The maximal cell size in the disc (lowest level of
refinement) is 0.01, the minimal cell size is 0.00125 (4 levels of 
refinement, in practice corresponding to 1.8\% of the minimum Hill
sphere radius).

Further details about the numerical method are given in Paper~I.

\section{Inward migration~-~standard case}

\label{inward_mig}

In this section we describe the inward migration of our standard case:
a Jupiter-mass planet $M_\rmn{P} = 0.001$ with the circumplanetary
disc aspect ratio $h_\rmn{p}=0.4$ in a disc with $\mu_\rmn{D} = 0.005$
and $\alpha_\rmn{\Sigma} = -1.0$; the disc has no outer edge.  The
surface density is taken to be initially constant inside the
gravitational softening for the star, giving a small jump in
$\Sigma_\rmn{init}$ at $r=0.5$. However, this jump is too small and
located too close the star to influence the planet's migration.  In
the simulation the effective gravitational mass of the planet was
increased by the mass content within the smoothing length.

\begin{figure*}
\includegraphics[width=84mm]{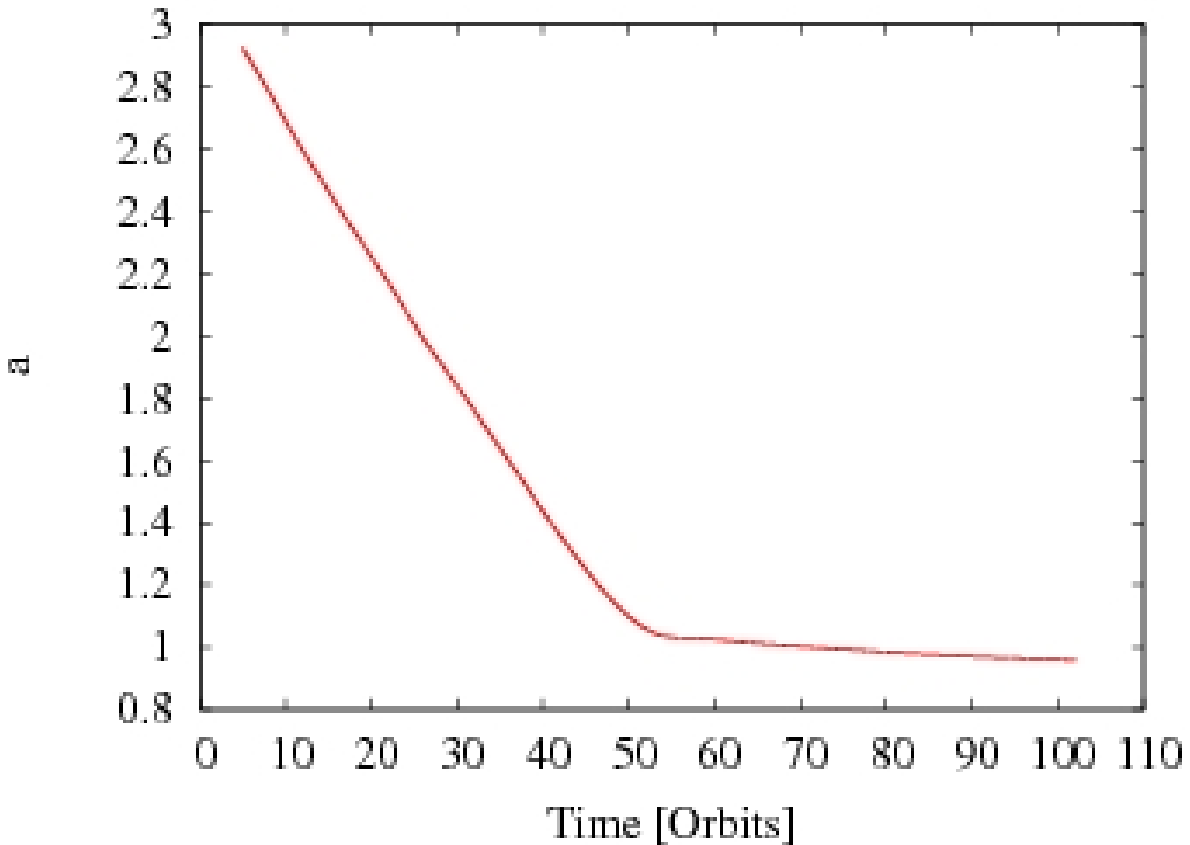}
\includegraphics[width=84mm]{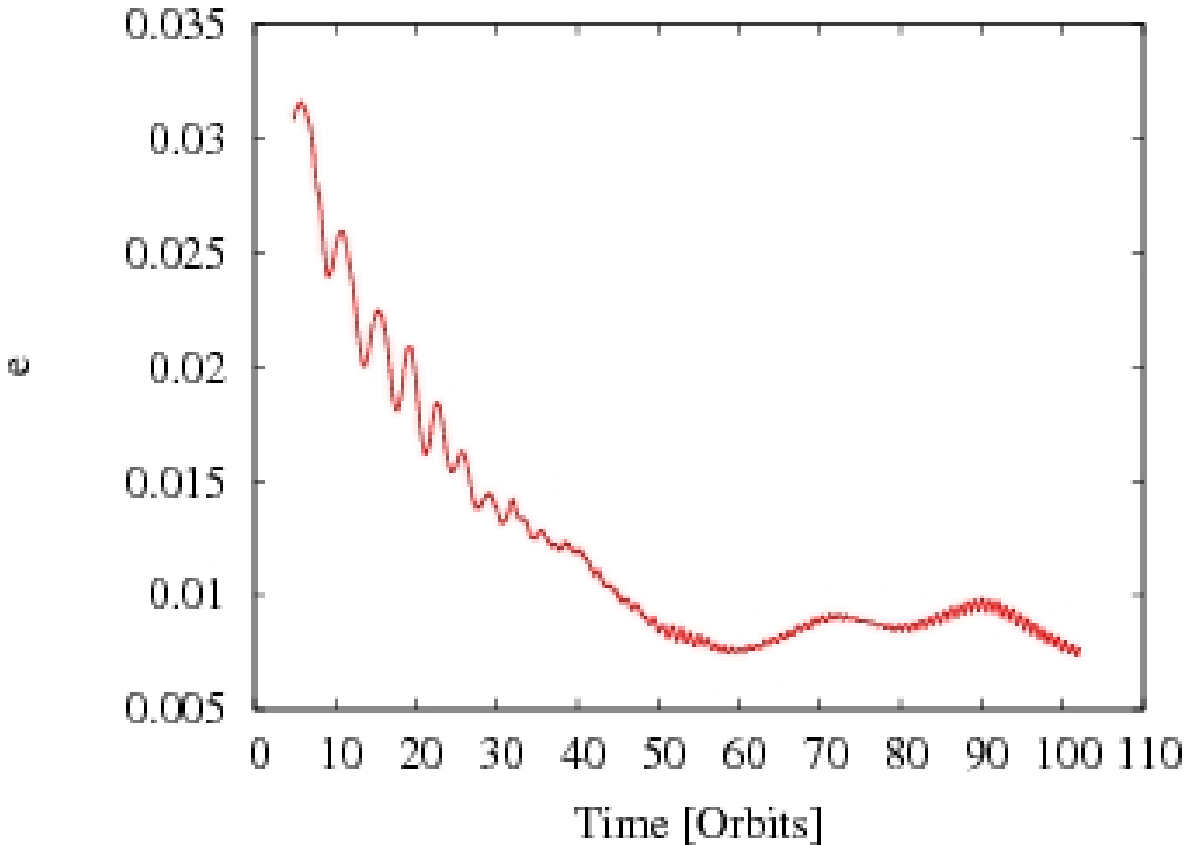}
\includegraphics[width=84mm]{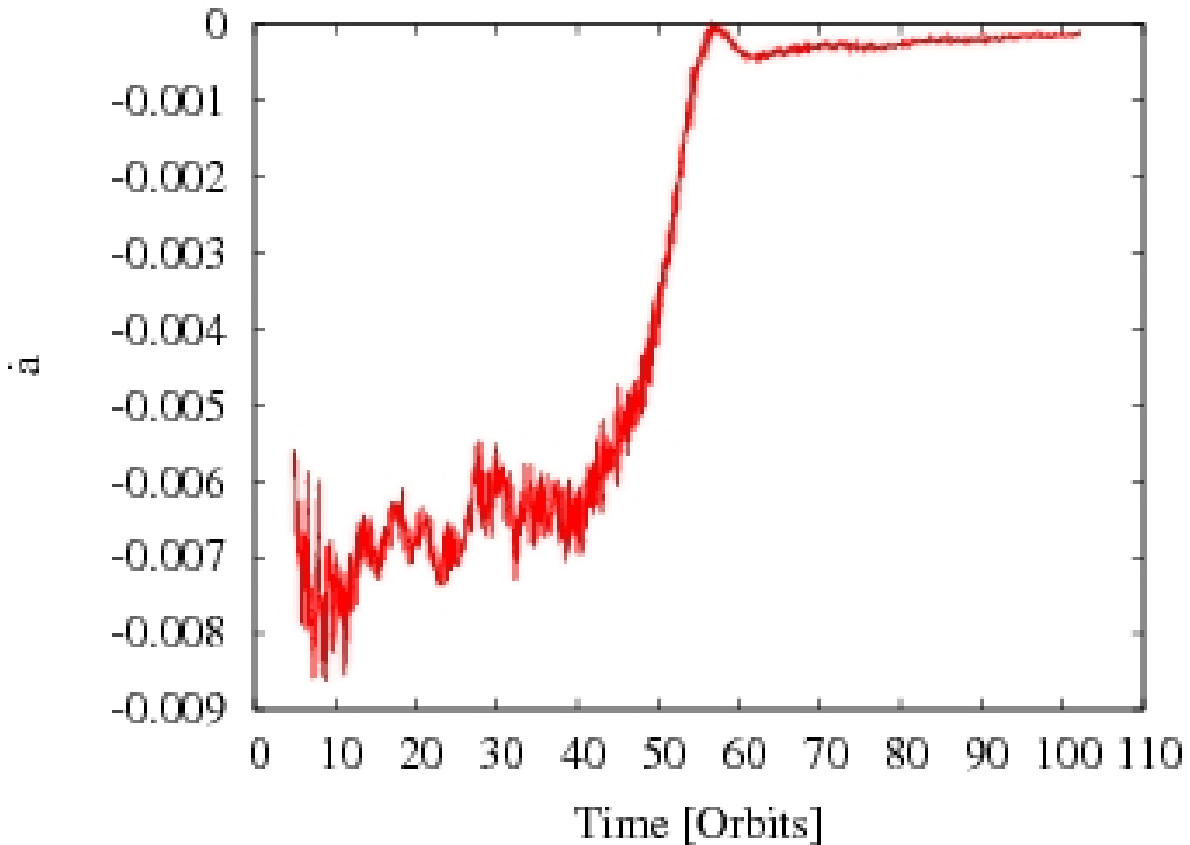}
\includegraphics[width=84mm]{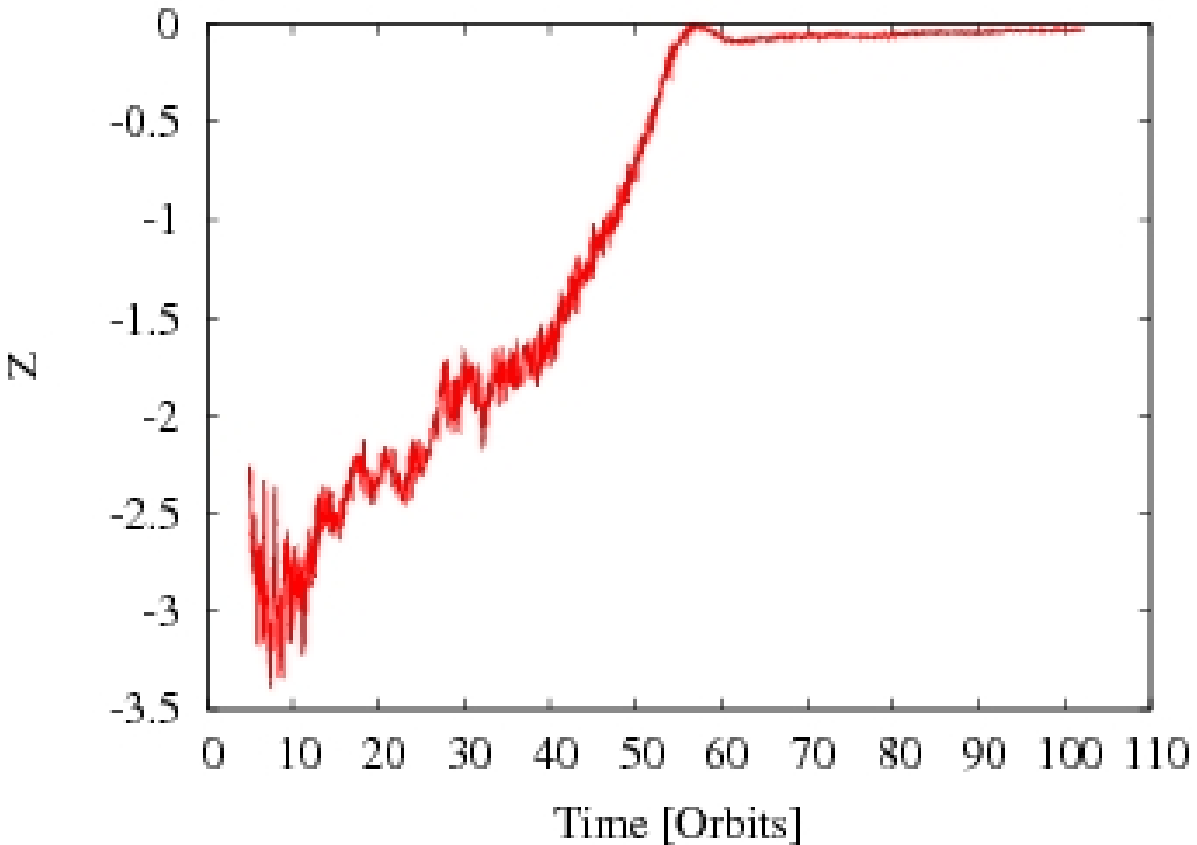}
\includegraphics[width=84mm]{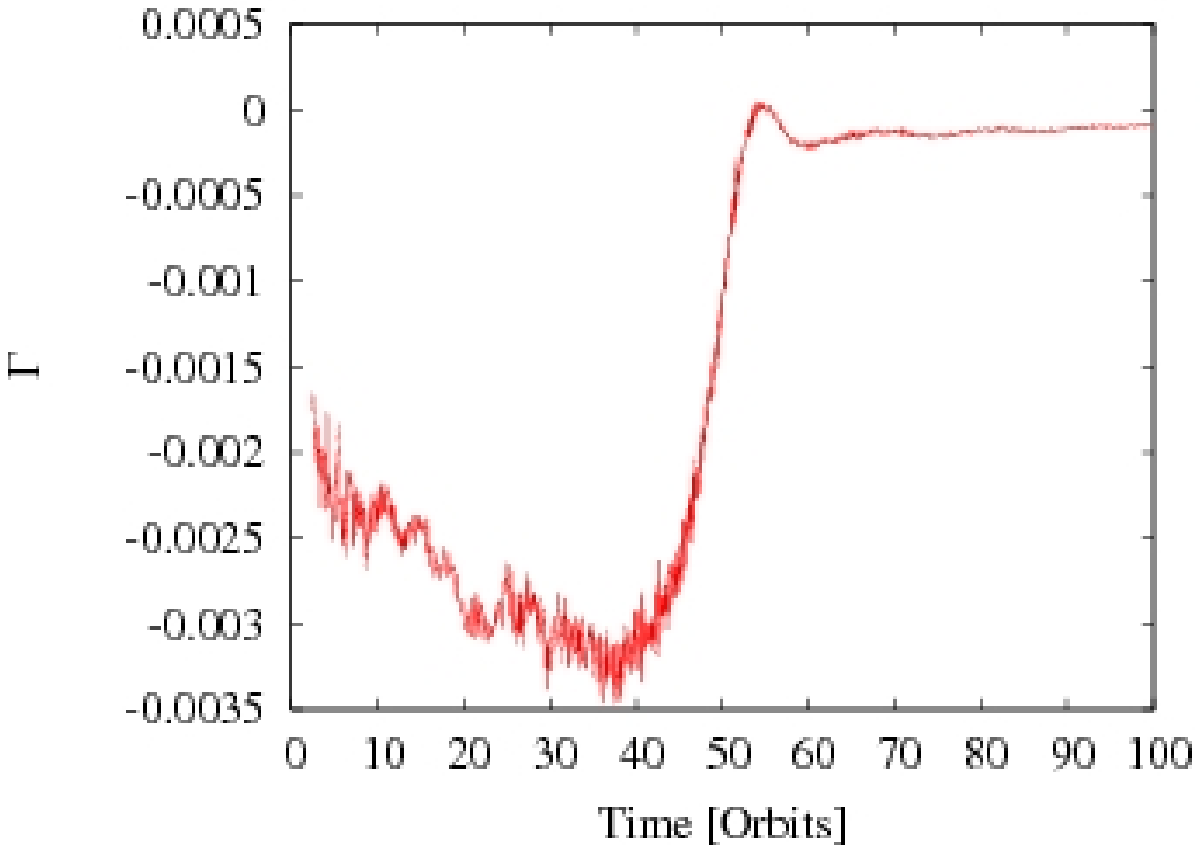}
\includegraphics[width=84mm]{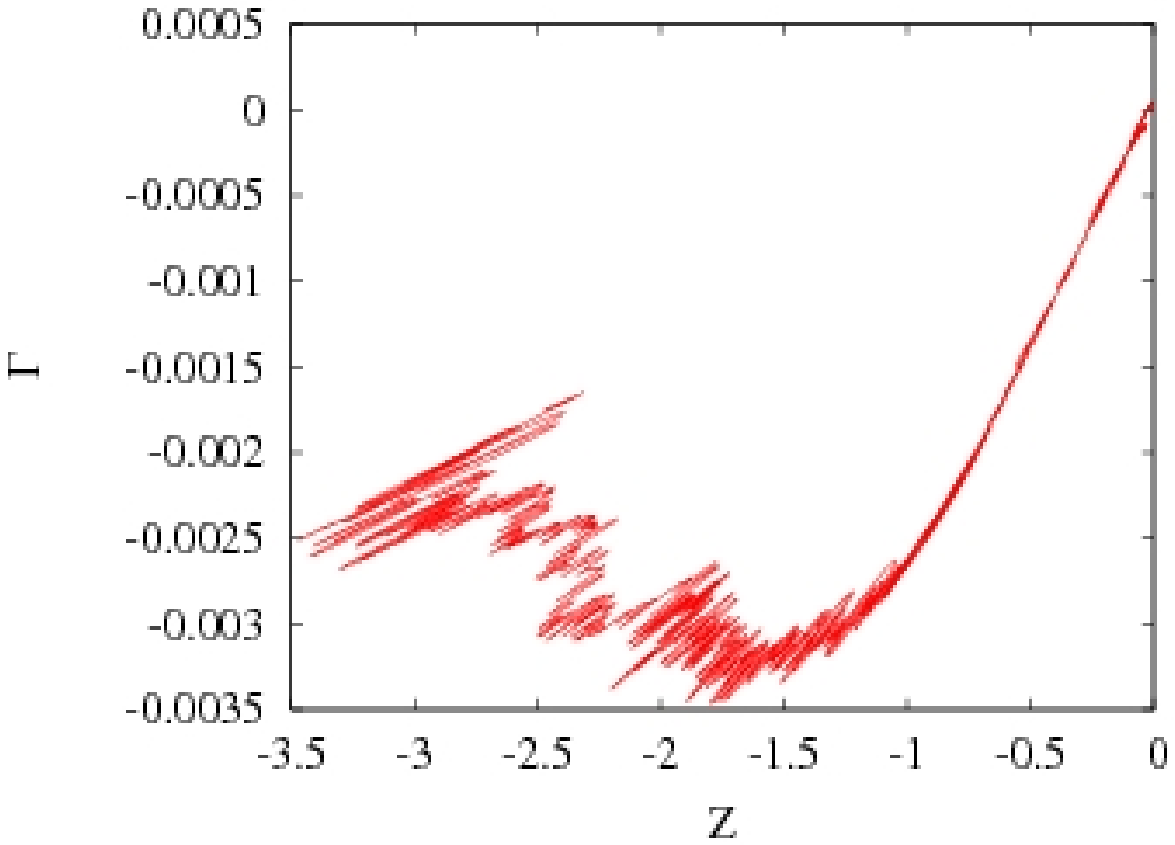}
\caption{Orbital evolution for the standard case. The upper left and
 upper right panels show the evolution of the planet's semi-major axis
 $a$ and eccentricity $e$. The evolution of the migration rate $\dot
 a$ and the non-dimensional migration rate $Z$ are presented in middle
 left and middle right panels.  The lower row shows the total torque
 $\Gamma$ exerted by the gas on the planet. Left and right panels
 present the torque as a function of time and as a function of the
 non-dimensional migration rate $Z$, respectively.}
\label{fim1_a}
\end{figure*}

\subsection{Orbital evolution}

The planet's orbital evolution is shown in
Fig.~\ref{fim1_a}\footnote{All plots in this paper present data
averaged over 5 periods.}.  Looking at the semi-major axis (upper left
panel) we can divide the planet's orbital evolution into two
stages. The first one is the {\it rapid migration stage}, which lasts for
about $60$ orbits and corresponds to the type III migration
regime. During this stage, the planet migrates rapidly due to the
co-orbital flow and is not able to clear a full gap. The migration
rate $\dot a$ (middle left panel) is initially close to constant
around $-0.0065$, but after $t=40$ orbits, $|\dot a|$ starts to
decrease rapidly. This behaviour is caused by the strong mass
accumulation in the planet's proximity (see
Sect.~\ref{sect_loc_flow_str} and Fig.~\ref{fim3_mp}). As the
migration slows down some of this mass is lost again, giving $\dot a$
close to zero near $t=55$ orbits. The dimensionless migration rate $Z$
(middle right panel) shows essentially the same behaviour but since
$\dot a_\rmn{f} \sim a^{-0.5}({\widetilde
M}_\rmn{P}/M_\rmn{S})^{2/3}$, it decreases even when $\dot a$ is
almost constant. The important transition from the fast ($|Z|>1$)
to the slow ($|Z|<1$) migration limit takes place at about
$47$ orbits.

The migration slows down as $a=1$ is approached. In the second stage
(after $60$ orbits) the planet migrates slowly enough to allow a gap
to open up. To avoid confusion with the concept of the slow migration
limit (defined as $|Z|<1$), we will refer to this second stage as the
{\it gap opening stage}. The $|\dot a|$ decreases slowly and is equal
$1.5 \times 10^{-4}$ at $t=100$ orbits, giving a migration time-scale
$\tau_\rmn{M} = a/\dot a$ equal $1.3 \times 10^4$~yr. Note that this
is not classical type II migration, as the total torque is not
dominated by the the differential Lindblad torque. However, as the
corotation torque $|\Gamma|$ slowly decreases with time, the
system may ultimately make the transition to type II migration.

The time evolution of the total torque $\Gamma$ exerted by the gas on
the planet is plotted in the lower left panel. In the rapid migration
stage the migration is driven by the corotation torque
$\Gamma_\rmn{CR}$ and the total torque $\Gamma$ is dominated by
$\Gamma_\rmn{CR}$, the differential Lindblad only playing a negligible
role. Unlike the migration rate, $|\Gamma|$ grows during the first
$40$ orbits, and then rapidly drops due to the strong mass
accumulation in the planet's proximity and the transition from the
fast to the slow migration limit. The corotation torque is
closely related to the migration rate, as illustrated in the lower
right panel. The plot shows $\Gamma$ versus the non-dimensional
migration rate $Z$. What is most notable is that the total torque
$|\Gamma|$ grows linearly with $|Z|$ for the slow migration limit. In
the fast migration limit the relation between $Z$ and $\Gamma$ is
non-linear and the torque saturates around $Z \approx
-1.5$. $|\Gamma|$ reaches its maximum value at $Z\approx -1.7$. It is 
consistent with Fig.~9 in \citet{2003ApJ...588..494M}.

In the gap opening stage $\Gamma$ evolves toward zero (it is equal
$-1.8 \times 10^{-4}$ and $-1.1 \times 10^{-4}$ at $75$ and $100$
orbits respectively), but is still larger than the predicted
differential Lindblad torque. This is caused by a small mass outflow
from the interior of the Roche lobe, which increases
$|\Gamma|$. In models where mass accretion on the planet is
allowed, this mass outflow is much more limited and the final torque
is an order of magnitude smaller.

The upper right panel in Fig.~\ref{fim1_a} shows the changes of the
eccentricity $e$.  The eccentricity grows rapidly during the first few
orbits and reaches 0.03, after that decreasing with time. In the gap
opening stage $e$ becomes $e<0.01$. We postpone further discussion of
the eccentricity evolution to Sect.~\ref{sec_ecc_ev}.

\begin{figure*}
\includegraphics[width=84mm]{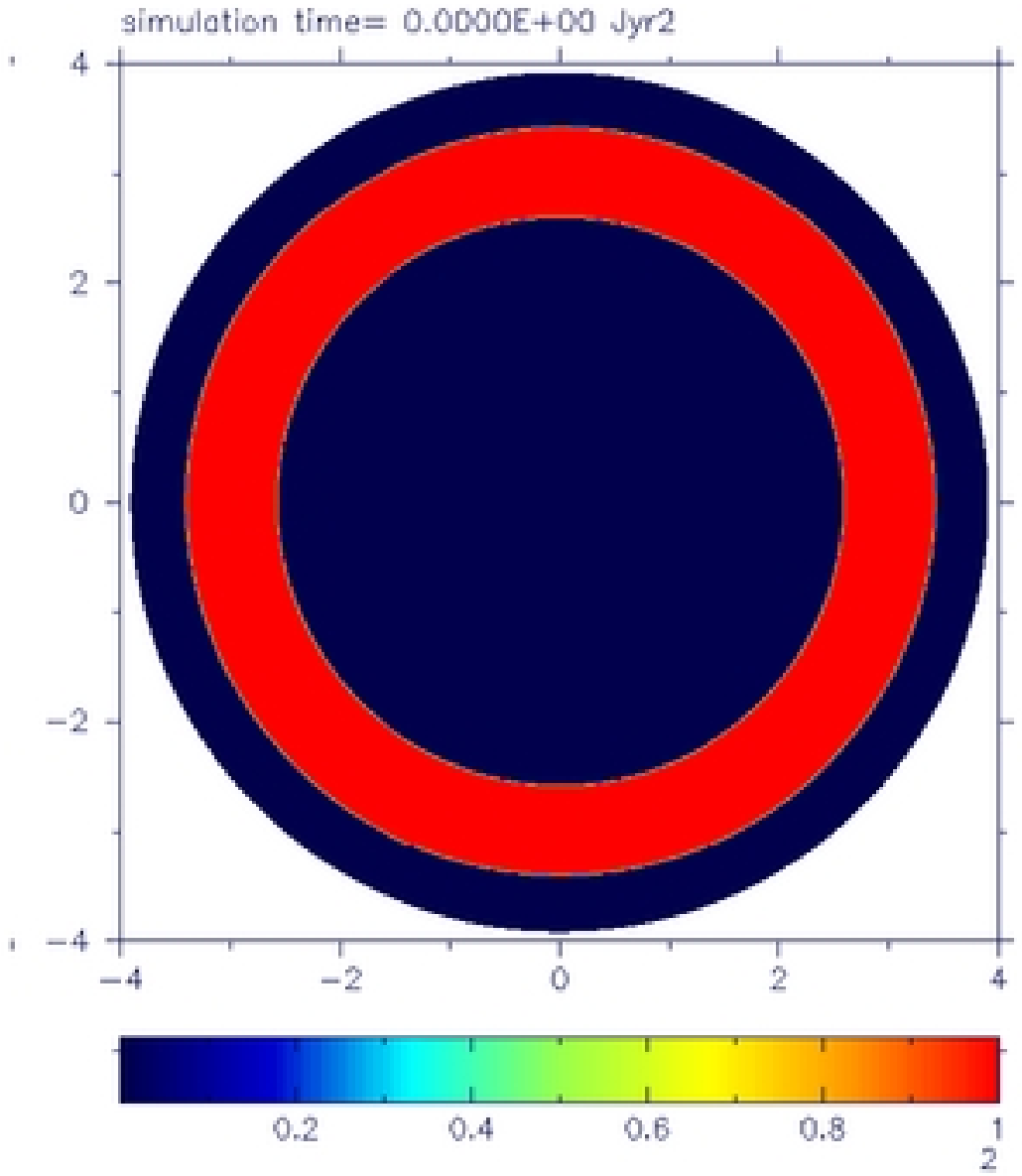}
\includegraphics[width=84mm]{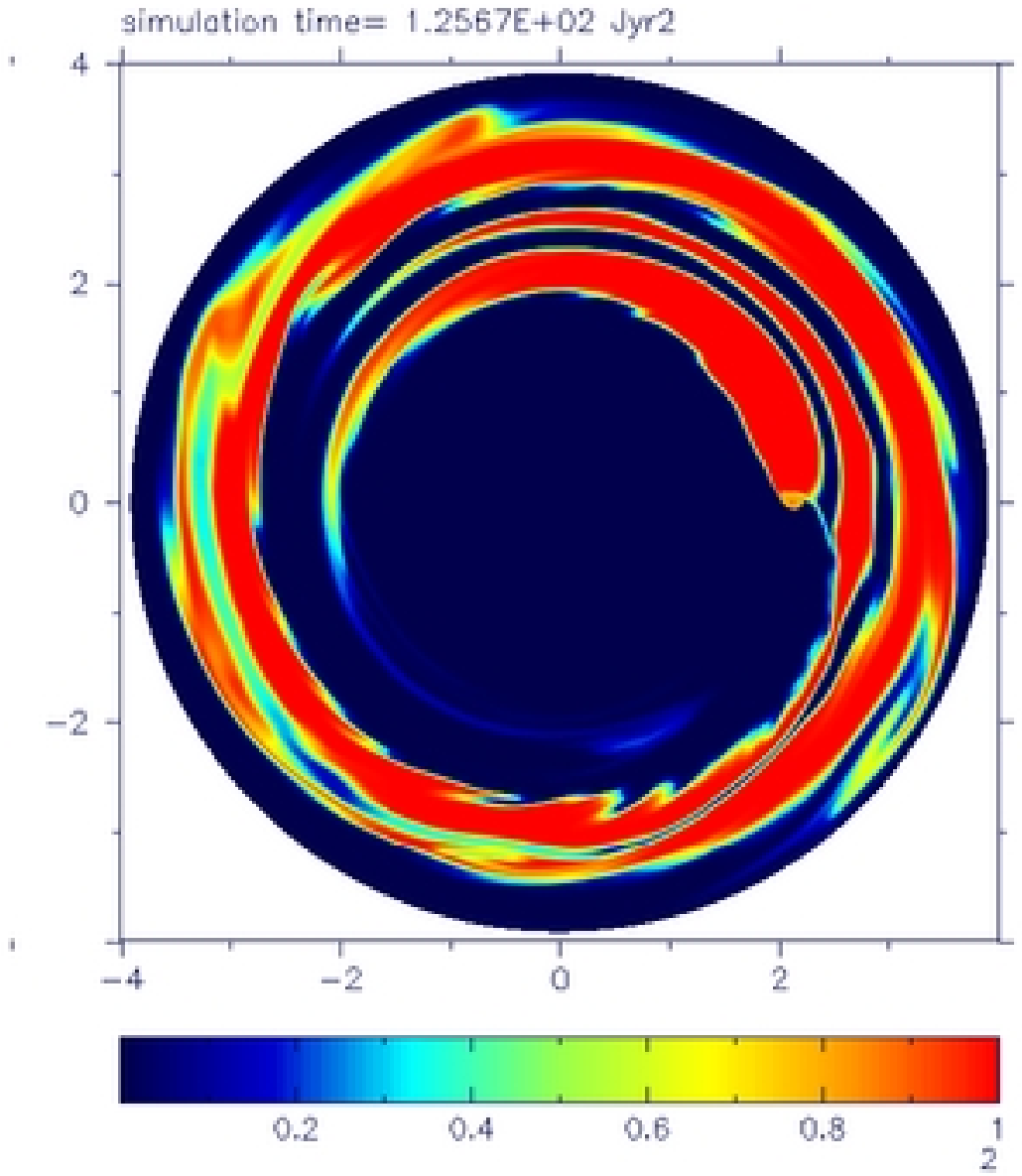}
\includegraphics[width=84mm]{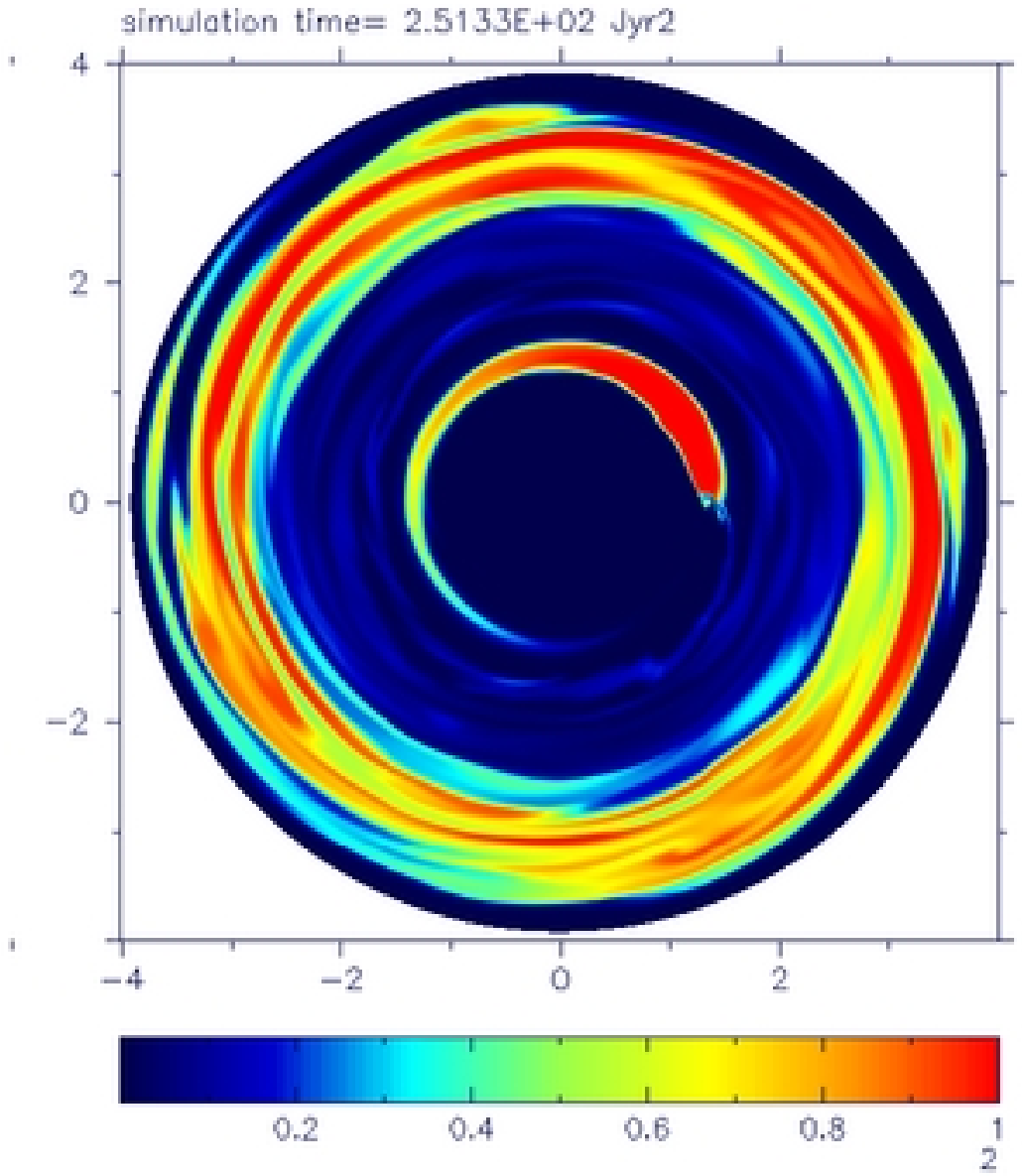}
\includegraphics[width=84mm]{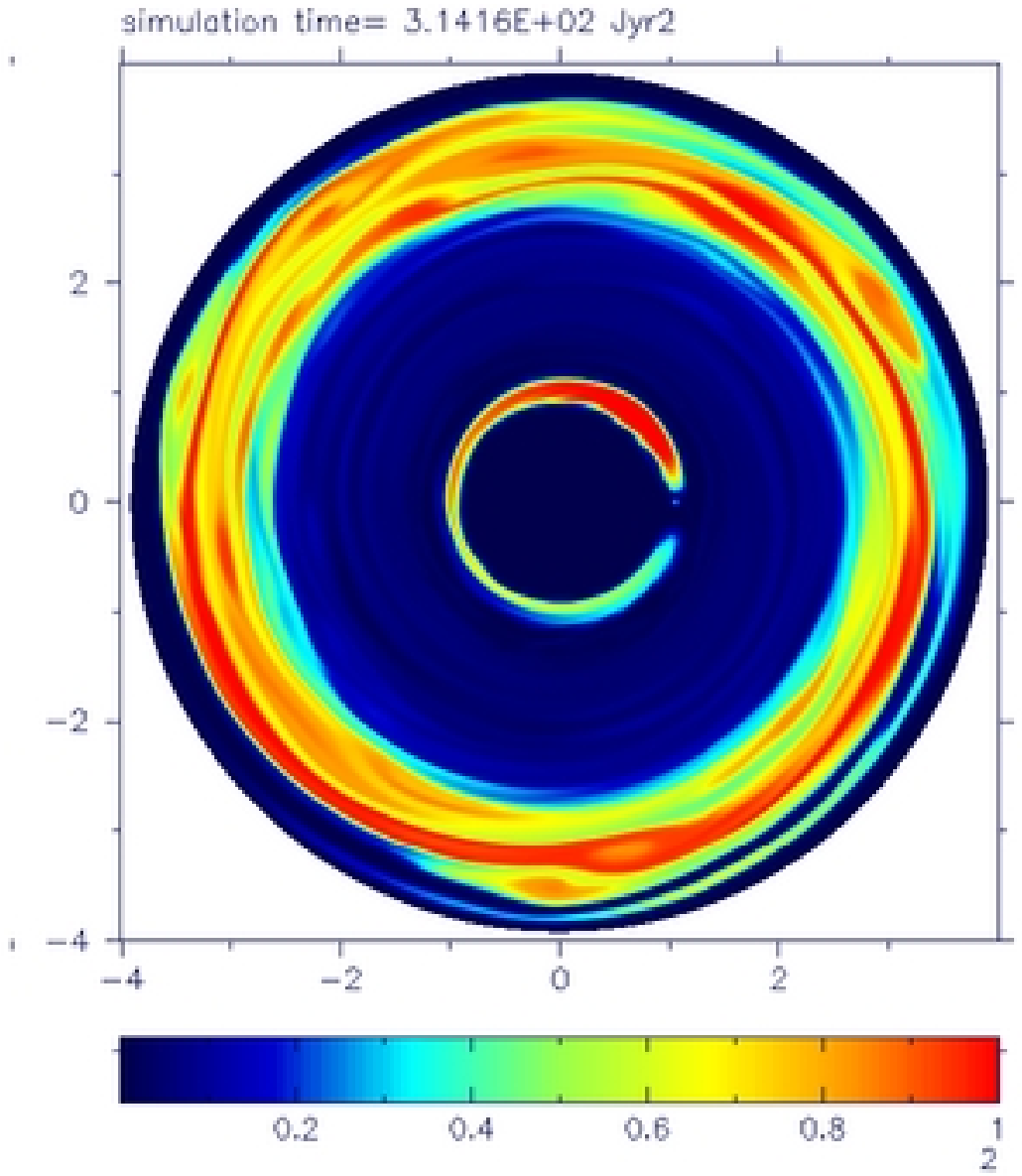}
\includegraphics[width=84mm]{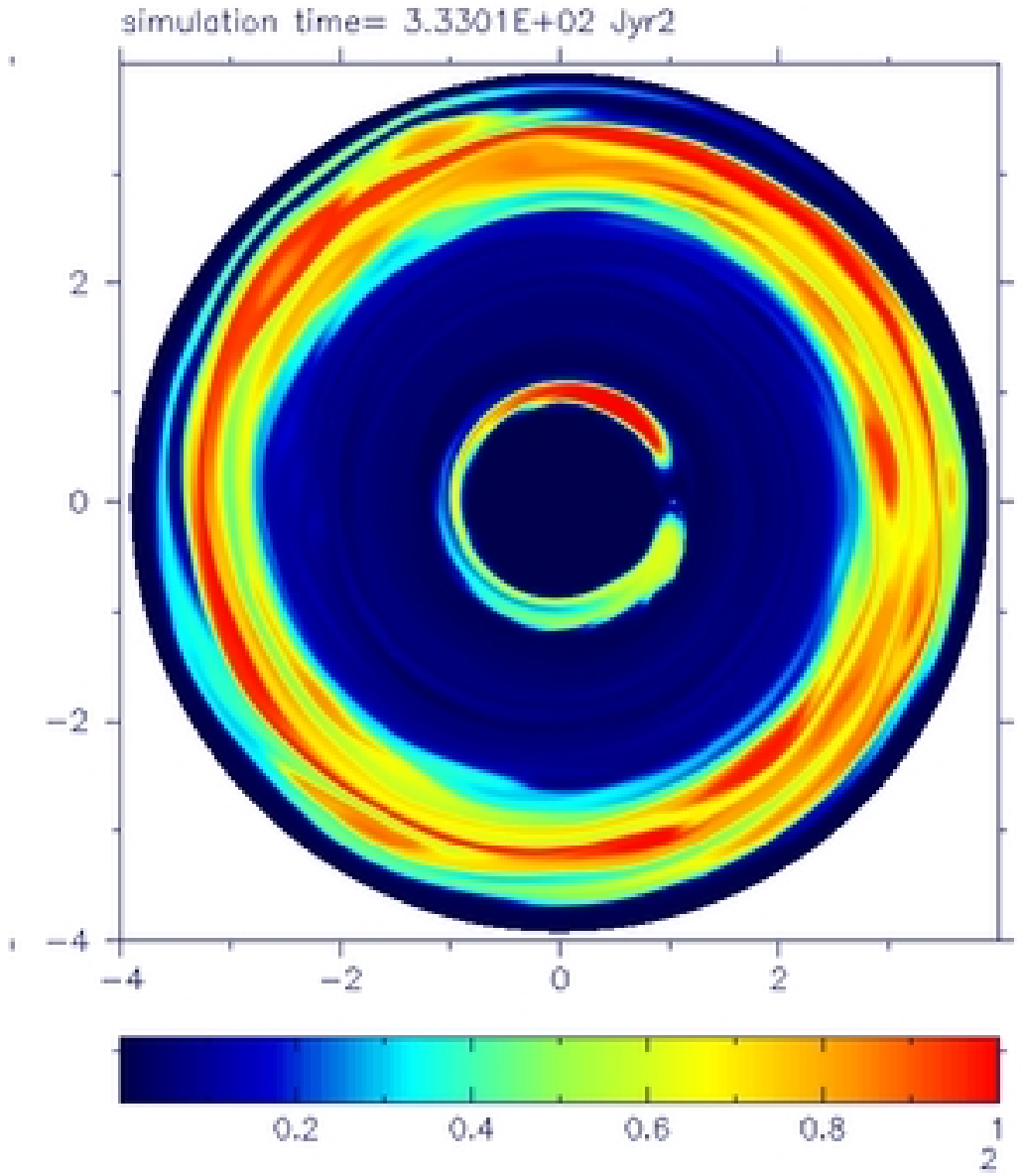}
\includegraphics[width=84mm]{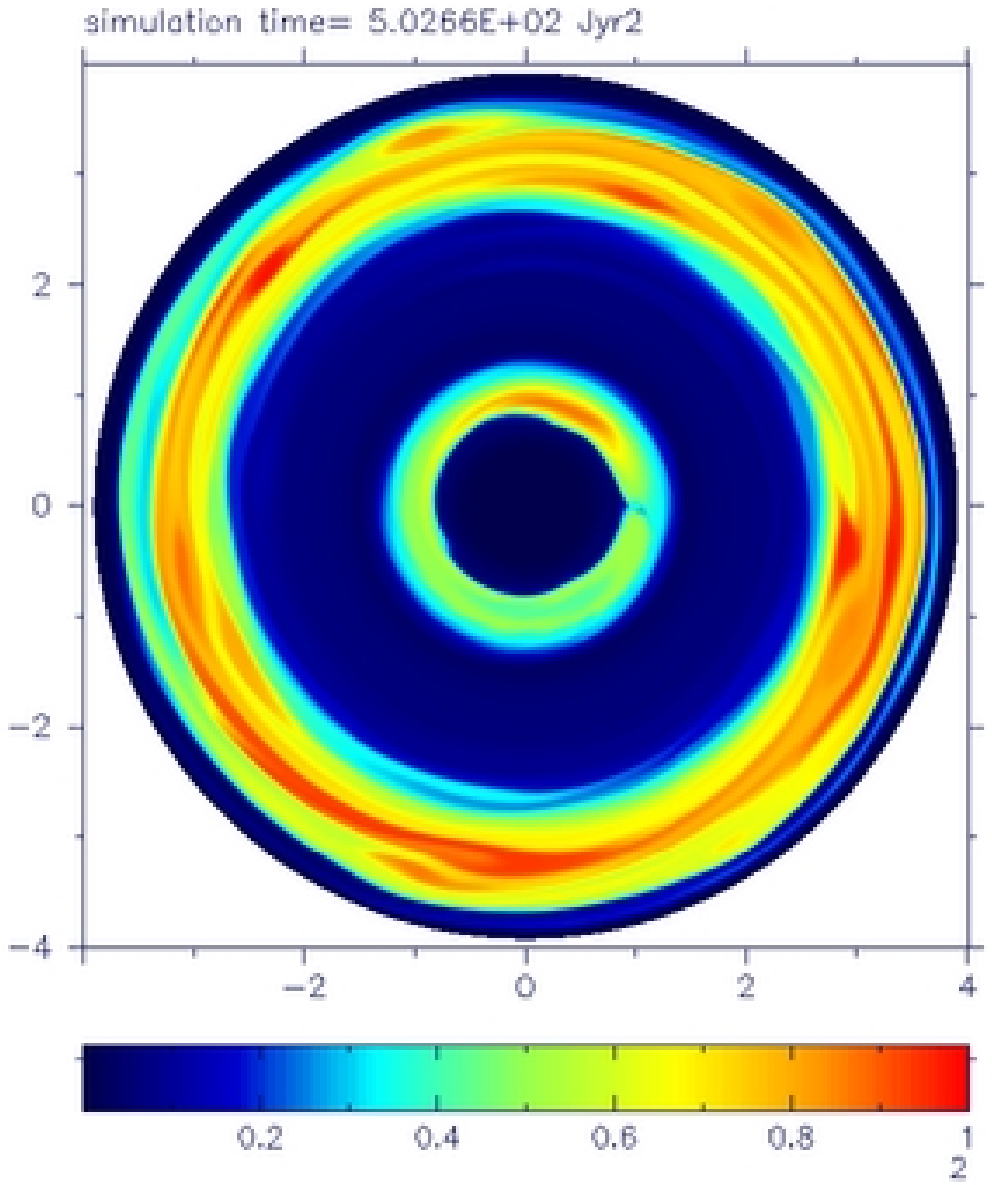}
\caption{The mass fraction of the gas that initially was placed in the
corotational region $a_\rmn{init}-2 R_\rmn{H} < r_\rmn{s}< a_\rmn{init}+2 R_\rmn{H}$,
where $r_\rmn{s}$ is the distance to the star and $a_\rmn{init}$ is
the planet's initial semi-major axis. The mass fraction goes from $0$ 
(gas from the inner disc
only) up to $1$ (gas from corotation only). The top left panel shows
the initial condition. The top right ($t=20$ orbits, $Z\approx-2.35$)
and a middle left ($t=40$ orbits, $Z\approx-1.4$) present the fast
migration limit. The middle right ($t=50$ orbits, $Z\approx-0.4$) and
a bottom left ($t=53$ orbits, $Z\approx-0.05$) show the slow migration
limit. The last panel displays the mass fraction during the gap
opening stage ($t=80$ orbits, $Z\approx-0.046$).  All plots are in a
co-moving reference frame with the planet at $(x,y)=(3,0)$,
$(2.25,0)$, $(1.44,0)$, $(1.1,0)$, $(1.05,0)$ and $(0.99,0)$
respectively.}
\label{fim8_comp_disc}
\end{figure*}

\subsection{Flow structure in the co-orbital region}
\label{in_flow_str}

In this section we discuss the relation between the corotational
torque $\Gamma_\rmn{CR}$ and the asymmetry of the horseshoe
region. The latter is connected to the the non-dimensional migration
rate $Z$, since $Z$ expresses the ratio of the migration to libration
time scales. In the slow migration limit ($|Z| < 1$) $T_\rmn{migr}$
is longer than $T_\rmn{lib}$ and the horseshoe region can extend
the full $2\pi$ in azimuth. For $|Z| > 1$ the horseshoe
region shrinks into a single tadpole-like region. Once this happens
the rest of the co-orbital region is taken up by a flow from the inner
to the outer disk. We will denote this flow as well as the region
it goes through as the {\it co-orbital flow}.

In the simulation $Z$ evolves from $-3$ to $0$ (middle right panel in
Fig.~\ref{fim1_a}), and thus the morphology of the co-orbital region
should show an evolution from tadpole to a regular horseshoe shape. To
study this we use the tracer fluid to follow the gas that was placed
in the initial corotational region ($a_\rmn{init}-2 R_\rmn{H} <
r_\rmn{s} < a_\rmn{init}+2 R_\rmn{H}$). Figure~\ref{fim8_comp_disc}
shows the evolution of this tracer fluid, with the top left panel
displaying the initial condition.

The top right and middle left panels present the mass fraction
during the fast migration phase for $Z=-2.35$ ($t=20$ orbits) and
$Z=-1.4$ ($t=40$ orbits). Due to the planet's fast inward migration,
most of the gas placed initially in the co-orbital region is left in
its initial position and only a small fraction of mass is captured by
the planet in the horseshoe region. The gas flow there is strongly
asymmetric and the region shrinks to a single tadpole-like region
around the L4 libration point. It has relatively sharp edges and the
gas captured by planet almost does not mix with the gas crossing the
co-orbital region. The azimuthal extent of the horseshoe region grows
slowly with decreasing $|Z|$ and is about $\pi$ and $1.2\pi$ for $Z$
equal $-2.35$ and $-1.4$ respectively. This means that the co-orbital
flow takes up a wide range in azimuth, that is only weakly
dependent on $Z$. Consequently, the mass flux of the gas crossing
corotation depends mostly on the initial density profile, and the
co-orbital torque $\Gamma_\rmn{CR}$ is a linear function of
$M_\rmn{\Delta}$, almost insensitive to the changes of $Z$. We should
keep in mind that $M_\rmn{\Delta}$ is not a really an independent
variable, since it depends on the way the horseshoe region is
populated, which depends on the migration history and thus indirectly
on the previous values of $Z$.

The volume of the co-orbital region is given by $V_\rmn{CR}=4\pi a
x_\rmn{s}$, and thus decreases for inward migration, as can be seen in
Fig.~\ref{fim8_comp_disc}. This is important since the mass deficit is
a function of $V_\rmn{CR}$ and of the difference in the density flux
between the co-orbital flow ($\Sigma_\rmn{s}$) and the horseshoe
region that moves together with the planet ($\Sigma_\rmn{g}$, also see
Eq.~\ref{eqn_m_delta_small}).  For a constant value of
$\Sigma_\rmn{s}-\Sigma_\rmn{g}$, the torque would thus decrease as
$V_\rmn{CR}$, i.e.~ as $a^2$.  A slow down of inward migration can
thus be caused by this simple geometrical reason. We discuss this and
other stopping mechanisms in Sect.~\ref{stop_mig}.

During the entire phase of fast migration, there is a small amount of
the gas leaving the horseshoe region. This is visible in
Fig.~\ref{fim8_comp_disc} as a light-blue leading spiral.

As $|Z|$ drops, the horseshoe region grows in azimuth and the
co-orbital flow region shrinks. For $|Z| \approx 1$ the horseshoe
region fills the whole co-orbital region, and the simulation enters
the slow migration limit. In the presented simulation the surface
density at $5.2$~AU is equal $523$~g~cm$^{-2}$ and $\Sigma \sim
r^{-1}$, and the transition between fast and slow migration limit
happens when the planet is at $6.2$~AU.  In the slow migration limit
phase the flow through the corotation is limited to a narrow stream at
the boundary of the Roche lobe and the asymmetry in the horseshoe
region is relatively small. However, this small asymmetry is sensitive
to small changes in $Z$, giving a strong dependence of
$\Gamma_\rmn{CR}$ on $Z$. As the flow becomes more symmetric both
globally (shape of the horseshoe orbits) and locally (shape of orbits
in the planets vicinity; see Sect.~\ref{sect_loc_flow_str}), $\Gamma$
diminishes rapidly.

The middle right and bottom left panels present the mass fraction for
the slow migration limit for $Z=-0.4$ ($t=50$ orbits) and $Z=-0.05$
($t=53$ orbits).  In this phase the planet's radial motion is slow and
a gap is being cleared.  Although the azimuthal extent of the
horseshoe region is close to $2\pi$ for both values of $Z$, the shape
of the horseshoe streamlines differ between the two snapshots. In the
earlier case an asymmetry is still visible, but in the later one some
gas from the inner disc has been trapped in the tadpole region
around the L5 Lagrange point and the asymmetry is starting to disappear. In
this phase of migration the second tadpole region is created and the
gas from the co-orbital flow mixes with the gas carried along
with the planet. On both panels there is a visible flow through the
co-orbital region, but in the second case this flow is dominated by
gas leaving the planet's vicinity. This lowers $\Gamma$ and gives
$\dot a \approx 0$ at about $55$ orbits.

The last panel displays the mass fraction after the planet has almost
opened a gap ($t=80$ orbits, $Z=-0.046$). In this case the asymmetry
of horseshoe streamlines is almost invisible, and the gas flow through
the co-orbital region is very small. We can still see a difference
between the tadpole regions around the L4 and L5 Lagrange points. Most
of the gas around L4 has been carried along with the planet, but the
gas around L5 was mostly captured during the last phase of migration.

\begin{figure*}
\includegraphics[width=84mm]{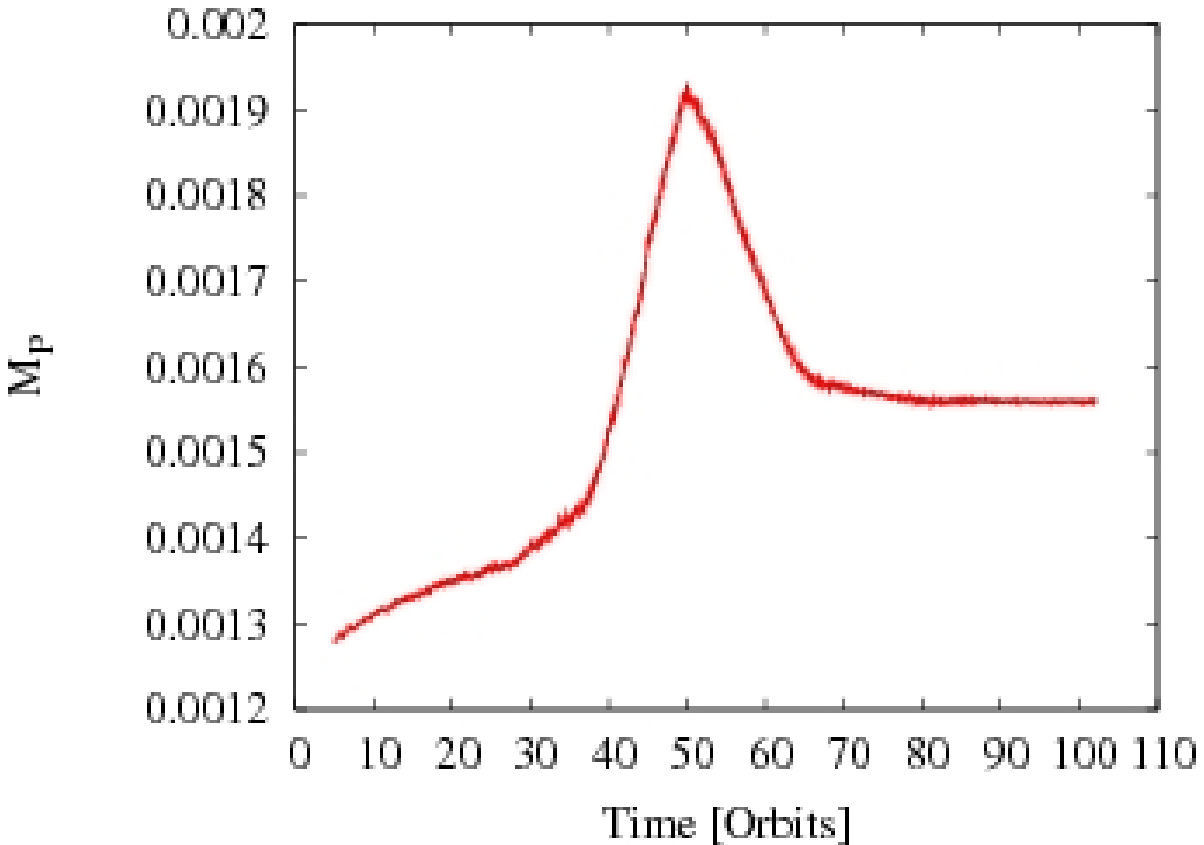}
\includegraphics[width=84mm]{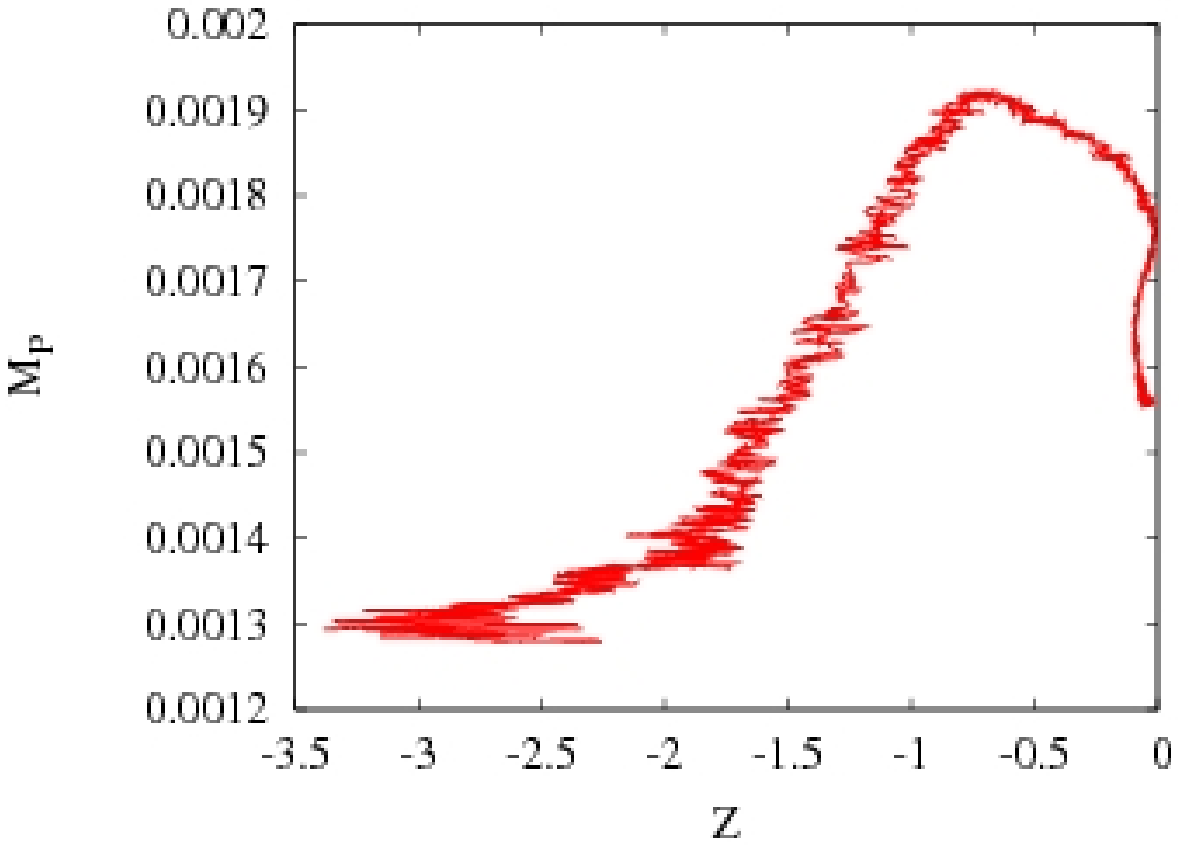}
\caption{The evolution of the effective planet mass for the standard
case. The left and right panels present the effective planet mass as
a function of time and of the non-dimensional migration rate,
respectively. The initial planet mass is $M_\rmn{P} = 0.001$. During
the first $40$ orbits the migration rate $\dot a$ is almost constant
and the rate of the mass accumulation $\dot {\widetilde M}_\rmn{P}$
remains constant too. $\dot {\widetilde M}_\rmn{P}$ increases when
migration slows down. After the planet enters the gap opening stage at
about $60$ orbits the planet loses about 14\% of its mass and at the
end of simulation ${\widetilde M}_\rmn{P} = 0.00156$.}
\label{fim3_mp}
\end{figure*}

\begin{figure*}
\includegraphics[width=84mm]{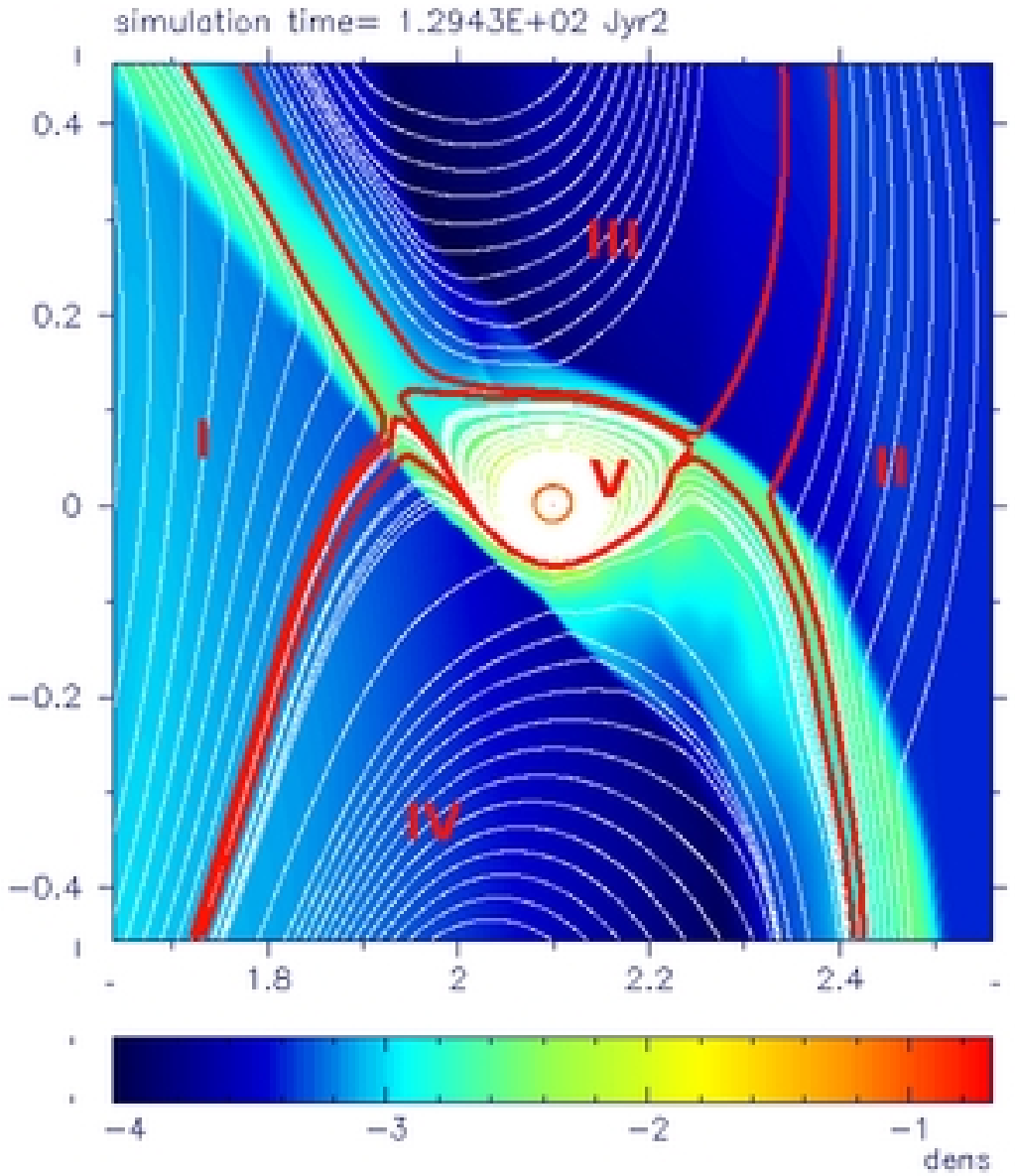}
\includegraphics[width=84mm]{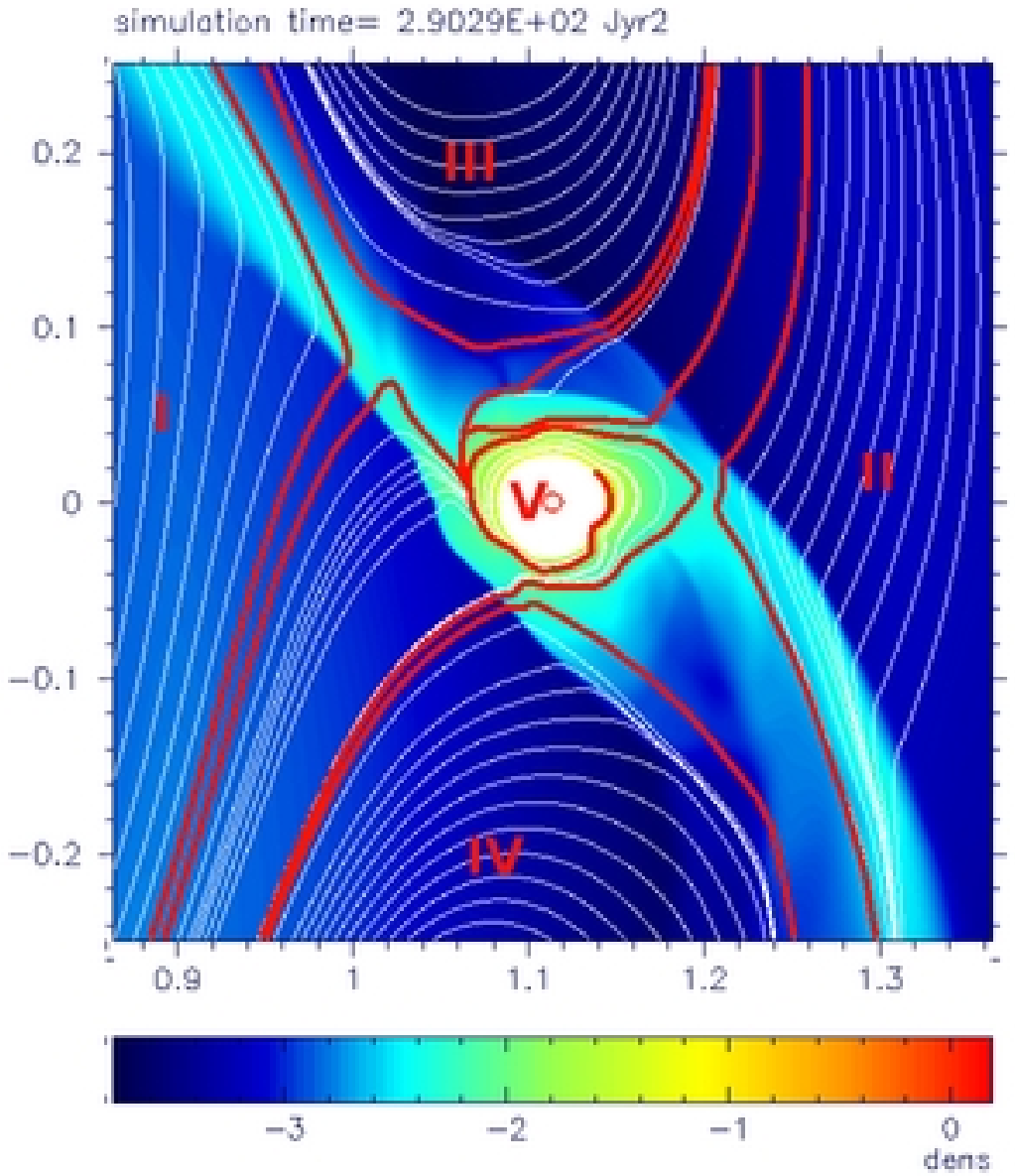}
\includegraphics[width=84mm]{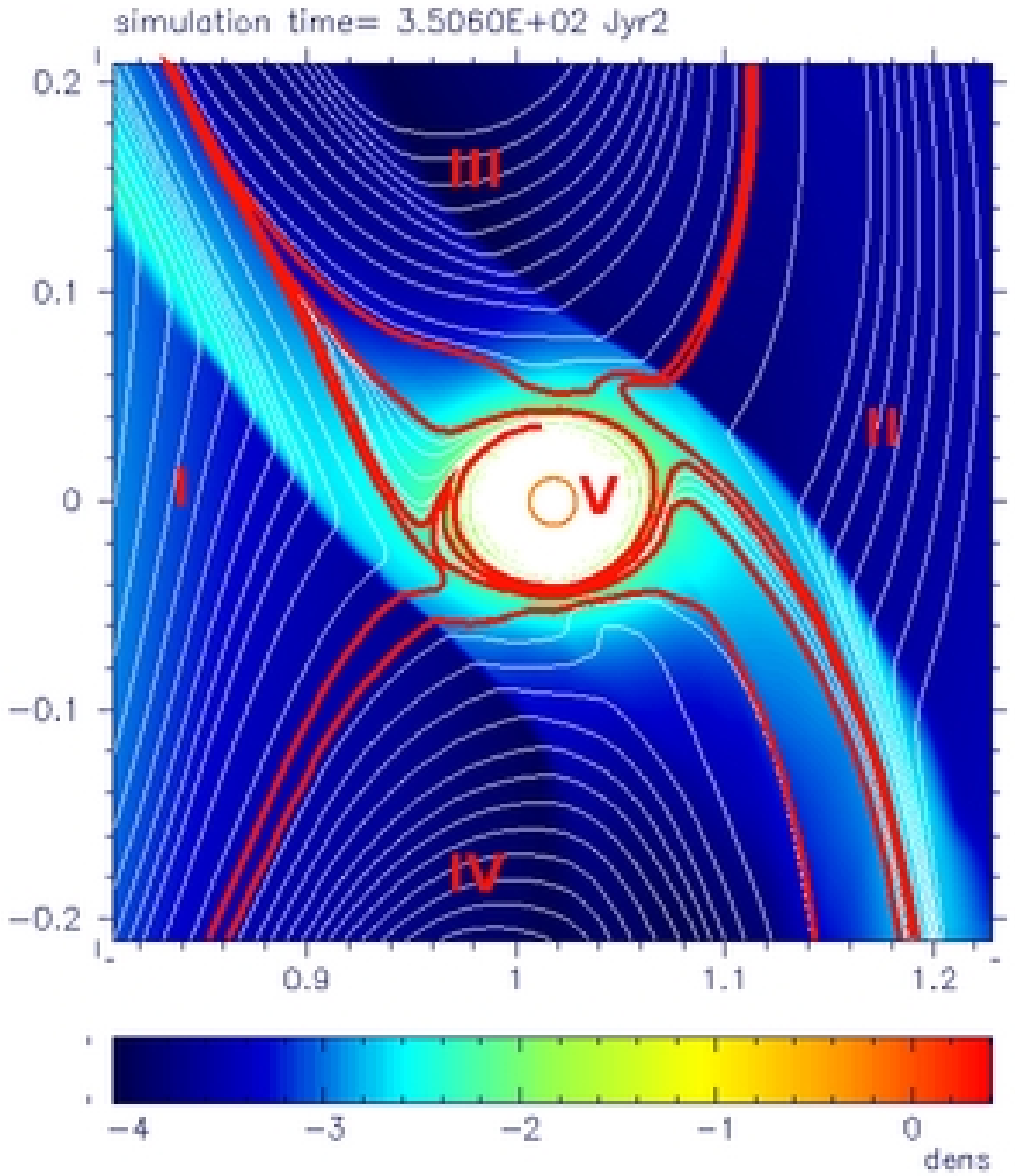}
\includegraphics[width=84mm]{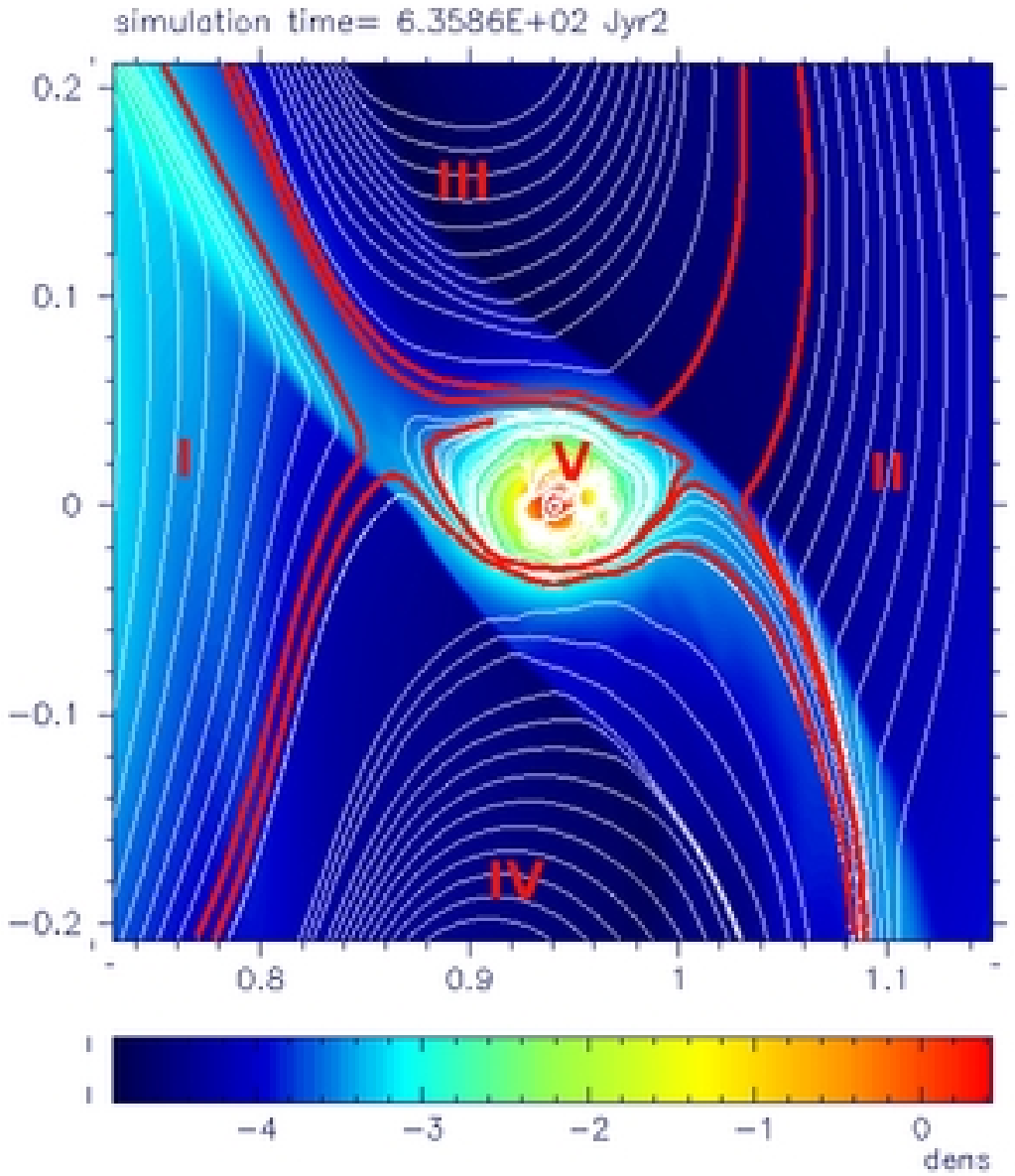}
\caption{The surface density and flow lines in the planet's vicinity
for the standard case at four different stages of migration. The upper
left panel corresponds to the fast migration limit with $Z \approx
-2.35$, $t=20$ orbits. The upper right and lower left panels show
slower migration ($Z \approx -1.1$, $t=46$ orbits and $Z \approx
-0.05$, $t=56$ orbits) with the circumplanetary disc quickly gaining
and losing mass respectively. The last panel (lower right) presents
the stage of the gap opening planet ($|Z| \ll 1$, $t=100$ orbits) with
an almost constant mass of the circumplanetary disc.  The colour scale is
logarithmic. The plotted domain is square region of the size of $6
R_\rmn{H}$. The flow lines close to the border of different regions
are made more visible. There are five regions marked in the plots:
inner (I) and outer (II) disc, the co-orbital region (III and IV) and
the circumplanetary disc (V). On the upper panels the regions III and
IV denote the horseshoe and co-orbital flow regions, respectively.}
\label{fim99_flow_local}
\end{figure*}

\begin{figure*}
\includegraphics[width=84mm]{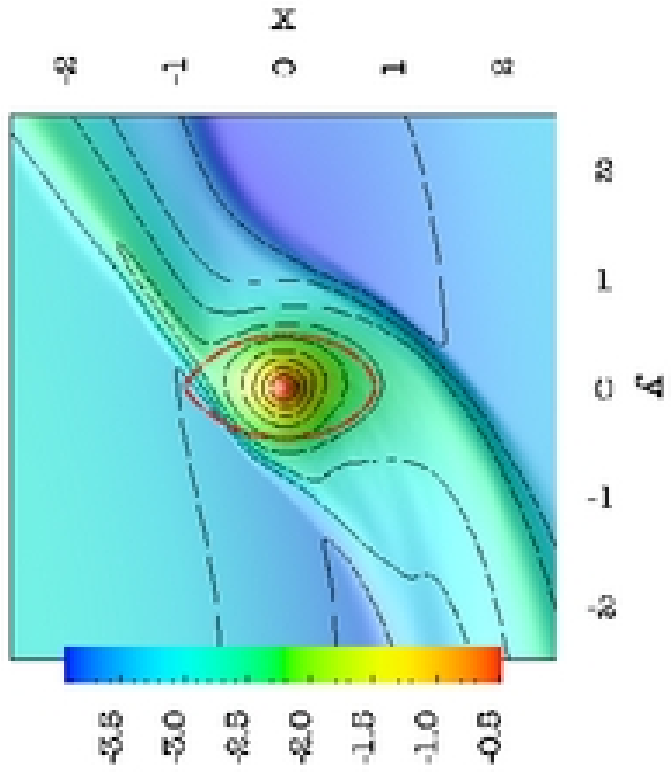}
\includegraphics[width=84mm]{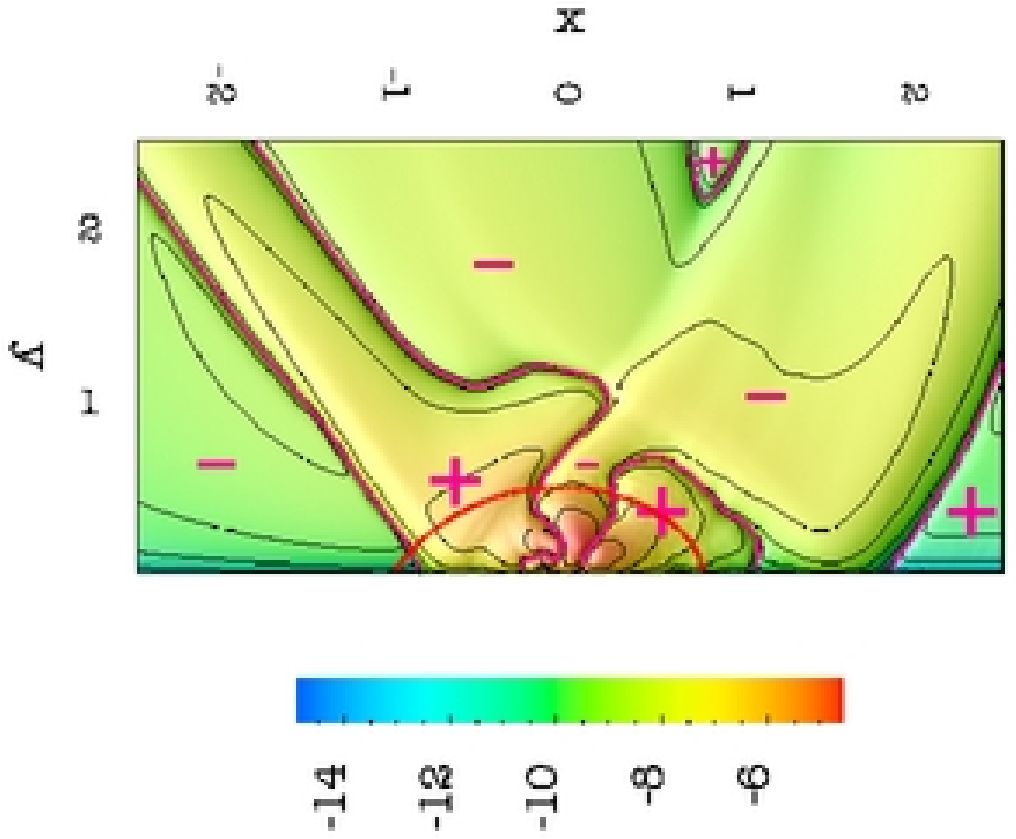}
\caption{The density distribution and the asymmetry in the torque
distribution in the planet's vicinity for the fast migration limit
with the low mass accumulation rate ($Z \approx -2.35$, $t=20$
orbits). The left panel shows the surface density averaged over 5
orbits. The right panel presents the absolute value of the sum
$\Gamma_\rmn{c}(x_\rmn{p},y_\rmn{p})+\Gamma_\rmn{c}(x_\rmn{p},-y_\rmn{p})$.
The borders of the regions giving positive (+) and negative (-)
contributions to the total torque are shown by thick pink lines. The red
line shows the Roche lobe size. The plots cover $5R_\rmn{H}$ and the
planet is placed at the centre. The colour scale is logarithmic.}
\label{fim9_comp_local}
\end{figure*}

\subsection{Mass accumulation and flow structure in the planet's vicinity}
\label{sect_loc_flow_str}

In the previous section we described the relation between the
non-dimensional migration rate and the asymmetry of the gas flow in
the full co-orbital region. We will now focus on the local gas flow near
to the planet. Just as we saw for the whole co-orbital region,
the planet's fast radial motion also strongly influences the gas
evolution inside its Roche lobe, causing the circumplanetary disc to
acquire an asymmetric shape. The co-orbital flow enters the Roche lobe
and after interacting with the gas in the planet's proximity, will
partly be captured into the circumplanetary disc. We have found this
mass accumulation to be an important ingredient in our simulations of
type III migration, determining the flow structure in the planet's
vicinity and the torque generated within the Roche lobe.

In our standard simulation, the effective gravitational planet mass
was increased by the mass content within the smoothing length
$\widetilde M_\rmn{P} = M_\rmn{P}+ M_\rmn{soft}$, where the smoothing
length corresponds to the ``surface'' of the planet
\citep{PaperI}. The time evolution of $\widetilde M_\rmn{P}$ is
presented in the left panel of Fig.~\ref{fim3_mp}.  During the first
$40$ orbits the migration rate is approximately constant, as is the
mass accumulation rate. During this time the mass contained within the
smoothing length $M_\rmn{soft}$ reaches about 45\% of the initial
planet mass. The mass of the gas inside the whole Hill sphere amounts
to about 55\% of the initial mass, so most of this material is found
inside the planet's gravitational smoothing length.

The mass accumulation rate increases sharply when the planet starts to
slow down at $t=40$ orbits. In the relatively short time of $10$
orbits $M_\rmn{soft}$ increases up to 90\% of the initial planet's
mass and a strong pressure gradient forms around the planet. This
pressure gradient is supported by the gas inflow into the
circumplanetary disc. The large mass accumulation rate is made
possible by our use of a locally isothermal equation of state that
makes the circumplanetary disc aspect ratio independent of the gas
density.  For the same reason the accumulated mass leaves the Hill
sphere shortly after the rapid migration has stopped. This affects the
total torque $\Gamma$ giving the positive value of $\dot a$ during
this short period (Fig.~\ref{fim1_a}, middle panels).

When the simulation reaches the second, gap opening stage,
$M_\rmn{soft}$ remains approximately constant at 56\% of $M_\rmn{P}$.

The relation between the planet's effective mass ${\widetilde
M}_\rmn{P}$ and $Z$ is presented in right panel of
Fig.~\ref{fim3_mp}. $\dot {\widetilde M}_\rmn{P}$ is low and almost
constant during the fast migration limit phase and rapidly grows when
$Z$ reaches $-1.7$, causing the total torque to decrease (lower right
panel in Fig.~\ref{fim1_a}). The next important change takes place in
the slow migration limit phase at $Z \approx -0.6$, when the planet
starts to open a gap and $\widetilde M_\rmn{P}$ starts to decrease.
We note that the evolution of $\widetilde M_\rmn{P}$ is strongly
dependent on choice for the circumplanetary disc aspect ratio
$h_\rmn{p}$, but for the given value of $h_\rmn{p}=0.4$ the evolution
of $M_\rmn{P}$ is almost independent of the other parameters, such as
$r_\rmn{soft}$. For more discussion on this, see
Sect.~\ref{sect_dep_on_par} and Paper~I.

Since the mass accumulation rate determines the structure inside the
Roche lobe, we now discuss the flow in the planet's vicinity for four
different stages of the simulation: the rapid migration stage with and
without the fast mass accumulation in the circumplanetary disc, the
stage of mass outflow from the planet's proximity and the final gap
opening stage with an approximately constant effective planet
mass. They are presented in Fig.~\ref{fim99_flow_local}. The plots
show the surface density and the flow lines. The flow lines close to
the border of different regions have been made more visible. There are
five regions marked in the plots: inner (I) and outer (II) disc, the
co-orbital region (III and IV) and the circumplanetary disc (V). It
should be realized that the gas evolution inside the Roche lobe is
variable and we present just a few snapshots to describe important
differences between different stages of simulation.

The upper left panel of Fig.~\ref{fim99_flow_local} corresponds to the
fast migration limit with $Z \approx -2.35$ and a relatively low mass
accumulation rate $\dot {\widetilde M}_\rmn{P}$. In this case region
III is the horseshoe region and IV the co-orbital flow. The orbits in
the circumplanetary disc are found to be strongly asymmetric,
compressed at the side of the co-orbital flow and stretched at the
side of the horseshoe region. The asymmetry of the bow shocks and the
flow lines in the co-orbital region is also visible.  The whole
horseshoe region is shifted backwards with respect to the planet's
radial motion, and the left bow shock is compressed into a straight
line, whereas the right bow shock is stretched and more curved. In
this phase the mass accumulation is still rather modest; closer
investigation shows the circumplanetary disc to be fed by both regions
III and IV. The averaged density and the torque asymmetry distribution
for this stage of migration are presented in
Fig.~\ref{fim9_comp_local}.

The stage of strong mass accumulation near the transition from the
fast to the slow migration limit ($Z \approx -1.1$) is presented in
upper right panel in Fig.~\ref{fim99_flow_local}. The asymmetry of the
bow shocks and the flow lines in the co-orbital region is still
visible, however the orbits in the circumplanetary disc have become
less regular. The interior of the Roche lobe is strongly disturbed by
the gas inflow from the inner disc~I and the horseshoe region III and
the flow lines are very time variable. In spite of this, the
circumplanetary disc no longer shows the asymmetry seen in the
previous phase, which explains the lower value for the torque
$\Gamma$.

The asymmetry of the bow shocks and the circumplanetary disc
disappears in the next stage. The lower left panel shows the flow
lines near the planet during the slow migration limit phase ($Z
\approx -0.05$) and strong mass outflow. As we saw in the previous
section, in this stage the horseshoe region extends over the whole
azimuthal range and the planet starts to open a gap. The whole
co-orbital region also becomes more symmetric and regular. The gas
leaves the circumplanetary disc flowing along the bow shocks and
influences the flow lines at the border of the Roche lobe, but the
orbits inside the circumplanetary disc are regular.

The last panel (lower right) presents the stage of the gap opening
planet $|Z| \ll 1$, with an almost constant mass in the circumplanetary
disc. In this stage the gap is almost cleared and the co-orbital
region is symmetric. Although the mass inside the Hill sphere is
almost constant some outflow from the Roche lobe can still be
seen. The most important difference with the previous stages is the
shape of the flow lines in the planet's vicinity. In the previous
stages the gas was orbiting the planet creating a circumplanetary
disc, however in this last stage the large pressure gradient prevents
circular orbits in the planet's proximity. We have to note that this
stage looks different for models with an accreting planet (see
Paper~I, Sect.~5.1.2). In that case the planet is surrounded by the
Keplerian-like sub-disk and there is no gas outflow from the Roche
lobe during the last phase of migration. This is consistent with what
was found by \citet{2003ApJ...599..548D}.

After having described the flow structure we now consider the torque
distribution in the planet's vicinity. We focus on the first stage,
i.e.\ rapid migration with a low mass accumulation rate
($t=20$~orbits). This corresponds to the upper left panel of
Fig.~\ref{fim99_flow_local}. In the left panel of
Fig.~\ref{fim9_comp_local} we present the surface density averaged
over 5 orbits for $Z\approx -2$. The plot covers $5R_\rmn{H}$ and the
planet is placed at the centre. The red line shows the Roche lobe
size. The asymmetry between the left and right bow shock which we
described above, is clearly visible. Although the left bow shock is
sharper and has a higher density close to the planet, the overall
amount of mass at the right bow shock is higher, since the density
increase close to the right bow shock extends over a much wider
area. The plot also shows the averaged surface density to be lower in
the horseshoe region (positive values of $y$) than in the co-orbital
flow (negative values of $y$). This asymmetry causes the right bow
shock to have a larger contribution to the total torque $\Gamma$ than
the left one. This is further illustrated in the right panel of
Fig.~\ref{fim9_comp_local}, where we show the absolute value of the sum
$\Gamma_\rmn{c}(x_\rmn{p},y_\rmn{p})+\Gamma_\rmn{c}(x_\rmn{p},-y_\rmn{p})$,
where $\Gamma_\rmn{c}$ is the torque generated by a single grid cell
and $(x_\rmn{p},y_\rmn{p})$ is the position of the cell with respect
to the planet. The borders of the regions giving positive~(+) and
negative~(-) contributions to the total torque are shown with pink
lines. The left and the right bow shocks give positive and negative
contributions, respectively. Although the absolute value of the torque
sum is similar for both shocks, the right bow shock is much wider and
its contribution to $\Gamma$ dominates. The region between the two
shocks gives a negative contribution showing that the torque generated
by the co-orbital flow is stronger than the torque generated by the
horseshoe region.

The interior of the Roche lobe shows three extremes. The two outer
ones give a positive contribution and the middle one a negative. These
extrema are the result of the asymmetry of the circumplanetary disc,
which as we saw above, is compressed at the side of the co-orbital
flow and stretched at the side of the horseshoe region. However the
sum of these extremes nearly cancels, and the overall torque generated
within the Roche lobe is found to be negative. This once more
illustrates the point made in Paper~I, that the interior of the Roche lobe cannot be
treated as a system isolated from the co-orbital flow and that the
torques from this inner region have to be taken into account when
studying type III migration (for more discussion, see \citet{PaperI}).

We would like to repeat here that we only show one snapshot here, but
that the gas flow in the planet's vicinity is highly variable, with
changes in the shapes of bow shocks and the circumplanetary disc,
causing the large variations in $\Gamma$ seen in the lower panels of
Fig.~\ref{fim1_a}

\begin{figure}
\includegraphics[width=84mm]{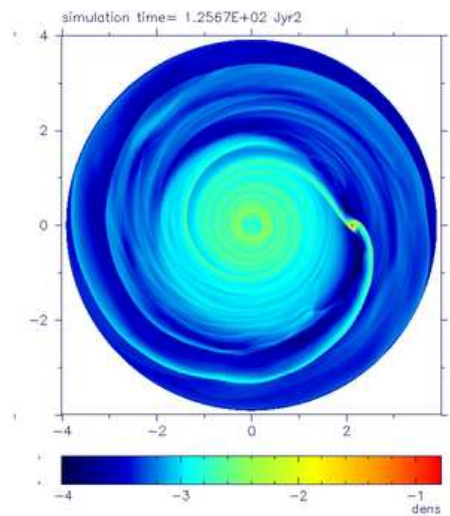}
\caption{Global surface density for the standard case at $t=20$
orbits. The planet has migrated from $a = 3$ to $a= 2$. The colour
scale is logarithmic. The planet's location is $(x,y)=(2.3,0)$}
\label{fim1_dens_disc}
\end{figure}

\begin{figure}
\includegraphics[width=84mm]{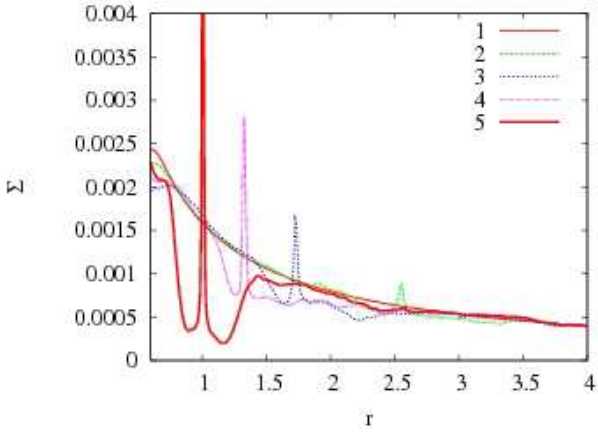}
\caption{The azimuthal average of the surface density $\Sigma$ for
times $t=0$, $10$, $30$, $50$ and $60$ orbits (curves 1, 2, 3, 4 and 5
respectively).  During the rapid migration phase no gap is formed
(curves 1 to 4). The last curve corresponds to the gap opening stage
The planet position is visible as a spike in the density profile.}
\label{fim11_sigma_time}
\end{figure}

\begin{figure}
\includegraphics[width=84mm]{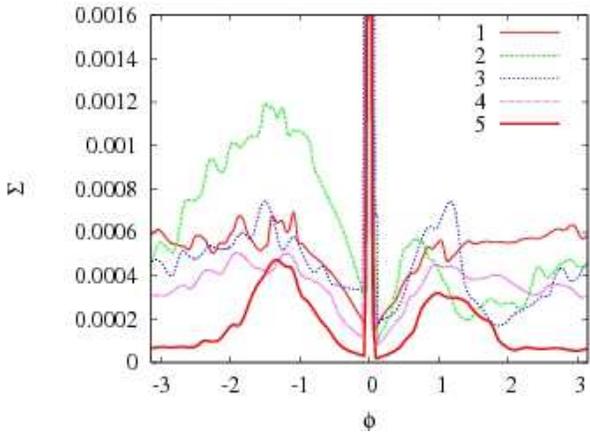}
\caption{The surface density $\Sigma$ along an azimuthal cut through
the planet position for times $t=10$, $30$, $50$, $60$ and $100$
orbits (curves 1, 2, 3, 4 and 5 respectively). The negative and
positive values of $\phi$ correspond to the co-orbital flow and
horseshoe region, respectively. The planet is located at $\phi=0$.}
\label{fim12_sigma_comp}
\end{figure}

\subsection{Surface density}
\label{sect_surf_dens}

To gain further inside into the migration process, we now study the
surface density distribution in the disc. A colour scale plot of the
global surface density distribution for the fast migration limit phase
at $t=20$ orbits ($Z\approx -2.35$) is displayed in
Fig.~\ref{fim1_dens_disc}. In this phase the planet does not yet open
a gap due to its fast radial motion. For the same reason the surface
density in the inner disc does not show significant differences from
the initial condition. There are two lower density regions at the
planetary radius corresponding to the horseshoe region (positive $y$)
and the co-orbital flow region (negative $y$).  The detailed structure
of the surface density distribution in the planet vicinity is shown
in Fig.~\ref{fim9_comp_local}.

To study the time evolution of the surface density $\Sigma$ we use a
series of plots showing this quantity in the radial and azimuthal
direction.  Fig.~\ref{fim11_sigma_time} presents the azimuthal average
of $\Sigma$ over $2\pi$ at times $t=0$, $10$, $30$, $50$ and
$60$ orbits (curves 1, 2, 3, 4 and 5 respectively). The average
surface density is weakly modified outside the planet's co-orbital
region. During the rapid migration phase (curves 1, 2, 3 and~4) no gap
is cleared, but the planet is followed by a lower density region. The
main reason for this depletion in $\Sigma$ in the planet's
neighbourhood is the mass accumulation in the planet vicinity and the
fact, that after an interaction with the planet the gas crossing the
co-orbital region is spread over a region with a radial size of a
few $x_\rmn{s}$. The last curve in Fig.~\ref{fim11_sigma_time}
corresponds to the gap opening stage ($\dot a$ approaching the value
for type II migration).

Fig.~\ref{fim12_sigma_comp} shows $\Sigma$ along an azimuthal cut
through the planet position for time $t=10$, $30$, $50$, $60$ and
$100$ orbits (curves 1, 2, 3, 4 and 5 respectively). It illustrates
the difference between the density of the horseshoe region
($\Sigma_\rmn{g}$) and the co-orbital flow region ($\Sigma_\rmn{s}$),
that drives type III migration. The region of
co-orbital flow and the horseshoe region are found at the negative and
positive values of $\phi$, respectively. The fast migration limit
phase is represented by curves 1 and 2. Initially the asymmetry in
$\Sigma$ is only located inside the Roche sphere (only barely visible
on the plot due to adopted scale), but in time it grows in azimuth
showing a large difference in $\Sigma$ between the L4 and L5 Lagrange
points. During the fast migration limit phase the L5 point is
moved in the direction of the planet (compare the position of the
density maximum for positive $\phi$ in curves 2 and 5). The asymmetry
decreases in the slow migration limit phase (curve 3) and becomes
unimportant when a gap starts opening up (curves 4 and 5). During the
whole simulation the surface density has maxima at the L4 and L5
points and decreases in the direction of the planet.

\begin{figure*}
\includegraphics[width=84mm]{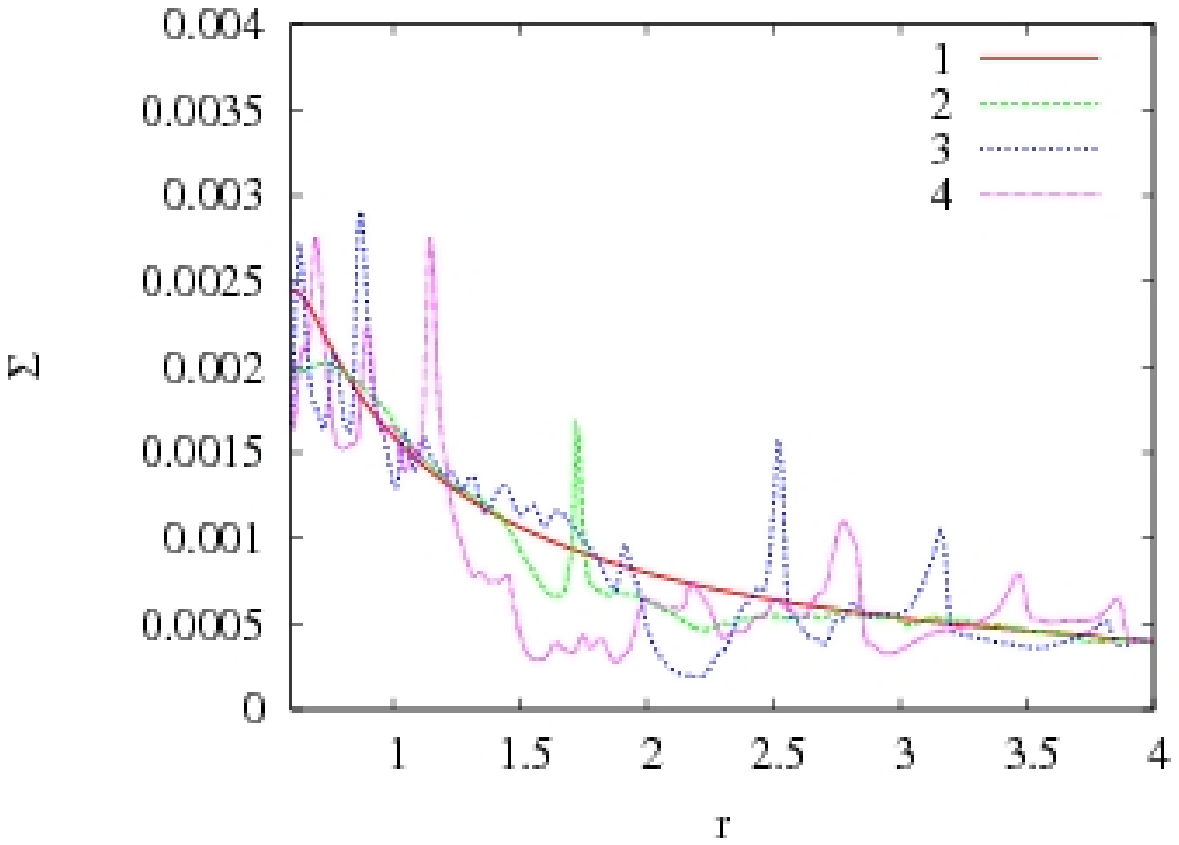}
\includegraphics[width=84mm]{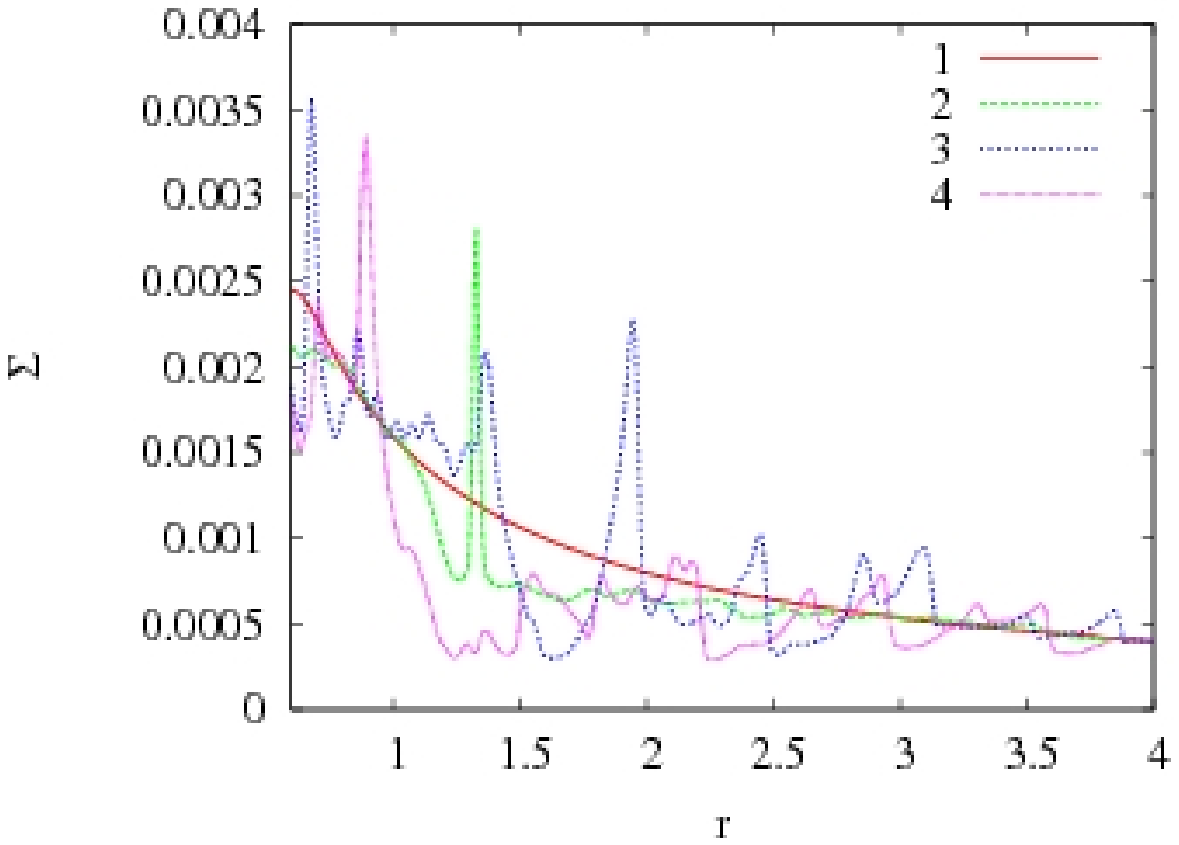}
\includegraphics[width=84mm]{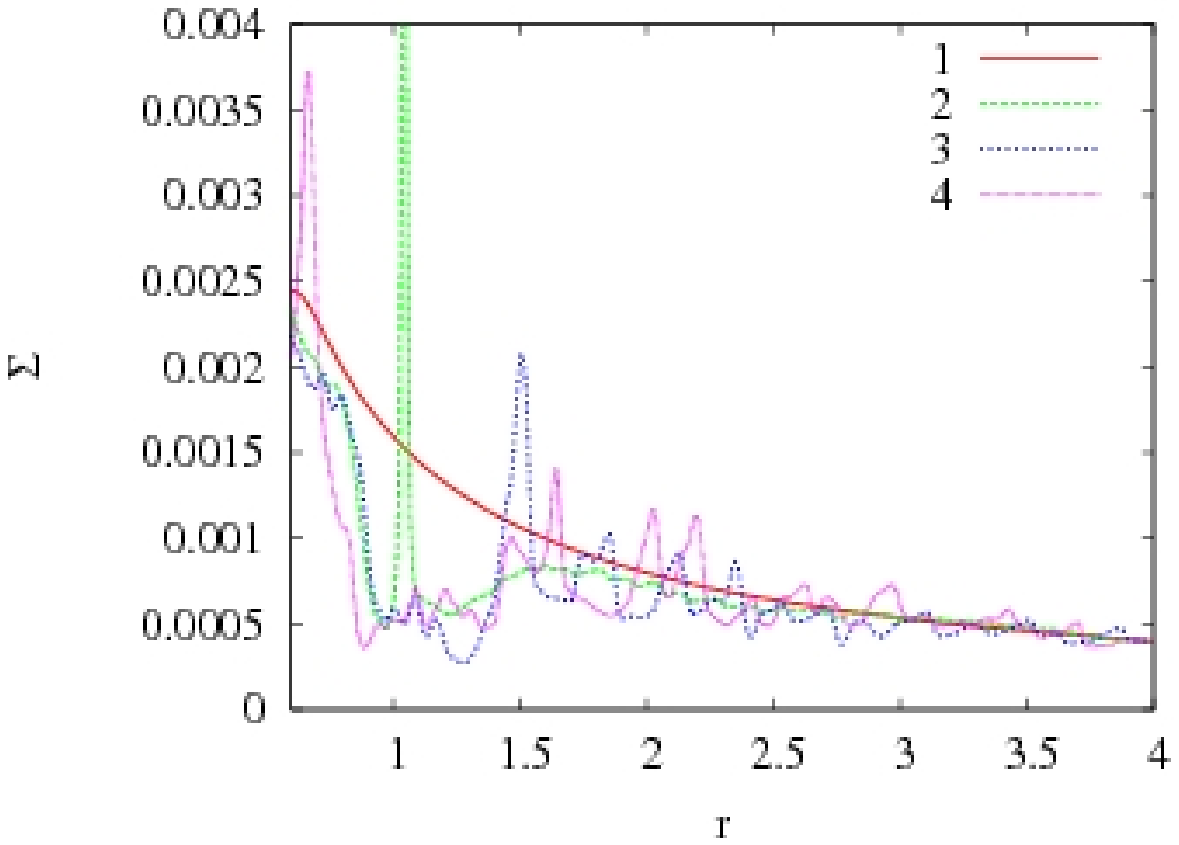}
\includegraphics[width=84mm]{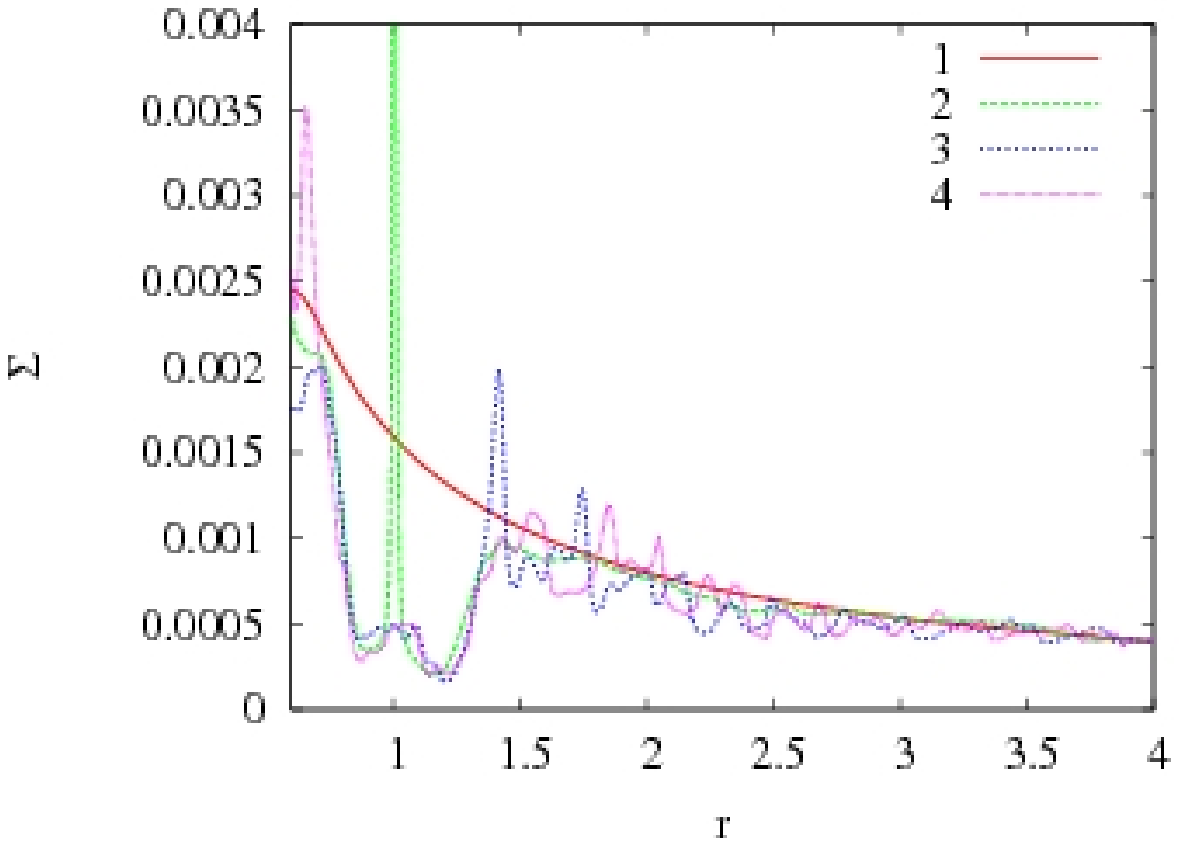}
\caption{The evolution of the density profiles in the standard
case. Each panel shows the initial density profile (curve 1), the
azimuthal average of the surface density (curve 2; the planet position
is visible as a spike) and the surface density cuts through the
Lagrange points L5 (curve 3) an L4 (curve 4). The panels upper left,
upper right, lower left and lower right show the density profiles
after $30$, $40$, $50$ and $60$ orbits, respectively.}
\label{fstop_sigma_no_edge}
\end{figure*}

To further illustrate the asymmetry, Fig.~\ref{fstop_sigma_no_edge}
shows radial surface density cuts through L5 (curve 3) and L4 (curve
4), together with the azimuthal average of the surface density (curve
2). Curve 1 shows the initial density profile. The upper left and
right panels show these quantities during the fast migration limit
phase ($t=30$ and $40$ orbits, respectively) and the lower left and
right panels the same during the slow migration limit phase ($t=50$
and $60$ orbits).  The cut through L4 gives the approximate width of
the co-orbital region and the density $\Sigma_\rmn{g}$ inside the
horseshoe region. Similarly the cut through L5 gives the approximate
value of the density of the gas crossing the co-orbital region
$\Sigma_\rmn{s}$. The difference between $\Sigma_\rmn{s}$ and
$\Sigma_\rmn{g}$ is the main reason for the rapid migration.

In the fast migration limit the planet carries along the low density
horseshoe region which has well defined, sharp edges and a width
$x_\rmn{w}\approx 2 R_\rmn{H}$.  $\Sigma_\rmn{g}$ is lower than the
initial density at the initial planet position $r=3$ and almost does
not change in time. In contrast, $\Sigma_\rmn{s}$ is close to the
initial density profile at the current planet's
position. $\Sigma_\rmn{g}$ and $\Sigma_\rmn{s}$ differ from each other
in the whole co-orbital region. The cut through L5 shows also the low
density region following the planet. It is caused by the spreading of
the gas over a few $x_\rmn{w}$ and compression at the two shocks in
the outer disc region (see Fig.~\ref{fim1_dens_disc}).

In the slow migration limit (lower left panel) the density around L5
drops significantly, as the horseshoe region gets more extended and
the co-orbital region becomes more symmetric. At the same time
the density inside the horseshoe region increases slowly. The asymmetry 
between the density
cuts through L4 and L5 is now only visible at the inner edge of the
co-orbital region. In the last, gap opening stage of migration $Z \ll
1$ (lower right panel) the co-orbital region becomes symmetric and the
density cuts through L4 and L5 are similar. The planet starts to opens
a gap and the density in the horseshoe region decreases.

\section{Dependence on the simulation parameters}
\label{sect_dep_on_par}

After having described type III migration in detail for one standard case,
we now consider the dependence of the planet's orbital evolution on the
most important simulation parameters.

\begin{figure*}
\includegraphics[width=84mm]{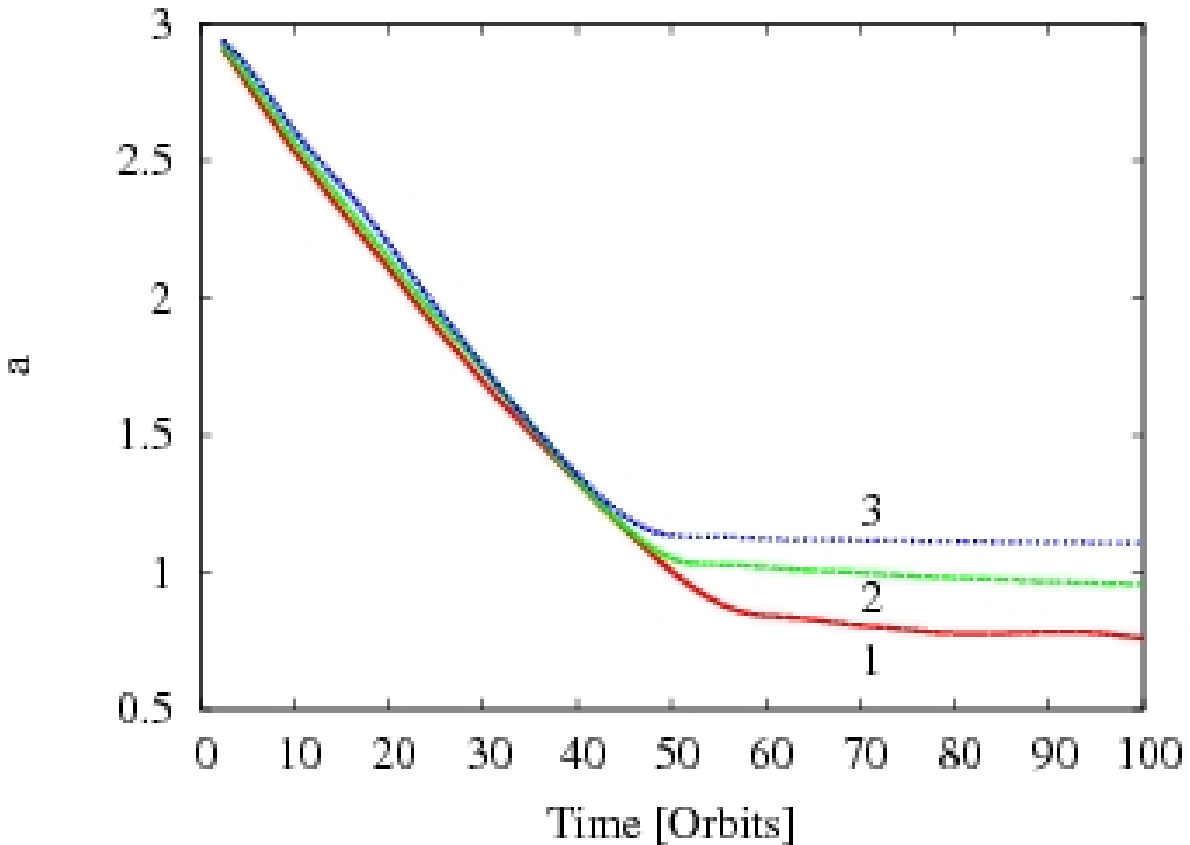}
\includegraphics[width=84mm]{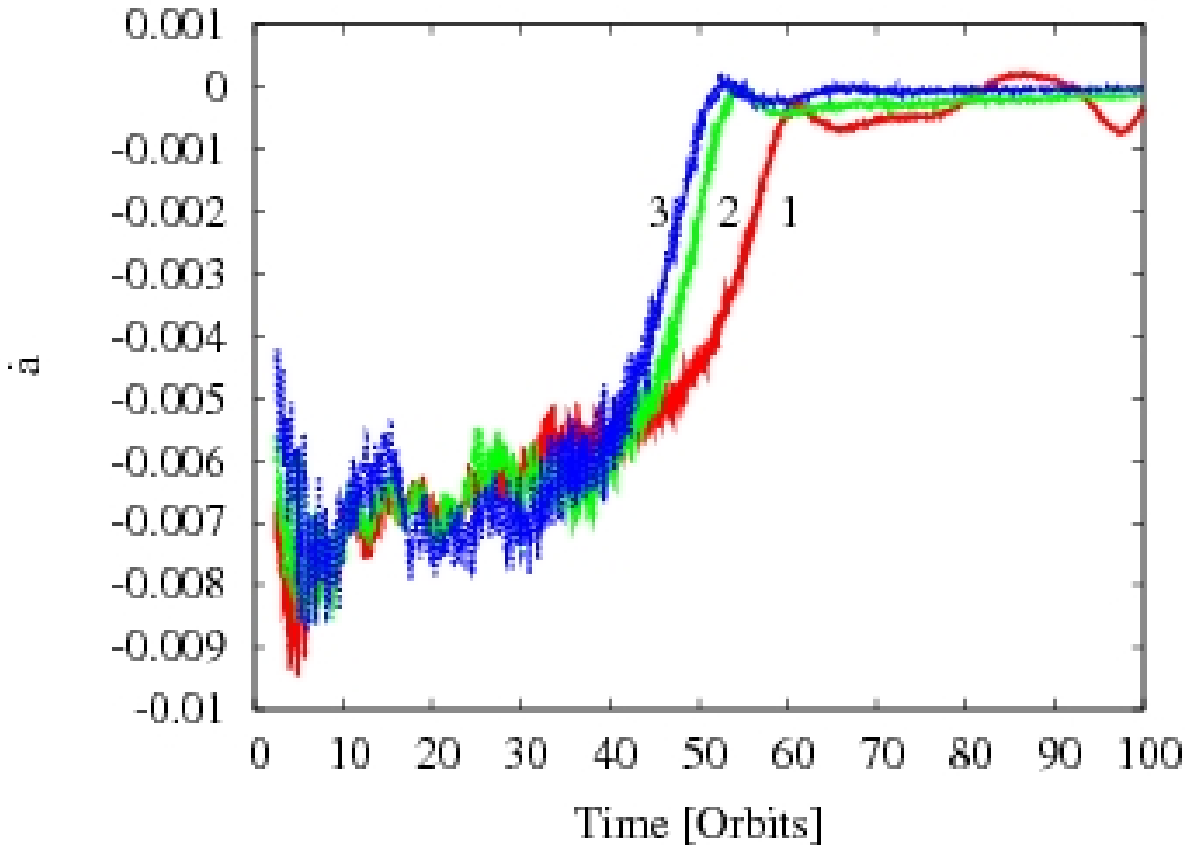}
\includegraphics[width=84mm]{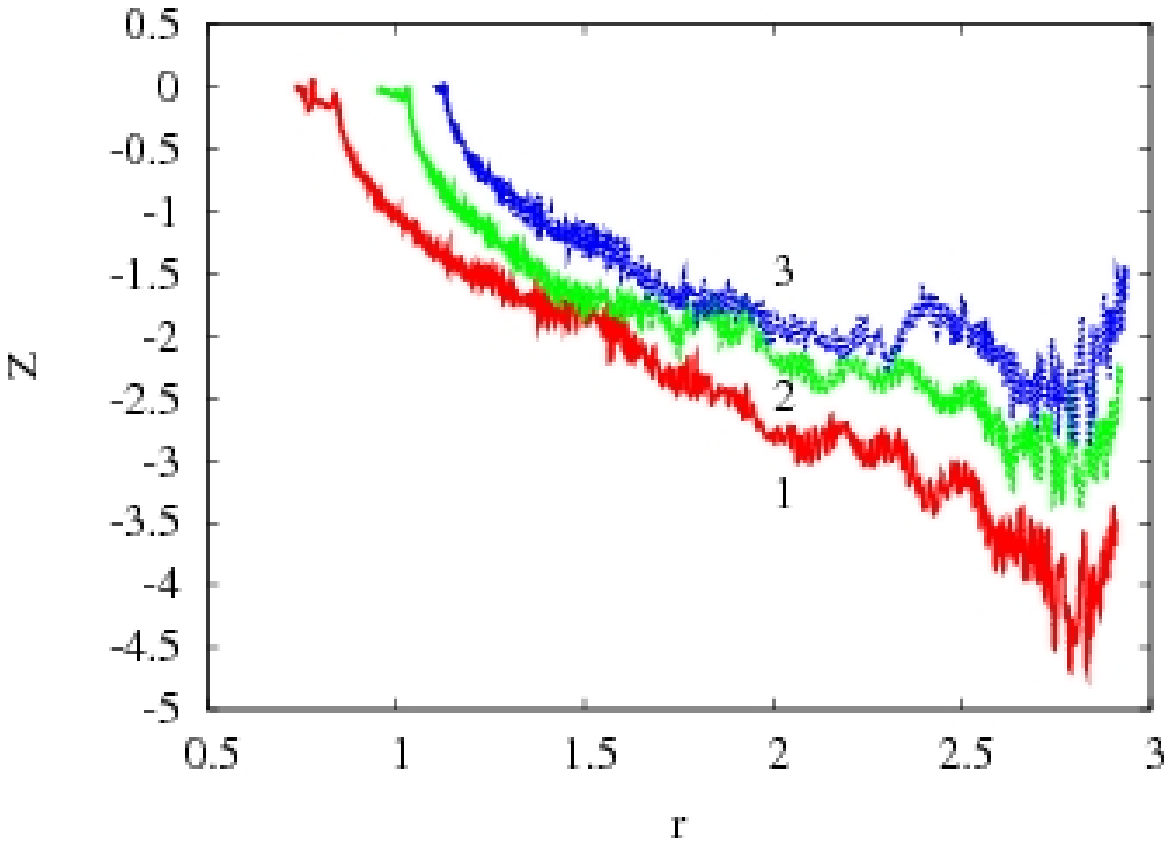}
\includegraphics[width=84mm]{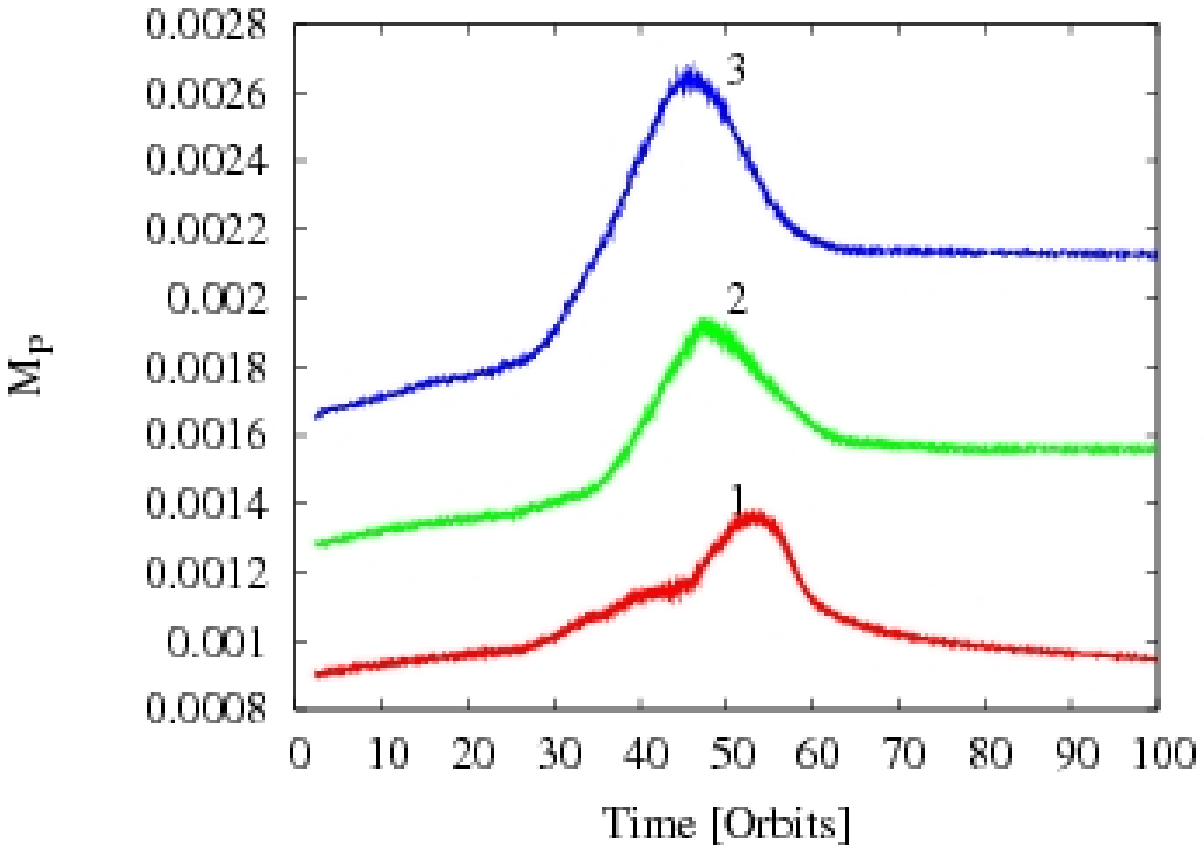}
\caption{Results of the simulations for different planet
masses. Curves 1, 2 and 3 correspond to the initial planet mass
$M_\rmn{P}$ equal $0.0007$, $0.001$ (standard case) and $0.0013$,
respectively. The upper left and upper right panels show the evolution
of the planet's semi-major axis $a$ and the migration rate $\dot
a$. The non-dimensional migration rate $Z$ as a function of the
planet's position in the disc and the evolution of the effective
planet mass $\widetilde M_\rmn{P}$ are presented in the lower left and
lower right panels. During the first $40$ orbits (fast migration limit
$|Z|>1$) all systems evolve almost exactly the same way even though
the (effective) planet mass may vary. The simulations start to differ
after reaching the slow migration limit. Even though the migration
rate $\dot a$ is almost identical for all simulations during the fast
migration limit, the non-dimensional migration rates $Z$ differ
slightly, since $\dot a_\rmn{f}$ depends on $\widetilde M_\rmn{P}$.}
\label{fmin_d_planet_mass}
\end{figure*}

\subsection{Planet's mass}

The first parameter we consider is the planet's mass. In our
investigation we concentrate on the orbital evolution of giant
planets. The results of the simulations with the initial mass
$M_\rmn{P}$ equal $0.0007$ and $0.0013$ (curves 1 and 3 respectively)
are presented in Fig.~\ref{fmin_d_planet_mass}, together with our
standard case of $M_\rmn{P}=0.001$

The upper left and upper right panels show the evolution of the
planet's semi-major axis $a$ and migration rate $\dot a$. During the
first $40$ orbits all systems evolve almost exactly the same way, even
though the effective planet masses (lower right panel) differ
significantly. This period corresponds to the fast migration limit
$|Z|>1$ for all three models. In this phase the orbital evolution is
independent of the planet mass since $Z \sim M_\rmn{\Delta}$, and both
the non-dimensional migration rate $Z$ and the mass deficit
$M_\rmn{\Delta}$ are proportional to $\widetilde M_\rmn{P}^{-2/3}$.

Once the planets enter the slow migration limit, the relation between
$Z$ and $M_\rmn{\Delta}$ becomes more complicated and their evolution
starts to differ.  In this phase, $|\dot a|$ decreases rapidly and the
planet is gradually settles into a type II like migration. Since $Z$
diminishes faster for more massive planets, this transition to the
slow migration limit happens sooner for more massive planets, and as a
result more massive planets stop their type III migration at larger
orbital radii. The final planet's semi-major axes are $0.75$, $0.95$
and $1.11$ for the initial mass $M_\rmn{P}$ equal $0.0007$, $0.001$
and $0.0013$, respectively.

The relation between the planet mass and its orbital evolution is
further illustrated in the lower left panel, where the non-dimensional
migration rate $Z$ is plotted as a function of the planet's position
in the disc. A larger planet mass corresponds to a smaller $|Z|$ and
thus displaces the orbital evolution curve to the right.

The lower right panel shows the effective mass evolution.  For a
constant circumplanetary disc aspect ratio $h_\rmn{p}$, higher mass
planets accumulate a larger amount of gas in the circumplanetary
disc. The mass accumulation rate is relatively low during the fast
migration limit and increases with decreasing $|Z|$. There is a visible
increase of $\dot {\widetilde M}_\rmn{P}$ for $Z \approx -1.7$ in all
three cases. Similarly the mass outflow from the Roche lobe starts at
$Z \approx -0.6$ independently of the planet mass (and thus happens
earlier for more massive planets). The final planet masses are
$0.00095$, $0.00155$, $0.0023$ for $M_\rmn{P}$ equal $0.0007$, $0.001$
and $0.0013$ respectively. This corresponds to a relative increase of
$35\%$, $55\%$ and $76\%$ after $100$ orbits respectively, and thus
the relation between $M_\rmn{P}$ and $\widetilde M_\rmn{P}$ is not
linear.

The lowest mass planet shows some evolutionary features not seen in
the other two. First, its mass accumulation rate shows {\it two}\/
instances of rapid growth, the first one at about $27$ orbits ($Z
\approx -2.5$), and the second one at $45$ orbits ($Z \approx
-1.7$). Second, after $80$ orbits the lowest mass planet shows
oscillations in its migration rate. After $120$ orbits these
oscillations have damped out, and the migration rate $\dot a$ becomes
similar to that of the more massive planets. These oscillations are
caused by a change of shape of the flow lines in the planet's vicinity
from regular orbits in the circumplanetary disc to a non regular flow
modified by the strong pressure gradient.

\begin{figure*}
\includegraphics[width=84mm]{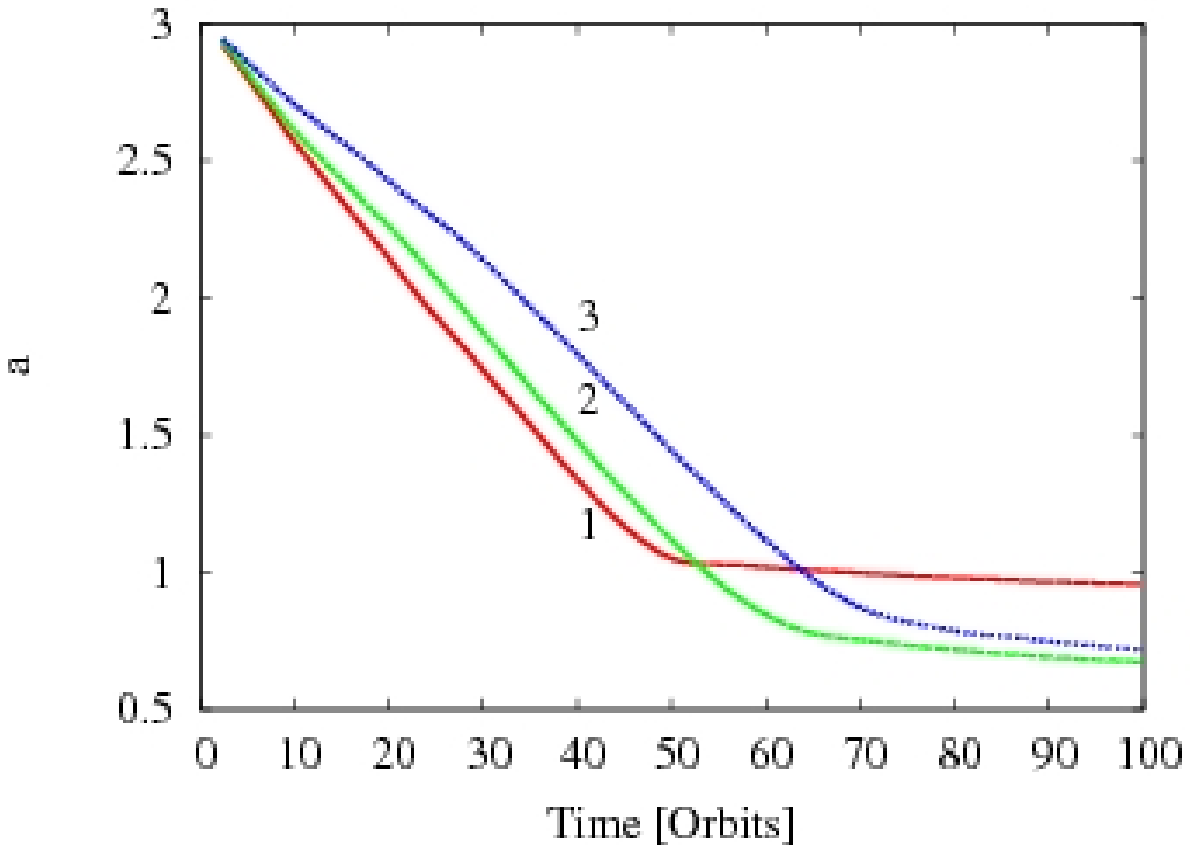}
\includegraphics[width=84mm]{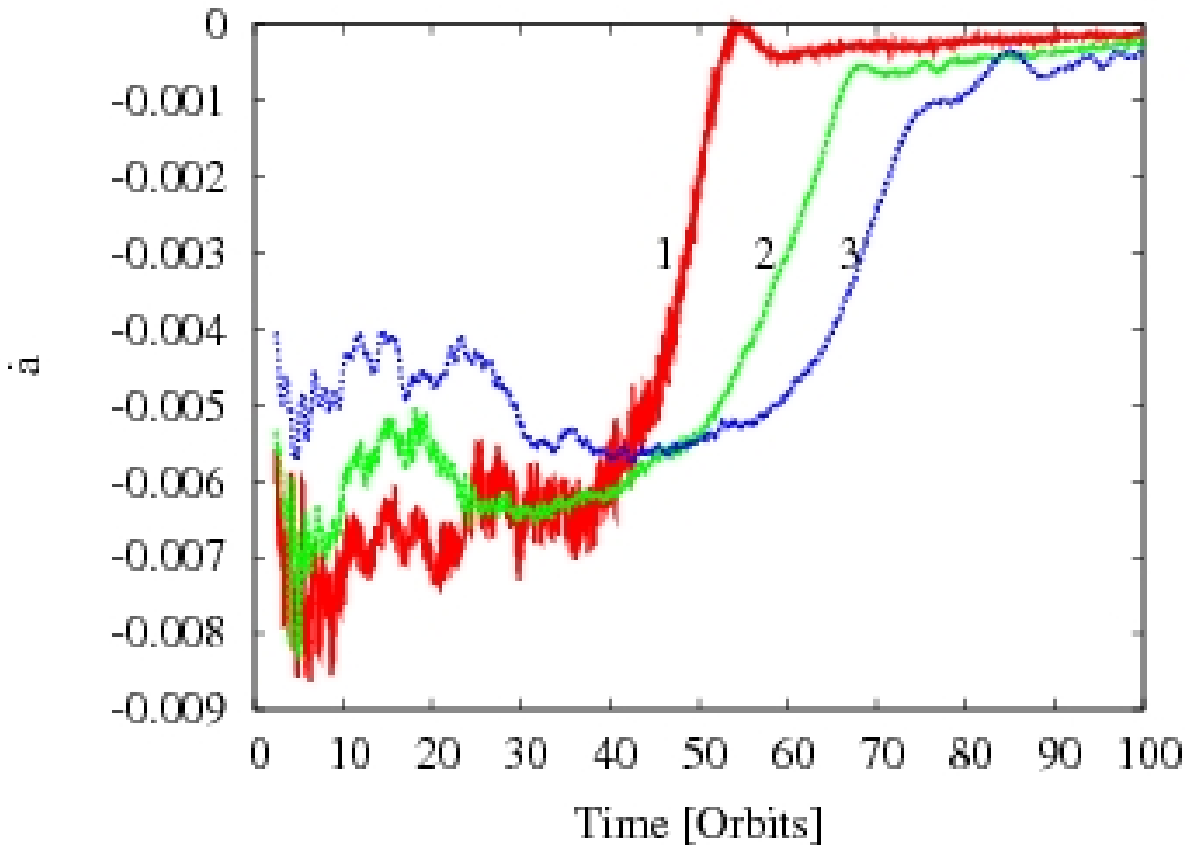}
\includegraphics[width=84mm]{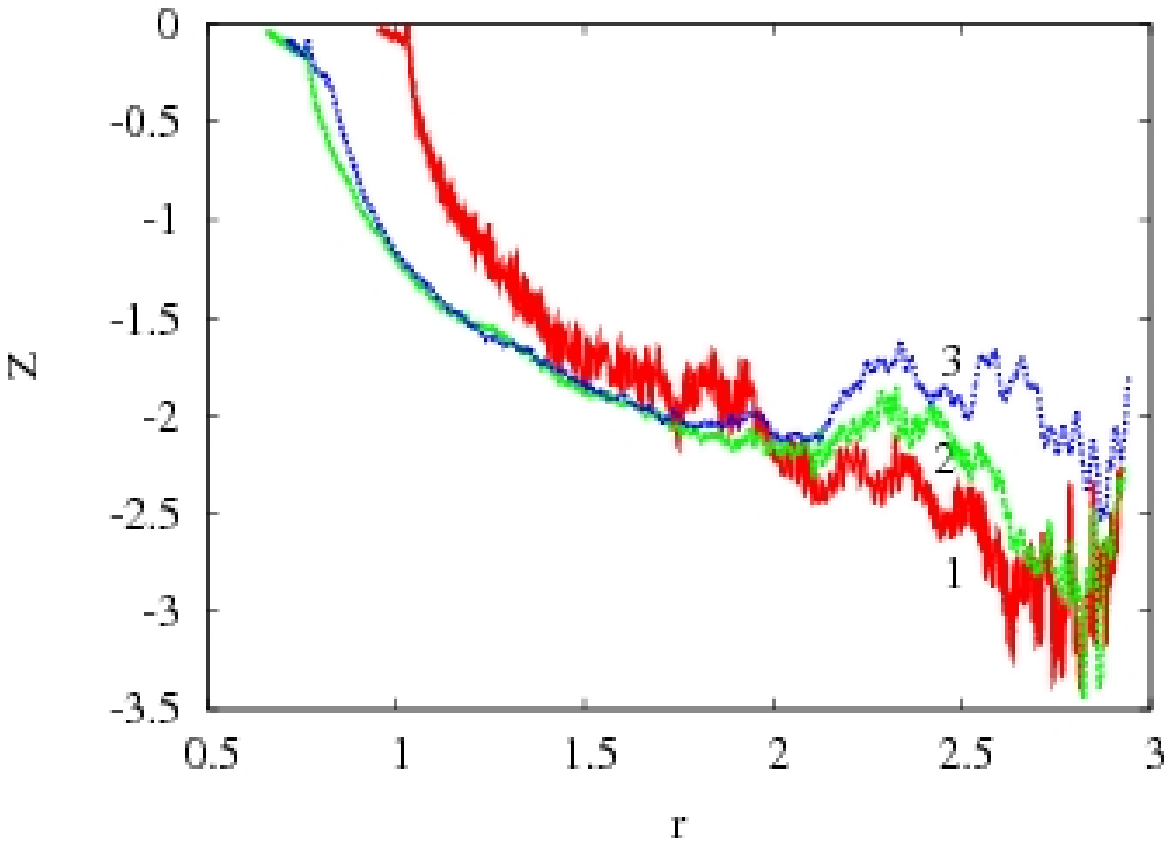}
\includegraphics[width=84mm]{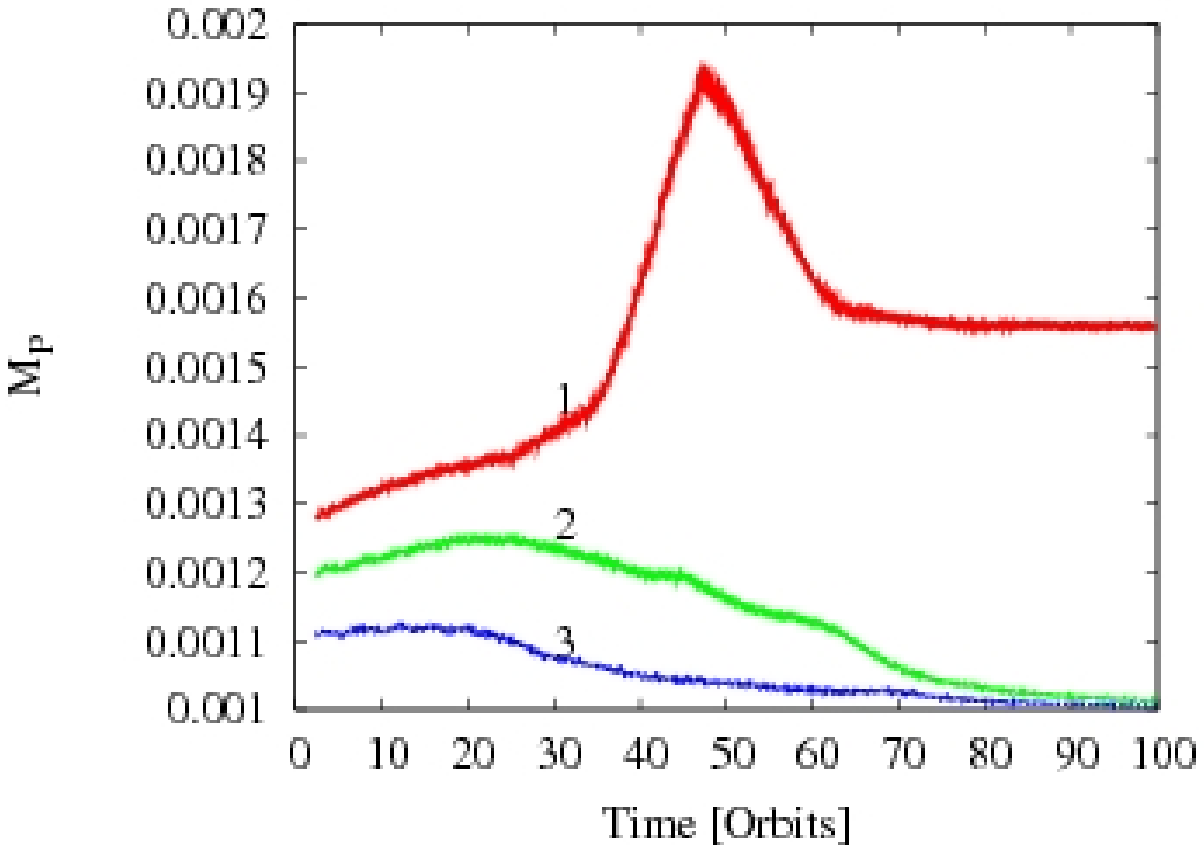}
\caption{Results of the simulations for the different circumplanetary
disc aspect ratios $h_\rmn{p}$. Curves 1, 2 and 3 correspond to
$h_\rmn{p}$ equal $0.4$, $0.5$ and $0.6$ respectively. The panels are
the same as in Fig.~\ref{fmin_d_planet_mass}.}
\label{fmin_d_planet_h}
\end{figure*}

\subsection{Effects of circumplanetary disc aspect ratio}

The next parameter we consider is the circumplanetary disc aspect
ratio $h_\rmn{p}$. As shown in Paper~I, this parameter corresponds
to the temperature of the circumplanetary disc, and determines how
much material can be accumulated there. Since the mass accumulation
influences type III migration, it is interesting to study the effect
of this parameter.

In addition to our standard case of $h_\rmn{p}=0.4$, we ran
simulations using values of $0.5$ and $0.6$. We do not use a value
lower than 0.4, since the tests presented in Paper~I showed that
a model with 0.3 does not fully converge for the resolution used in
our simulations\footnote{The results for $h_\rmn{p}=0.3$ are actually
a more extreme version of the 0.4 results, see Paper I.}.

The results for all three cases (labelled 1, 2 and 3) are presented in
Fig.~\ref{fmin_d_planet_h}. We use the same set of plots as in
Fig.~\ref{fmin_d_planet_mass}: the evolution of the planet's
semi-major axis $a$ and the migration rate $\dot a$ in the upper
panels, and the non-dimensional migration rate $Z$ as a function of
the planet's position in the disc together with the evolution of the
effective planet mass in the lower panels.

As expected, changing $h_\rmn{p}$ has a strong effect on the evolution
of the effective planet mass $\widetilde M_\rmn{P}$. As we saw
above, the standard case (0.4, curve 1) shows a phase of strong mass
accumulation, followed by mass loss from the circumplanetary
environment after migration has slowed down. The other two cases do
not show such strong mass accumulation, and the higher $h_\rmn{p}$,
the less mass is accumulated. In fact curves 2 and 3 show that mass
outflow already starts at $t\approx 20$ orbits, i.e.\ during the rapid
migration stage.

Considering the evolution of $a$ and $\dot a$ we notice some small
differences. For $h_\rmn{p}=0.5$ and 0.6 the flow asymmetries in the
planet's vicinity during the fast migration limit phase become less
pronounced, leading to a lower total torque $|\Gamma|$. This results in a
somewhat lower value for the migration rate $\dot a$ during this
phase. The start of mass outflow at $t\approx 20$ orbits in these two
models leads to a small increase of the migration rate. Consistent
with the results from the previous section, the lower final values of
$\widetilde M_\rmn{P}$ cause the two thicker disc cases to migrate
closer to the star. Their type III migration stops at $a \approx 0.7$.
It is noticeable that the evolution of the system in the slow
migration limit is almost independent of the circumplanetary disc
aspect ratio for $h_\rmn{p} \ge 0.5$.

Tracing the evolution in the $Z-r$ plane (lower left panel), all three
simulations show similar behaviour until $Z \approx -2$ is reached
(lower left panel). After $|Z|$ drops below $1.7$ the model with
$h_\rmn{p} = 0.4$ starts rapid mass accumulation in the
circumplanetary disc, causing it to evolve differently from the other
two models.

We have to keep in mind that the assumption of a constant
circumplanetary disc aspect ratio is a rather crude one, and in
reality this parameter will react to the mass accretion into this
disc. More realistic calculations would either need to treat the
thermal evolution of the circumplanetary disc self-consistently, or
use some approximate recipe relating $h_\rmn{p}$ to $Z$ and the mass
accumulation rate. However, the results in this section show that
large changes in $h_\rmn{p}$ only have a limited effect on the
migration behaviour of the planet.


\begin{figure*}
\includegraphics[width=84mm]{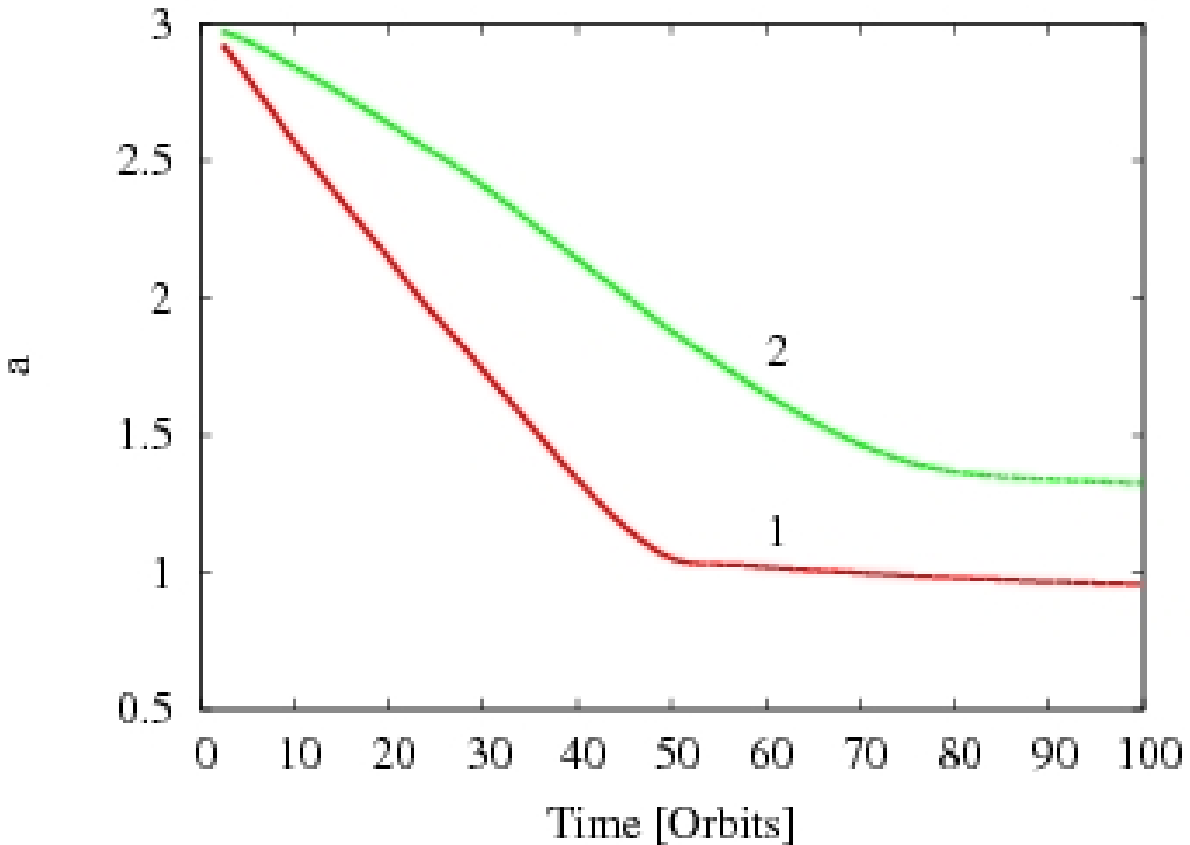}
\includegraphics[width=84mm]{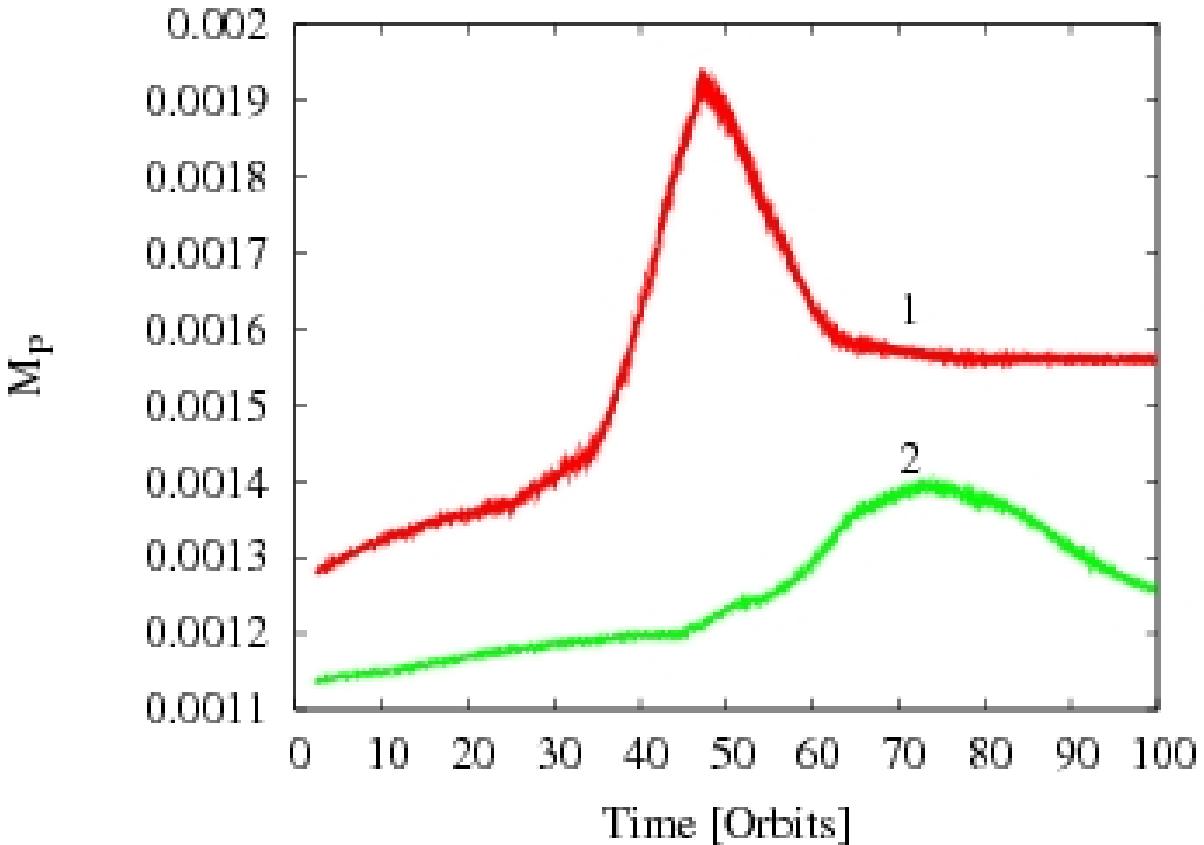}
\includegraphics[width=84mm]{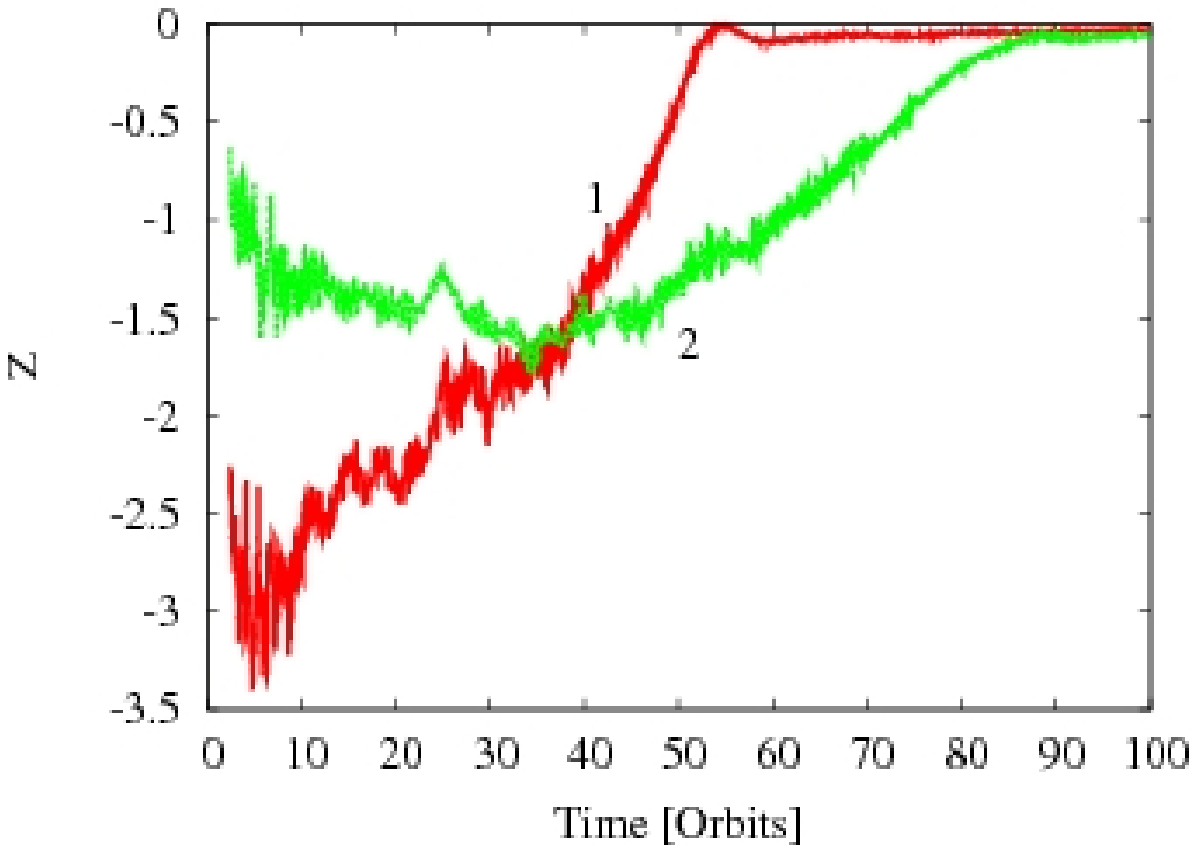}
\includegraphics[width=84mm]{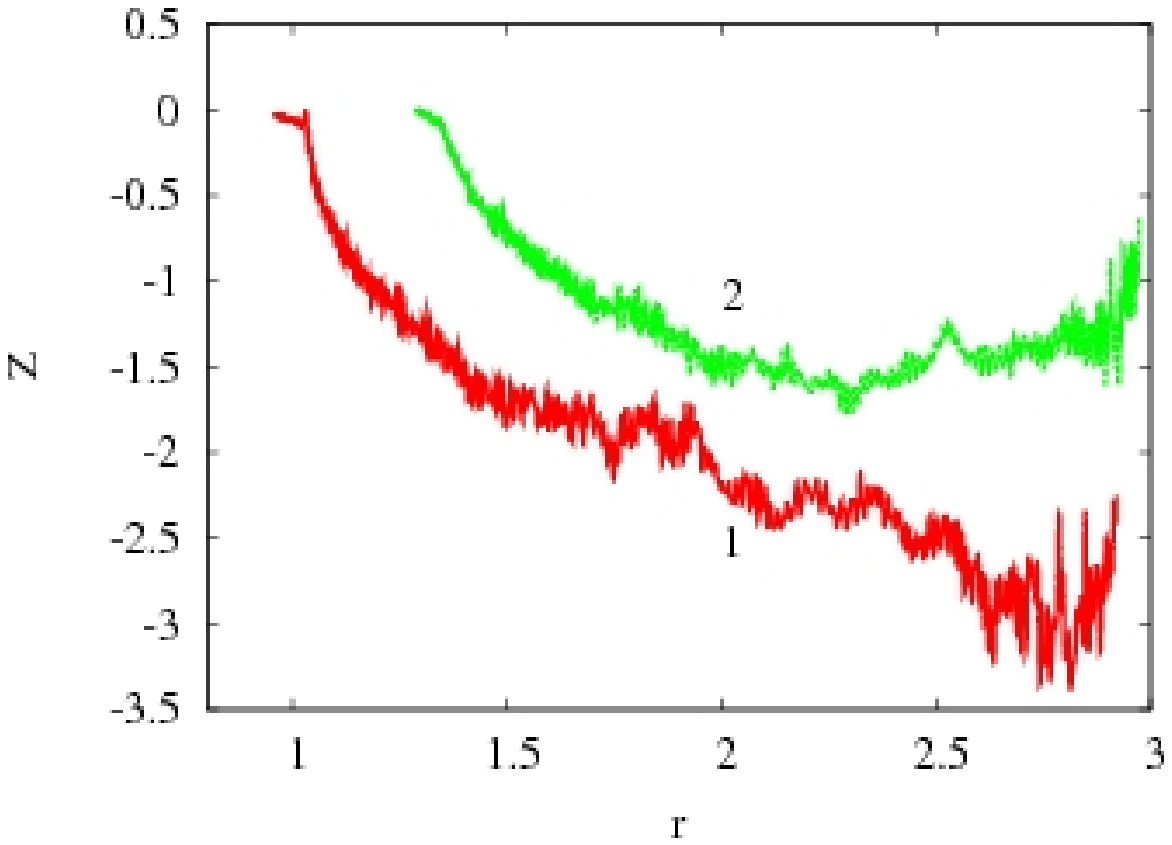}
\caption{Results of the simulations for different total disc
masses. Curves 1 and 2 correspond to the disc to the primary mass
ratio $\mu_\rmn{D}$ equal $0.005$ (standard case) and $0.0025$
respectively. The upper left and right panels show the evolution of
the semi-major axis and the effective mass; the non-dimensional
migration rate $Z$ as a function of time and of radius are presented
in the lower left and lower right panels.}
\label{fmin_d_disc_mass}
\end{figure*}

\begin{figure*}
\includegraphics[width=84mm]{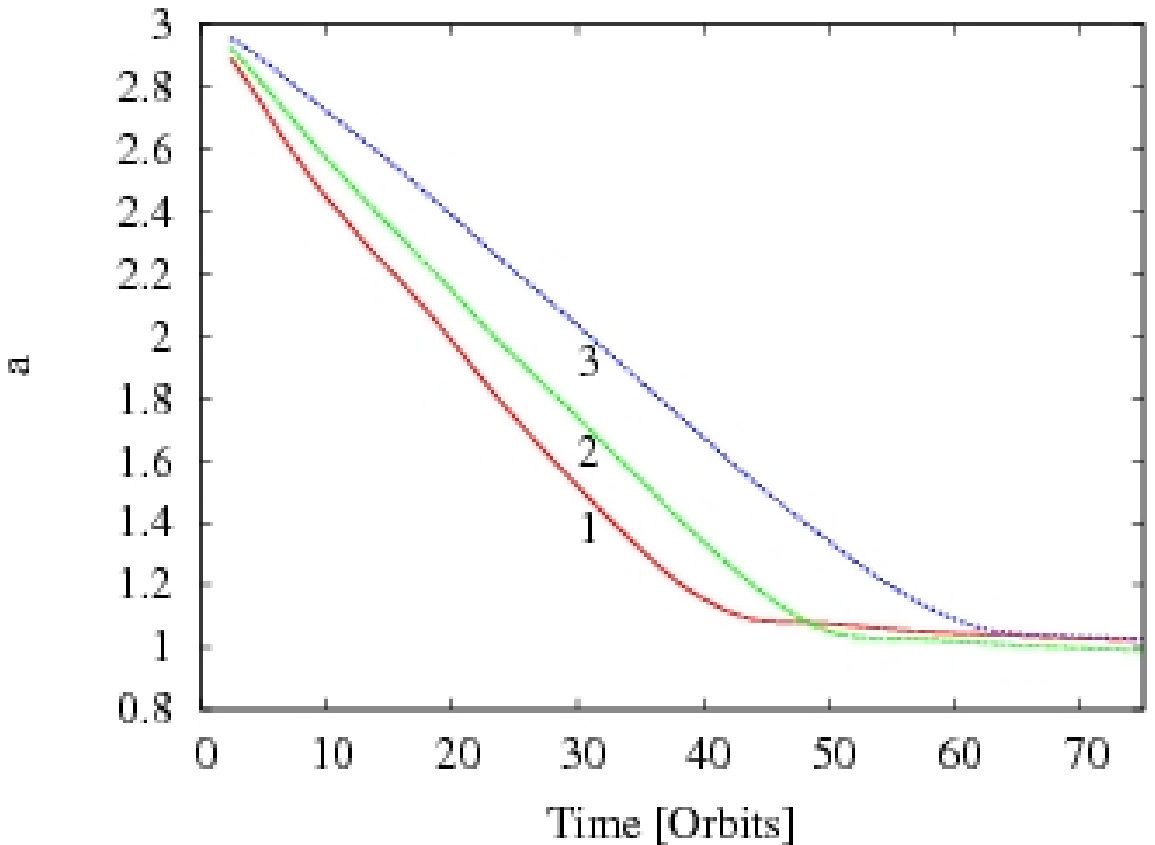}
\includegraphics[width=84mm]{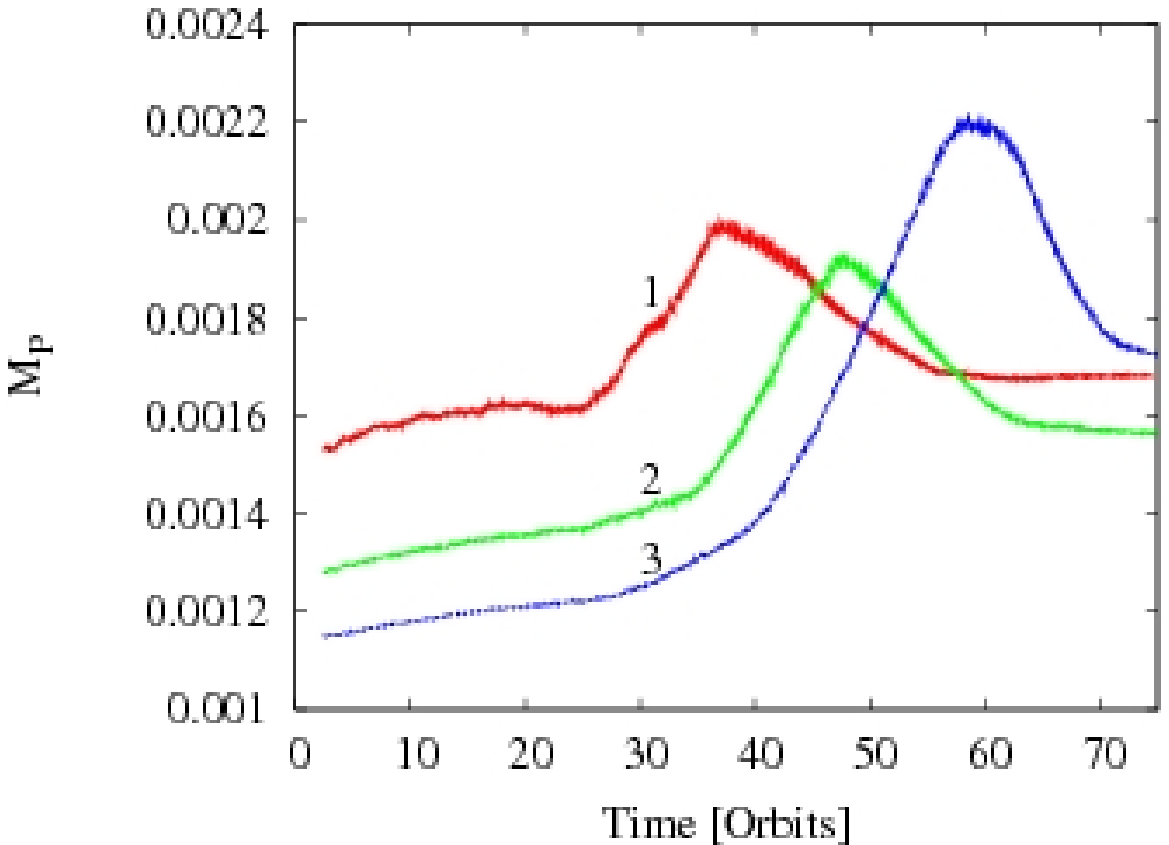}
\includegraphics[width=84mm]{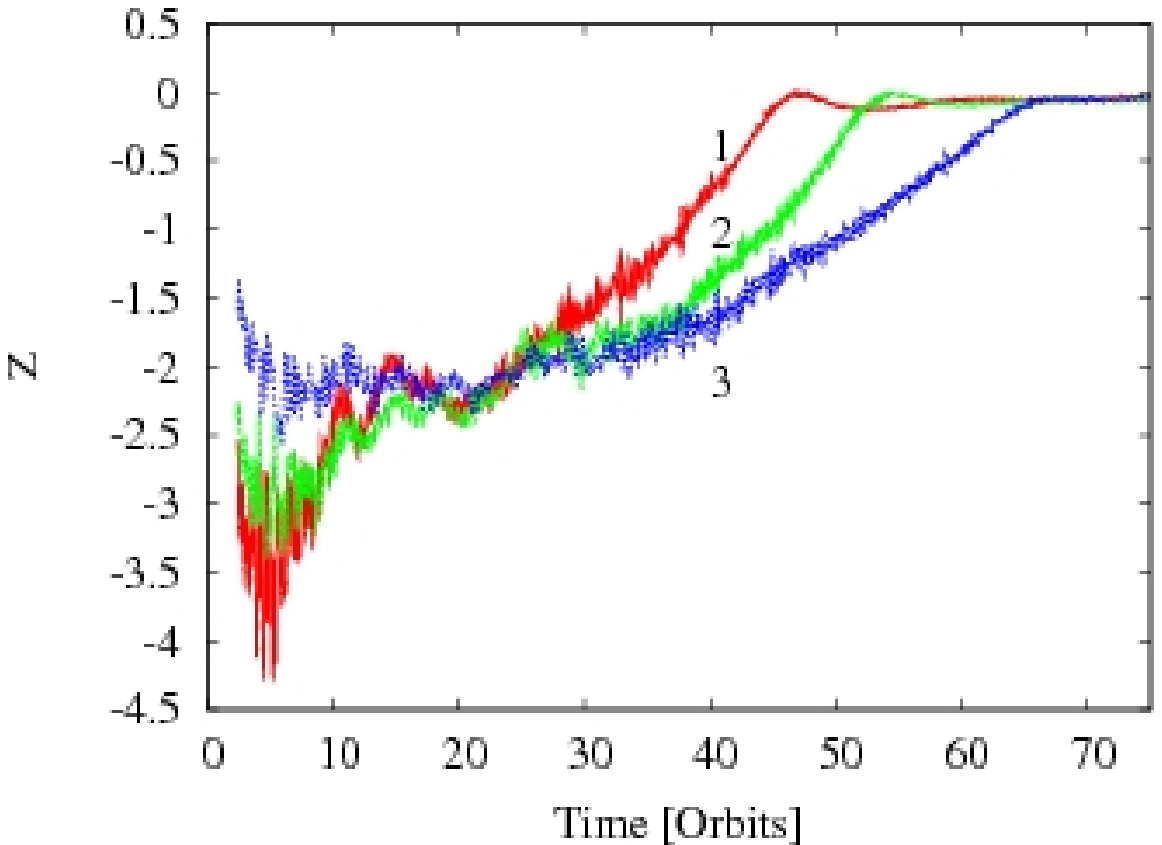}
\includegraphics[width=84mm]{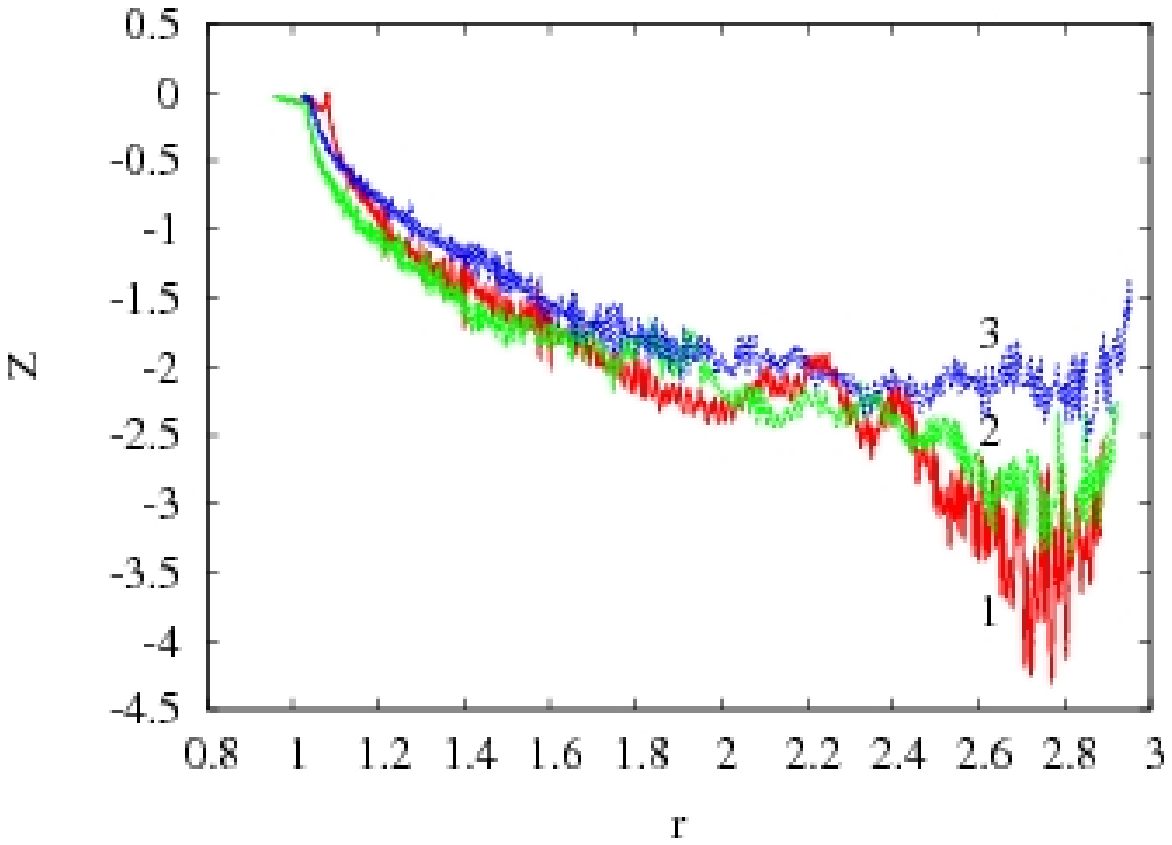}
\caption{Results of the simulations for the different initial disc
profiles with $M_\rmn{P}^* = \widetilde M_\rmn{P}$. Curves 1, 2 and 3
correspond to $\alpha_\rmn{\Sigma}$ equal $-0.5$, $-1$ (standard case)
and $-1.5$ respectively. The layout of the panels is the same as in
Fig.~\ref{fmin_d_disc_mass}.}
\label{fmin_d_disc_prof}
\end{figure*}

\begin{figure*}
\includegraphics[width=84mm]{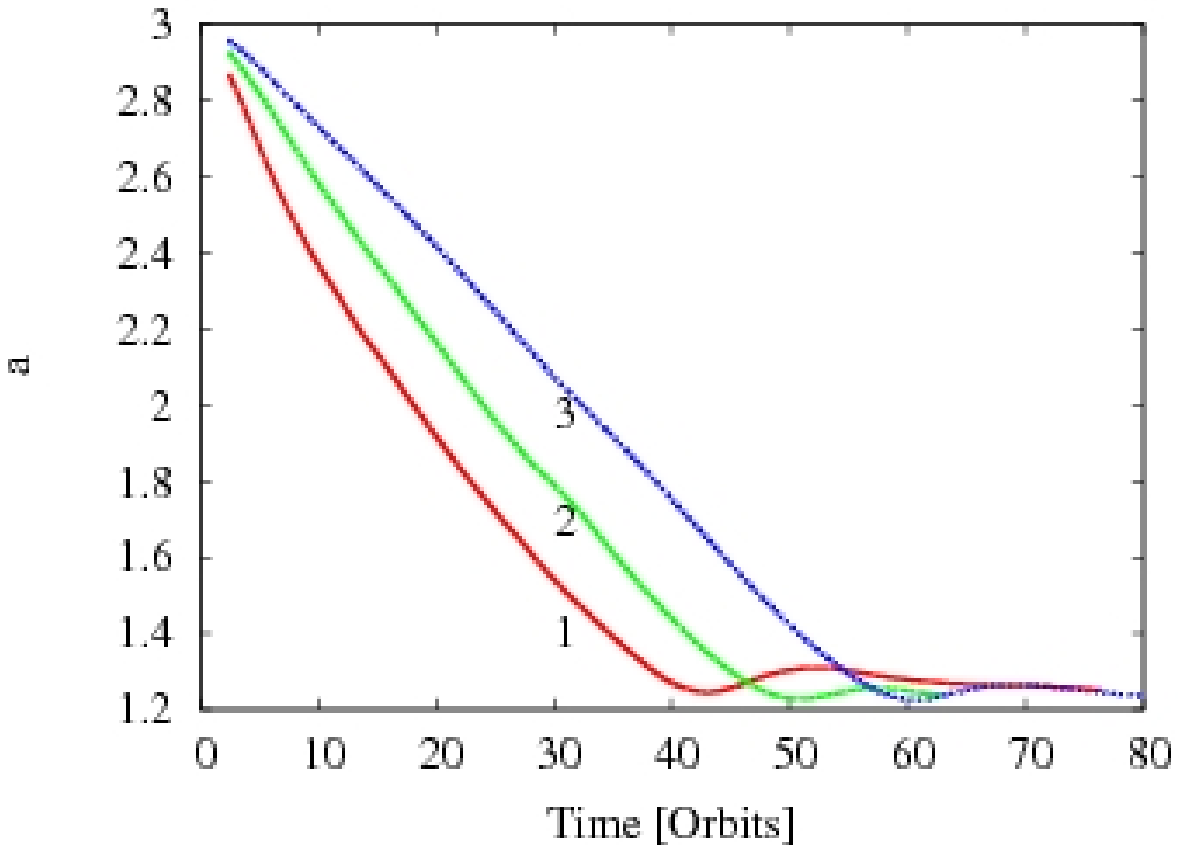}
\includegraphics[width=84mm]{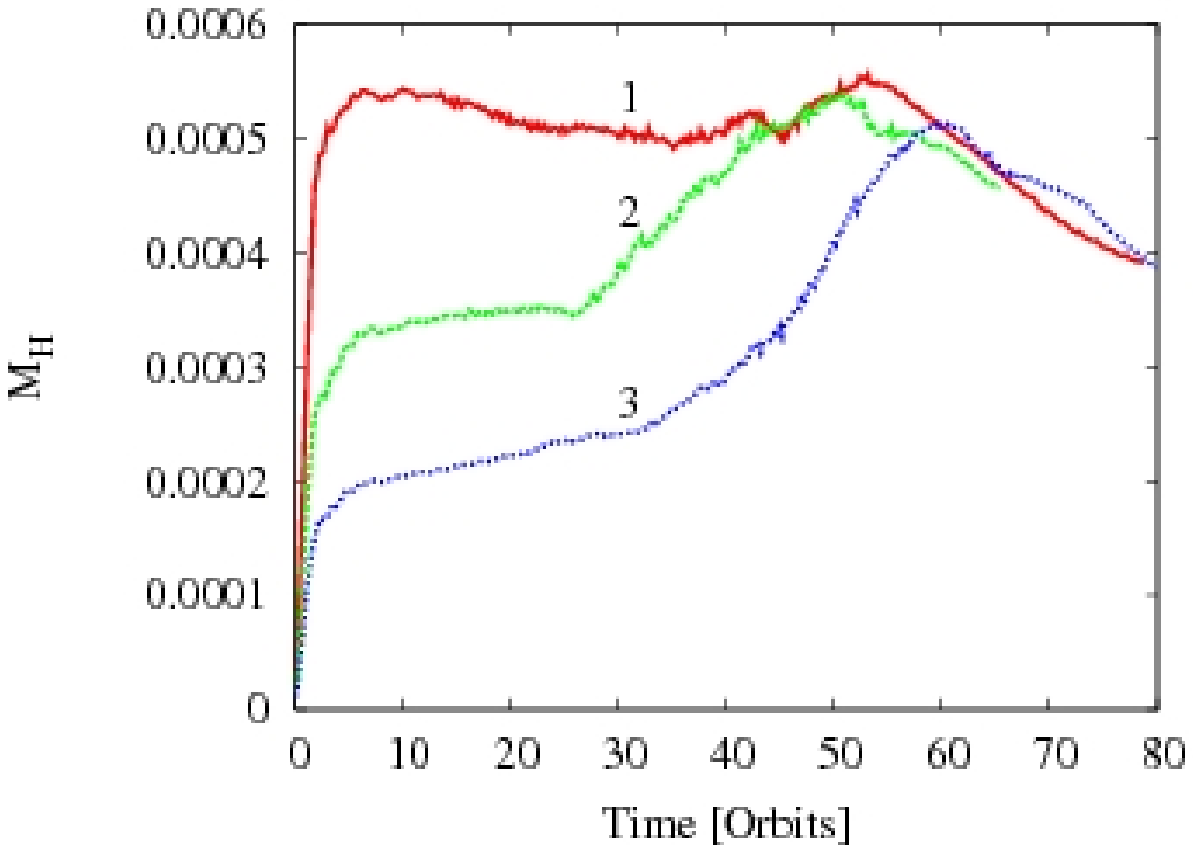}
\includegraphics[width=84mm]{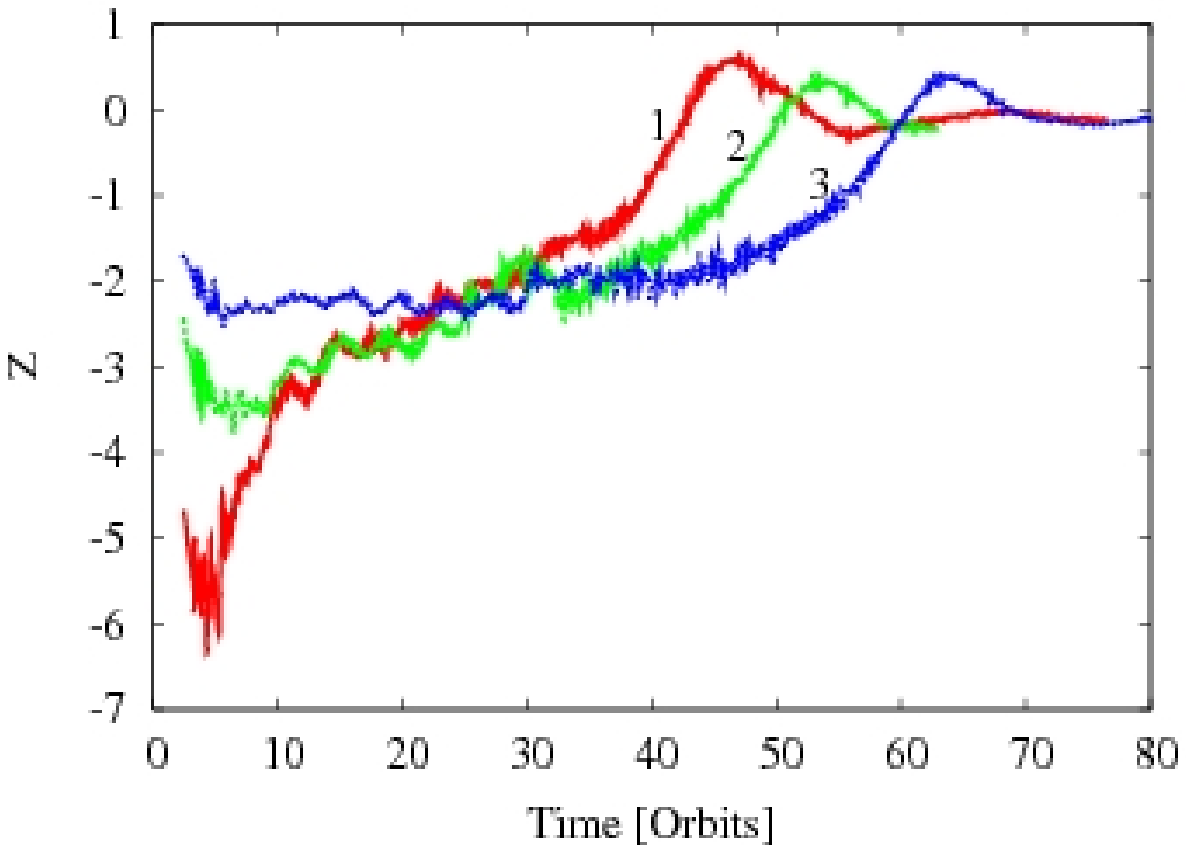}
\includegraphics[width=84mm]{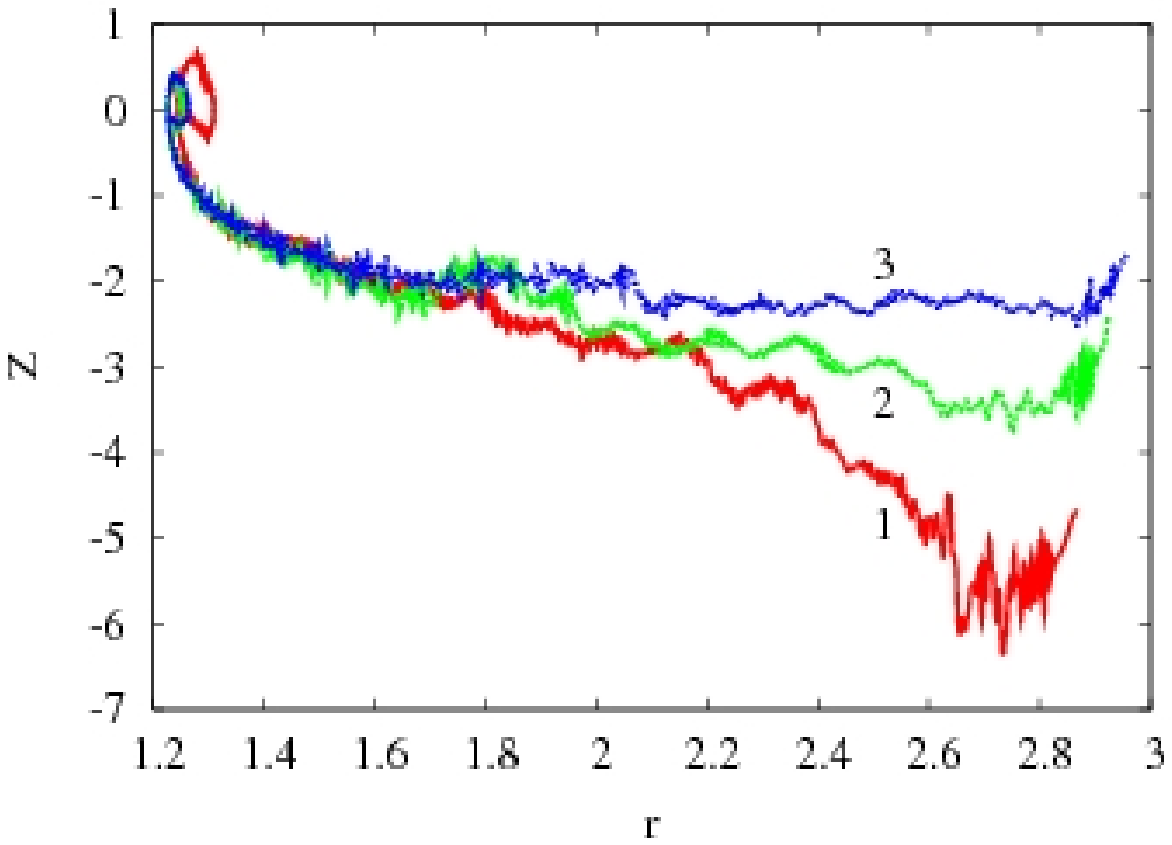}
\caption{Results of the simulations for the different initial disc
profiles with constant $M_\rmn{P}^* = M_\rmn{P}= 0.001$. Curves 1, 2 and 3
correspond to $\alpha_\rmn{\Sigma}$ equal $-0.5$, $-1$ and $-1.5$
respectively. The layout of the panels is the same as in
Fig.~\ref{fmin_d_disc_mass}, except that here the upper right panel shows
the evolution of the mass inside the Hill sphere.}
\label{fmin_d_disc_prof2}
\end{figure*}

\subsection{Total disc mass}
\label{sub_dep_dens_prof_tot_mass}

In the previous two sections we discussed parameters connected to the
planet and its circumplanetary disc. In this section and the next we
consider parameters defining the circumstellar disc, namely the total
disc mass (given by the disc to the primary mass ratio $\mu_\rmn{D}$)
and the slope of the initial density profile (described by the
exponent $\alpha_\rmn{\Sigma}$), see Sect.~\ref{setc_init_cond}. 

To study the relation between the total disc
mass and the planet migration we compare a simulation with
$\mu_\rmn{D}=0.0025$ to our standard case of $0.005$. The other
parameters are identical between these two simulations. The results
are presented in Fig.~\ref{fmin_d_disc_mass}.

As can be expected the planet migrates slower and less far in the
lower mass disc. The stage of the rapid migration lasts longer ($85$
orbits instead of $50$) and the transition between the fast and slow
migration limits happens at $60$ orbits, instead of $40$. After the
rapid migration ends the planet is locked in the disc at $a=1.3$
($a=0.95$ for the standard case).  

There is no simple relation between the total disc mass and the
position where the planet is locked into the disc, or the duration of
the rapid migration stage. This is because the migration rate in type
III migration depends both on the co-orbital mass deficit
$M_\mathrm{\Delta}$ and on the asymmetry in the co-orbital
region. Changing the total disc mass changes the mass inside the
co-orbital region, but the asymmetry depends on the current and
previous values of the migration rate. However, $\mu_\rmn{D}$ is an
important parameter, since the amount of the mass in the co-orbital
region defines the region where rapid migration can take place. Since
the volume of the co-orbital region is proportional to $a^2$, for the
standard smooth density profile with $\alpha_\rmn{\Sigma} \geq -1.5$
it is easier to start rapid migration at larger orbital radii (see the
discussion in \citealt{2003ApJ...588..494M}). The radii $r=0.95$ and
$r=1.3$ are the outer borders of the region, where rapid migration for
a Jupiter-mass planet cannot be started for $\mu_\rmn{D}$ equal
$0.005$ and $0.0025$ respectively.

The initial value of the migration rate is given by the amount of the
gas in the co-orbital region. After about $10$ orbits the
non-dimensional migration rates are $Z \approx -2.7$ and $Z \approx
-1.4$ (lower left panel in Fig.~\ref{fmin_d_disc_mass}), their ratio
indeed matching the ratio of the disk masses. Later both systems
evolve differently. The non-dimensional migration rate decreases
quickly during the whole stage of rapid migration for the more massive
disc, but for the lower mass disc $|Z|$ increases during the first
$35$ orbits and then decreases much slower. More similarities between
both simulations are visible after plotting $Z$ as a function of the
planet's position in the disc (lower right panel in
Fig.~\ref{fmin_d_disc_mass}). Neglecting the first $25$ orbits (region
outside $r \approx 1.9$ and $r \approx 2.5$ for the high and low mass
disc respectively) the shape of both curves is similar, but the curve
for the more massive disc is shifted inward by $0.3$ in radius. This
number corresponds exactly to the shift of the boundary of the region
where rapid migration is allowed. Decreasing the disc mass thus has a
similar effect as increasing the planet mass (lower right panel in
Fig.~\ref{fmin_d_planet_mass}).

The effective planet mass (upper right panel in
Fig.~\ref{fmin_d_disc_mass}) depends on the total mass of the disc and
on the non-dimensional migration rate. It grows slowly in the fast
migration limit phase until $|Z|$ drops below $1.7$ after which the
mass accumulation rate increases rapidly for both models. The maximum
value of the planet mass for both models is reached at $Z \approx
-0.6$ at which point the mass outflow from the circumplanetary disc
starts. The values are $1.56 M_\rmn{P}$ and $1.22 M_\rmn{P}$.

\subsection{Initial density profile}
\label{alpha_dependence}
The next parameter is the exponent $\alpha_\rmn{\Sigma}$. The standard
case has $\alpha_\rmn{\Sigma}=-1$ (curve 2 in
Fig.~\ref{fmin_d_disc_prof}), and we performed two additional
simulations using $-0.5$ and $-1.5$ (curves 1 and 3). All other
parameters were taken identical to the standard case. Because of our
definition of the surface density, the disc densities are equal at
$r=1$, but differ at the initial planet position $r=3$. The 
$\alpha_\rmn{\Sigma}=-0.5$ case
has the largest initial mass in the co-orbital region, and therefore
evolves the fastest. It reaches $a \approx 1$ after about $43$
orbits. For the $\alpha_\rmn{\Sigma}=-1.0$ and $-1.5$ cases the planet
settles at the same position after about $51$ and $64$ orbits
respectively. Because this position is near $r=1$, all three models
agree on the final orbital radius, since this is where the surface
density is identical between the models, and at a value that is too
low to support further rapid migration.

The exponent $\alpha_\rmn{\Sigma}$ determines the average rate with
which the non-dimensional migration rate is decreasing and influences
the relation between $Z$ and the planet's position in the disc, which
can be seen in the lower right panel. The evolution of the systems
differs during the first $16$ orbits only. Later on the relation
between $Z$ and $a$ becomes independent of
$\alpha_\rmn{\Sigma}$. However, as we will see below this is mostly
caused by the evolution of the effective planet mass which is
presented in the upper right panel. In the fast migration limit
$\widetilde M_\rmn{P}$ depends on the initial mass of the co-orbital
region (largest for $\alpha_\rmn{\Sigma}=-0.5$) and grows slowly. The
mass accumulation rate $\dot {\widetilde M_\rmn{P}}$ increases when
the planet makes the transition to the slow migration limit. The
increase is steep for $\alpha_\rmn{\Sigma}=-0.5$ and relatively smooth
for $\alpha_\rmn{\Sigma}=-1.5$. What is important is that $\dot
{\widetilde M_\rmn{P}}$ has similar values in all presented models and
the highest value of $\widetilde M_\rmn{P}$ is reached in the
simulation with steepest density profile, since the mass accumulation
phase takes the longest time there. The final mass of the planet is
equal $1.62 M_\rmn{P}$, $1.56 M_\rmn{P}$ and $1.72 M_\rmn{P}$ for
$\alpha_\rmn{\Sigma}$ equal $-0.5$, $-1.0$ and $-1.5$ respectively.

To test how the mass accumulation near the planet influences the
results we ran a set of simulations where the correction for
self-gravity is turned off by keeping the planet mass constant at
$M_\rmn{P}^*=M_\rmn{P}=0.001$.  In these simulations we also
introduced an inner disc edge (see Paper~I Fig.~2). This edge is not
significant for the subject discussed in this section, but it is
relevant for the discussion on stopping mechanisms given in the next
section. The inner gap starts at $r=1.3$ and the density is constant
at $\Sigma=0.00016$ inside $r=0.85$.

The results of these simulations are shown in
Fig.~\ref{fmin_d_disc_prof2}. The upper left panel shows the planet's
orbital evolution. In all simulations the planet stops its rapid
migration at the disc edge. Similarly the transition between the fast
and the slow migration limit takes place at the same position,
$a=1.28$ close to the initial density maximum. The interaction with
the disc edge also leads to a temporary outward directed migration,
although this effect was found to be sensitive to the resolution
used. The stopping of rapid migration at the disc edge is caused by
the loss of flow asymmetry in the co-orbital region due to the
interaction with the low density region. A more detailed discussion is
presented in the next section.

Like in the previous model the initial amount of the gas within the
co-orbital region determines the planet's orbital evolution and the
disc with the smaller density gradient gives faster migration. In a
similar way it influences the initial mass of the gas inside the Hill
sphere. The later evolution of $M_\rmn{H}$ depends on
$\alpha_\rmn{\Sigma}$. For $\alpha_\rmn{\Sigma} = -0.5$ $M_\rmn{H}$
remains almost constant until the planet reaches the disc edge, and
then starts to decrease. For the other models this mass remains
approximately constant until $Z > -1.7$, and then starts to
increase. When the planet reaches the disc edge $M_\rmn{H}$ has
a similar value for all models, and then decreases with the same rate.

The relation between the non-dimensional migration rate and
$\alpha_\rmn{\Sigma}$ is illustrated on lower right panel showing $Z$
as a function of the planet's position in the disc. Here we see that
without mass accumulation, the curves do not overlap. In the fast
migration limit $Z$ is approximately constant for
$\alpha_\rmn{\Sigma}=-1.5$ and changes fastest for
$\alpha_\rmn{\Sigma}=-0.5$.  This could be due to the fact that $Z$ is
related to the specific vorticity $\bmath{w} = (\mathrm{\Delta} \times
\bmath{v}) / \Sigma$, which has an almost constant value in the disc
for $\alpha_\rmn{\Sigma}=-1.5$ and decreases with the radius for
bigger $\alpha_\rmn{\Sigma}$. Alternatively, this relation may also be
explained by the fact, that in the fast migration limit $Z \sim
M_\mathrm{\Delta} \sim a^2 (\Sigma_\rmn{s}-\Sigma_\rmn{g})$.

All the curves merge at $Z \approx -2$ and $r \approx 1.45$, since
$\Sigma_\rmn{s}$ at becomes similar in all simulations and the
interaction with the disc edge starts to be important. Later on all
the systems evolve in a similar way due to the interaction with the
disc edge, however the temporary outward migration is stronger for the
model with $\alpha_\rmn{\Sigma} = -0.5$.


\section{Stopping type III migration}

\label{stop_mig}

\begin{figure*}
\includegraphics[width=84mm]{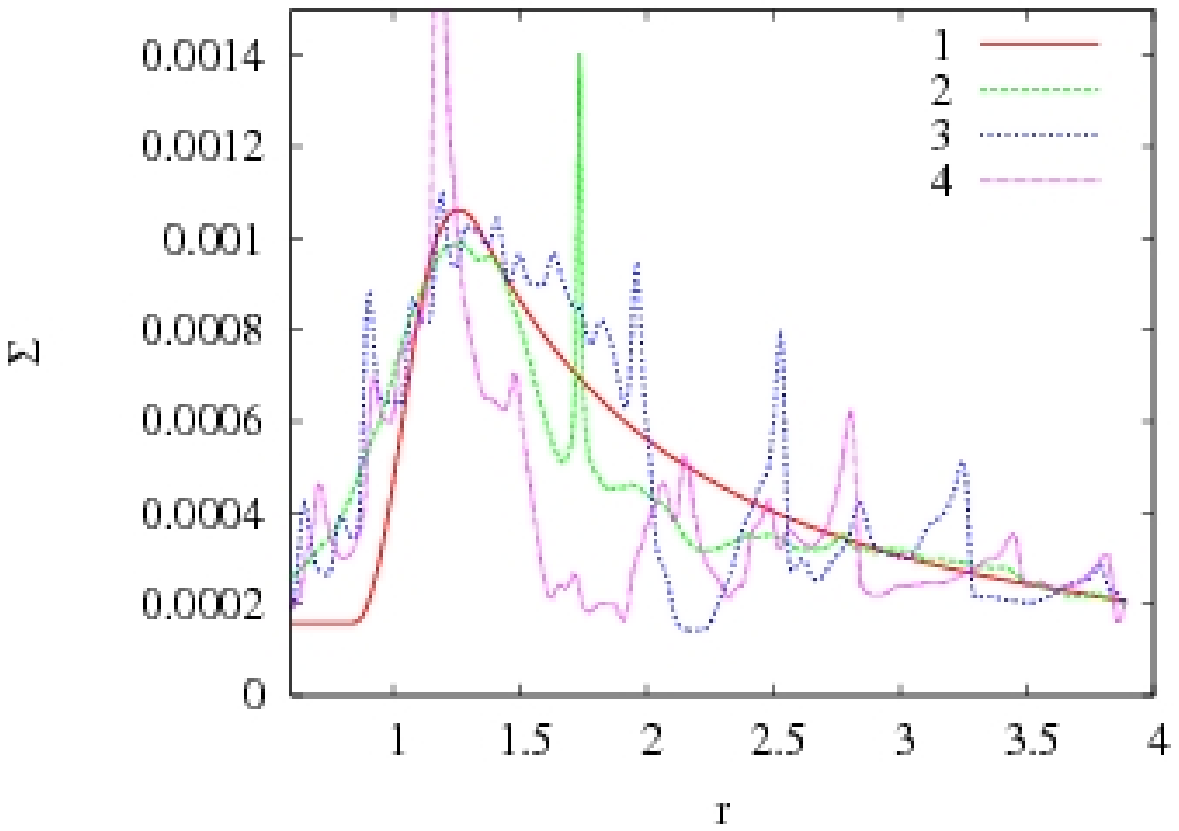}
\includegraphics[width=84mm]{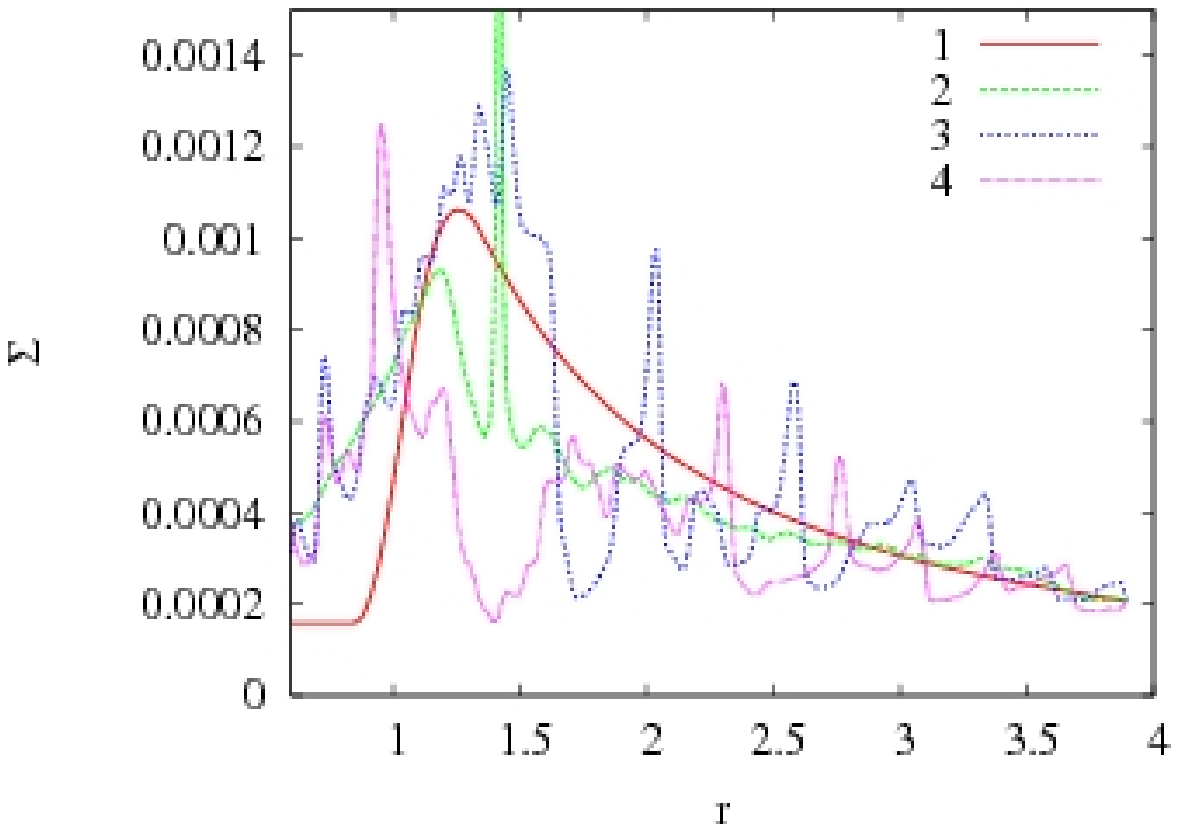}
\includegraphics[width=84mm]{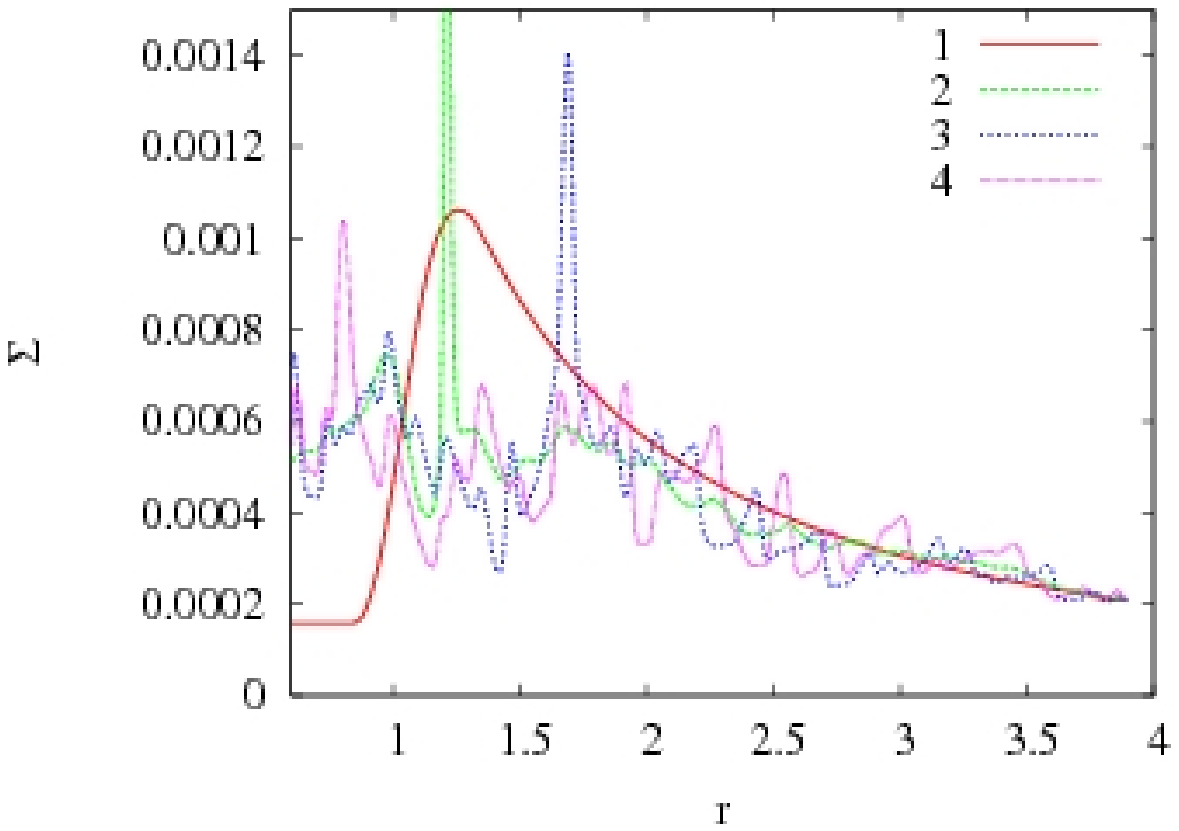}
\includegraphics[width=84mm]{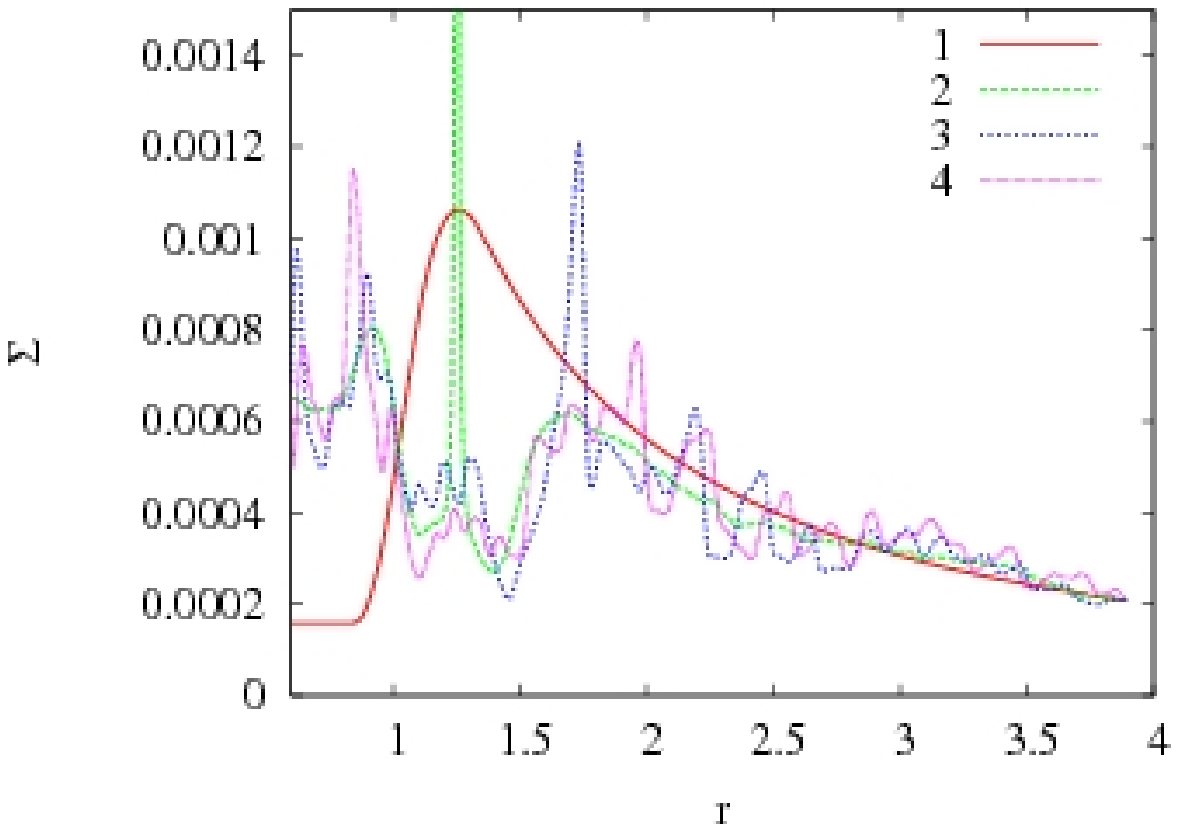}
\caption{The evolution of the density profiles during the planet
migration in a disc with an inner edge. Each panel shows the initial
density profile (curve 1), the azimuthal average of the surface
density (curve 2; the planet position is visible as a spike)
and the surface density cuts through the Lagrange points L5 (curve 3)
an L4 (curve 4). The panels upper left, upper right, lower left and
lower right show the density profiles after $40$, $50$, $60$ and $70$
orbits, respectively.}
\label{fstop_sigma_edge}
\end{figure*}

In this section we address the processes that stop rapid type III
migration.  We consider two possible causes. In the first one
migration slows down due to the decrease of the volume of the
co-orbital region, and in the second one the planet stops at an inner
disc edge. When discussing stopping of type III migration we mean a
radical slow-down of migration to a rate similar to type II
migration, or $|Z| \ll 1$. However, an exact stopping, $Z=0$, can be
observed in some cases.

Since the non-dimensional migration rate $Z$ is a function of the
co-orbital mass deficit $M_\mathrm{\Delta}$, to stop a type III
migration we have to diminish $M_\mathrm{\Delta}$. This can be
achieved by either decreasing the mass of the co-orbital region or by
removing the asymmetry in this region. The migration rate depends on
the amount of angular momentum transferred between the planet and the
gas crossing the planet's orbit. To start and maintain rapid migration
the impulse that the planet gets from the orbit crossing gas has to be
strong enough to move the planet into the undisturbed region of the
disc where it can interact with a new portion of gas. This impulse
depends on the change in the radial position of the gas element
crossing the orbit given by $x_\rmn{s}$ and the asymmetry in the
co-orbital region given by $\Sigma_\rmn{s} - \Sigma_\rmn{g}$
(Eq.~\ref{eqn_m_delta_small}). Assuming maximal asymmetry and putting
$\Sigma_\rmn{g} = 0$ leaves us with two parameters, $\Sigma_\rmn{s}$
and $x_\rmn{s}$, whose product is proportional to the mass of the
co-orbital region. This parameter allows us to qualitatively
understand the stopping mechanism for type III migration and the
dependence of $Z$ on the disc total mass and on
$\alpha_\rmn{\Sigma}$. It is however difficult to get quantitative
results, since $\Sigma_\rmn{g}$ in practice is not zero, and depends
on the migration history.

Our simulations show that fast migration is not allowed in the region
where the mass of the co-orbital region $M_\rmn{CR}$ is smaller than
the planet mass.\footnote{The exact calculations show, that at the 
stopping radius $M_\rmn{CR}$ ranges between $1.8$ and $2.3$ of 
$M^*_\rmn{P}$ taken at the stopping time.} 
Since for $\alpha_\rmn{\Sigma} > -2$ this parameter
decreases with decreasing distance to the star, it is easier to start
type III migration further out in the disc. The migration rate
decreases with distance to the star and thus there will be an inner
region where fast migration cannot occur. This means that in most
cases the planet will not be able to migrate inward all the way by
type III migration. The value of the radius where the planet stops
rapid migration depends on the surface density profile and the planet
mass. This radius decreases with increasing $\mu_\rmn{D}$ and
decreasing $\alpha_\rmn{\Sigma}$. Planets with smaller masses should
be able to migrate further in, but with a lower migration rate since
$\dot a_\rmn{f} \sim M_\rmn{P}^{-2/3}$. Interestingly, this analysis
implies that the situation is reversed in discs with steep density
profiles, $\alpha_\rmn{\Sigma} < -2$. In this case it is easier to
start fast migration close to the star and the inward migrating planet
should increase the migration rate with decreasing distance to the
star.

$M_\rmn{CR}$ decreases mostly due to the shrinking of the radial
extent of the co-orbital region. This can be seen in
Fig.~\ref{fstop_sigma_no_edge}, where the the azimuthal average of the
surface density (curve 2, and the surface density cuts through the L5
(curve 4) an L4 (curve 5) points are plotted. The planet moves through
the disc leaving the surface density profile slightly changed (curve 1
is the original density profile). After $30$ and $40$ orbits the
strong asymmetry between L5 and L4 that is driving migration is
visible. However, this asymmetry disappears after $50$ orbits even
though the mass difference between the inner and outer edge of the
co-orbital region still exists. The migration stops since the jump in
the radial position of gas crossing the co-orbital region is too small
to remove a substantial amount of angular momentum from the planet.

The other mechanism for stopping type III migration is the interaction
with a steep density change, such as a disc edge. At a significant
density jump $M_\rmn{CR}$ will diminish considerably, and asymmetry in
the co-orbital region will disappear. Figure~\ref{fstop_sigma_edge}
shows the evolution of the density profiles in the simulation
presented at the end of Sect.~\ref{alpha_dependence}.  The upper left,
upper right, lower left and lower right panels show the density
profiles at $t=40$, $50$, $60$ and $70$ orbits respectively. At these
times $Z\approx -2$, $-1.5$, $-0.05$, $0.0$. At $t=70$~orbits the
planet is locked in the disc and gap opening has started. The upper
left and upper right panels show the density distribution during the
fast migration limit $Z < -1$ and are similar to the upper left and
upper right panels in Fig.~\ref{fstop_sigma_no_edge}.  The lower left
panel shows the planet interacting with the disc edge. At this time
the mass difference between the inner and outer edge of the co-orbital
region disappears and the planet gets a positive value of $Z$. However
we found this temporary outward migration to be sensitive to the grid
resolution (decreasing for higher resolution).

The results show that a Jupiter is massive enough to destroy the
initial density jump and fill the inner empty region. That is why when
the planet finally settles, the inner and outer disc have almost equal
surface densities. This actually prevents the planet from stopping
completely. The situation is different for lower mass planets that
cannot change significantly the density profile. They can be trapped
close to the position of maximum density for a long time and grow
slowly \citep{2006ApJ...642..478M}. To have this happen to a
Jupiter-mass planet would require an additional mechanism that removes
material from the inner disc and keeps the radial density gradient
sharp.  Such a mechanism would keep the planet at a constant orbit due
to the gas flow from the co-orbital region until the horseshoe region
has completely cleared. However, if the rate of mass accretion onto
the star is too small to fill the co-orbital region and keep a
constant gas flow through the co-orbital region the planet would still
move away from its position at the disc edge.

In paper~III we will encounter two additional mechanisms that can stop
type III migration, but which are more evident in the outward migration
case.
\begin{figure*}
\includegraphics[width=84mm]{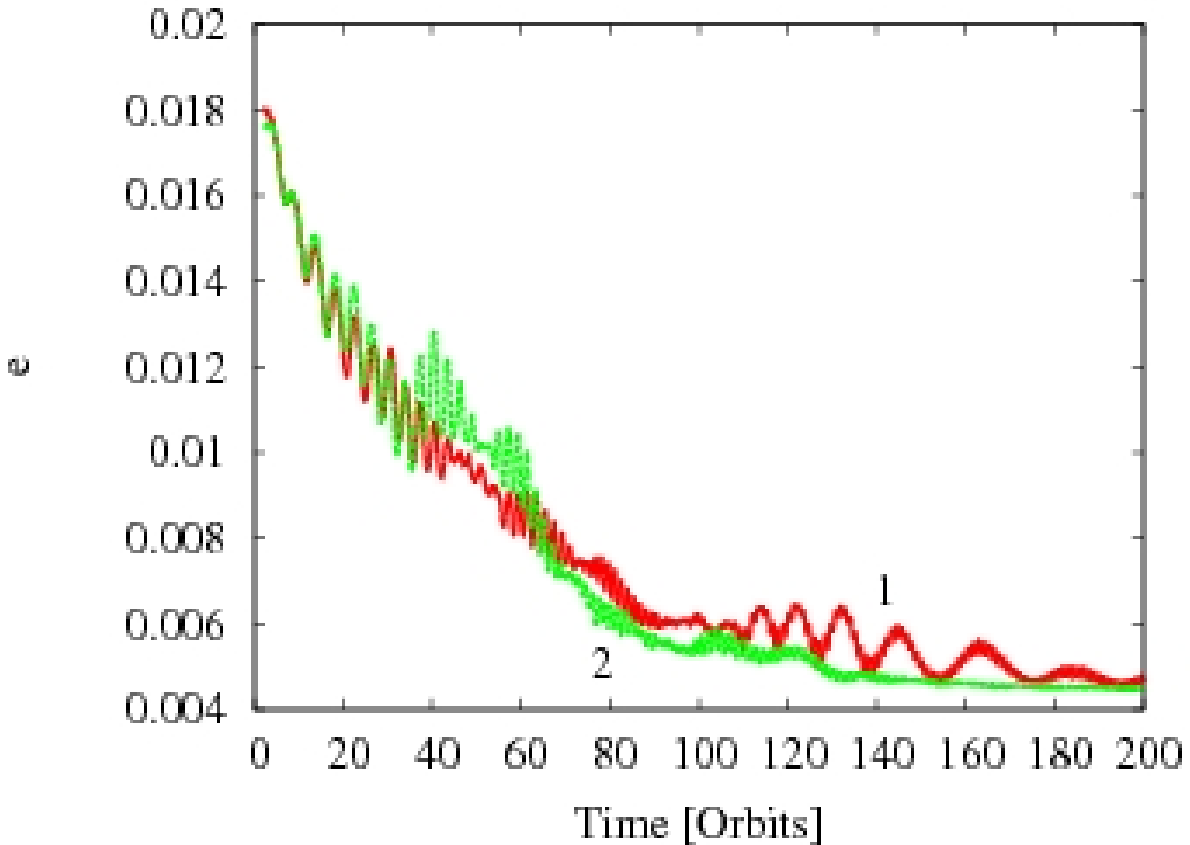}
\includegraphics[width=84mm]{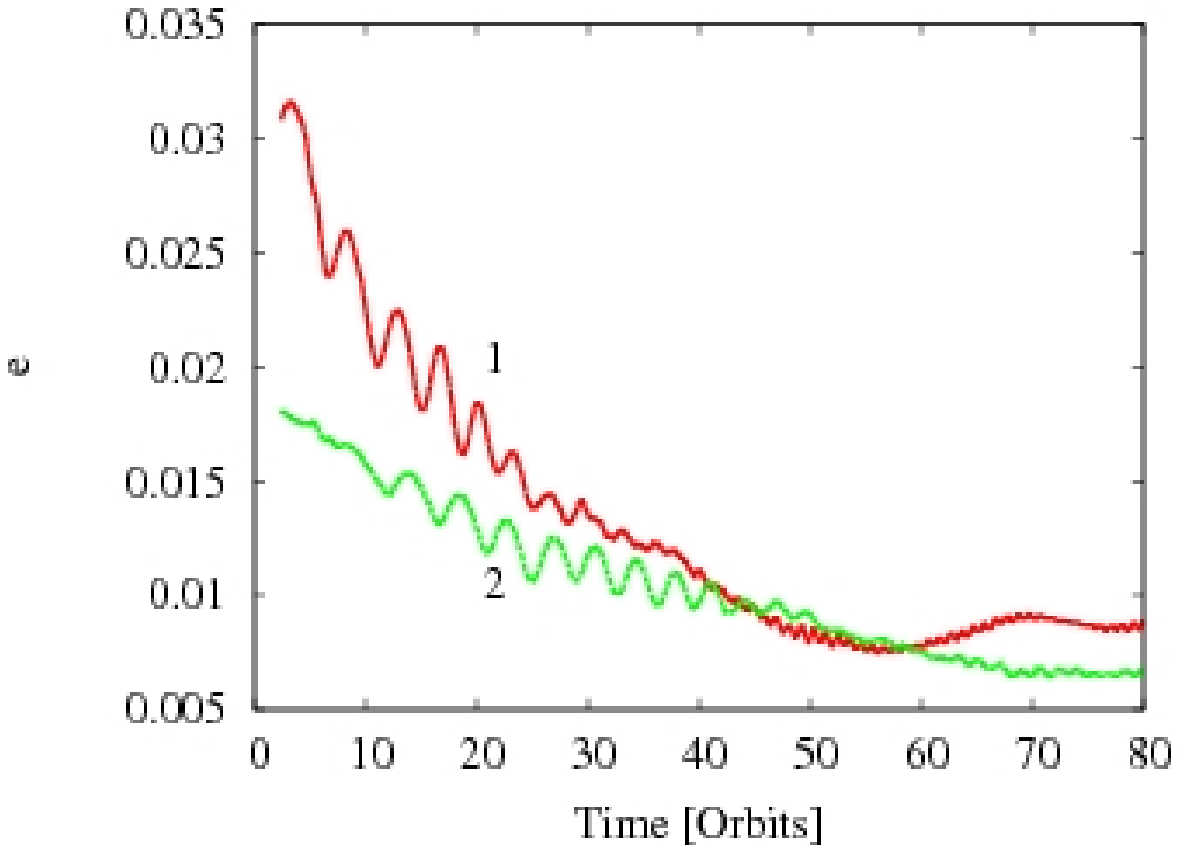}
\caption{The time evolution of the eccentricity $e$. The left panel
 shows the results of the models with the constant effective planet's
 mass $M_\rmn{P}^*=M_\rmn{P}=0.001$ (curve 1) and the model including
 the gas accretion on the planet (curve 2). The right panel presents
 the results of the models with different total disc masses. Curves 1
 and 2 correspond to $\mu_\rmn{D}$ equal $0.005$ (standard case) and
 $0.0025$ respectively.}
\label{comp_ecc}
\end{figure*}
\section{Eccentricity evolution}

\label{sec_ecc_ev}

In this section we discuss the eccentricity evolution and its
dependency on different simulation parameters. The results of
different models are presented in Fig.~\ref{comp_ecc}, see also
the top right panel of Fig.~\ref{fim1_a} for our standard case.

The left panel of Fig.~\ref{comp_ecc} shows the time evolution of the
eccentricity $e$ for the model with the constant effective planet's
mass $M_\rmn{P}^*=M_\rmn{P}=0.001$ (curve 1) and the model including
the gas accretion on the planet (curve 2). In the second model the
planet's mass was increased by the gas mass removed from the planet's
proximity (for more discussion see Paper~I Sect.~5.1.2). It shows that
during the rapid migration phase (lasting for about 70 orbits) the
evolution of the eccentricity is almost independent of the planet's
mass and the mass of the circumplanetary disc. The eccentricity grows
rapidly during the first few orbits and reaches $0.018$, and
diminishes with time on the timescale of migration or faster. The
differences between the models show up in the slow migration phase,
where the first model loses gas from the Hill sphere and the
eccentricity starts to oscillate due to these gas motions in the
circumplanetary disc. In the second model accretion has removed most
of the mass of the circumplanetary disc, and the eccentricity slowly
decays without oscillations. We found similar results in simulations
with the $\widetilde M_\rmn{P} = M_\rmn{P}+ M_\rmn{soft}$ and
different initial planet masses $M_\rmn{P}$. In all of these cases the
time-scale for the eccentricity damping during the rapid migration
phase is independent on the planet mass. We found it to be weakly
dependent on other simulation parameters such as the density profile
exponent $\alpha_\rmn{\Sigma}$.

However, the time-scale for the eccentricity damping does depends on
the total disc mass. The right panel in Fig.~\ref{comp_ecc} shows the
time evolution of $e$ for the models with $\mu_\rmn{D}$ equal $0.005$
and $0.0025$ (curves 1 and 2). The value of $e$ after a few first
orbits depends on $\mu_\rmn{D}$ and is about $0.03$ and $0.017$ for
the first and the second model respectively. In both models the
eccentricity is damped during the rapid migration phase, and the
corresponding damping time-scale is about $33$ and $69$ orbits. This
suggests that the damping time-scale is proportional to
$1/\mu_\rmn{D}$

A probable explanation for the eccentricity damping is the presence of
dense gas in the region of the strong $e$-damping co-orbital Lindblad
resonances.  Although the eccentricity driving theory of
\citet{1993ApJ...419..166A} does not explicitly account for the
modification of gas flow in the corotational region and is not meant
for gap-opening planets, it can provide an approximate upper limit to
the strength of eccentricity damping by density wave emission at the
Lindblad resonances. Eccentricity driving is a result of the
competition between external Lindblad resonances and co-orbital
Lindblad resonances, the latter being about 3 times stronger than the
former. Since type III migration implies that the gap around the
planet is asymmetric, let us assume that the mean gas density in the
co-orbital resonance region near the planet is reduced by a factor of 2
with respect to the case of a smooth disk, considered by
\citet{1993ApJ...419..166A}, equivalent to the simplified view of zero
gas density in the libration region and unperturbed values
elsewhere. Eccentricity damping occurs in direct proportion to the gas
density.  Therefore, the resulting eccentricity damping timescale
should in our standard case be larger than Artymowicz's estimate by a
factor of $(3 - 1)/(3/2-1) = 4$. In a disk with $\mu_D = 0.005$,
$h_\rmn{s}=0.05$ everywhere, and $\alpha_\Sigma = -3/2$, we obtain the
following lower limit on the damping timescale at $a=5$ AU:
\begin{equation}
e/\dot{e} > (0.008/\mu) {\rm yr}
\end{equation} 
(where $\mu=0.001$ for Jupiter), giving 8 years (about one orbit) at
Jupiter's radius. This is much shorter than the numerically determined
timescale for $e$ to drop from $e=0.02718$ to $e=0.01$, which equals
about 30 periods, as can be seen in Fig.~\ref{fim1_a}. We conclude that
the analytical description is consistent with our calculations but does
not provide a good guidance. A theory better adjusted to the
corotational gas flow streamlines and/or eccentric corotational
torques is needed. It is interesting to note that eccentricity damping
and migration can operate on a similar, short timescales.

\section{Conclusions}

\label{conclusions}

In this paper we presented a study of inward directed type III
migration of Jupiter-mass planets embedded in a disc. We used two
dimensional numerical simulations, performed in a Cartesian coordinate
system and the inertial reference frame. We focused on the detailed
flow structure in the planet's vicinity and used an adaptive mesh
refinement to achieve high resolution inside the Roche lobe. We used
a modified version of the usual local-isothermal approximation,
where the temperature depends on the distance to both the star and the
planet, and also added a correction for the gas self-gravity near to
the planet.

Type III migration is driven by the co-orbital torque
$\Gamma_\rmn{CR}$ exerted on the planet by the gas crossing the
co-orbital region. The value of this torque depends on the mass flux
crossing the planet's orbit and on the density asymmetry in the
co-orbital region. This asymmetry can be characterised by the
non-dimensional migration rate $Z = \dot a / \dot a_\rmn{f}$ which
gives the ratio between the the migration timescale (defined as
$x_\rmn{s}/{\dot a}$) and libration timescale. 
We can thus divide type III migration into a slow ($|Z| < 1$) and a
fast ($|Z| > 1$) migration limit. The shape and the structure of the
horseshoe region differ between these two limits as the azimuthal
extent of the horseshoe region decreases with $|Z|$. In the slow
migration limit the horseshoe region fills the whole co-orbital region
and the flow asymmetry is limited to the planet's direct
vicinity. Both tadpole regions are present, although their shape
changes with $Z$ and the positions of the stationary points move away
from the Lagrangian points L4 and L5. In the fast migration limit the
whole horseshoe region shrinks to a single tadpole-like region.

In the fast migration limit the planet's radial motion is too fast to
allow a gap to be opened (Fig.~\ref{fim1_dens_disc} and
\ref{fim11_sigma_time}). Instead there is a strong gas flow through
the corotation region, moving gas from the inner disc (the region with
$r < a-x_\rmn{s}$) to the outer disc (the region with $r >
a+x_\rmn{s}$). In the inertial reference frame the planet together
with the gas captured in the horseshoe region moves fast with respect
to the gas that was initially placed in the inner disc
(Fig.\ref{fim8_comp_disc}). Note that there is no global motion for
most of the gas in the disc in this case.  Although the fluid captured
by the planet in the horseshoe region moves together with the planet,
the gas crossing the corotation changes its initial position in the
disc only by order of $x_\rmn{s}$. This is an important difference
with type II migration, where the planet is locked in the disc and
moves on the viscous time scale.  In the slow migration limit, when
the migration timescale becomes longer than the libration timescale,
the planet starts clearing a gap, and gradually makes the transition
to type II migration.

Another difference between the slow and fast migration limits is the
relation between $\Gamma_\rmn{CR}$ and $Z$.  The torque exerted on the
planet depends mostly on the structure of the flow lines in the
horseshoe region and in the circumstellar disc. In the slow migration
limit the flow through the corotation is limited to the a narrow
stream at a boundary of the Roche lobe and the asymmetry of both the
circumplanetary disc and the horseshoe region is relatively small.
This asymmetry is sensitive to the small changes in $Z$, giving
a strong dependence of the total torque $\Gamma$ (dominated by the
corotational torque) on $Z$. On the other hand in the fast migration
limit the asymmetry of the circumplanetary disc and the horseshoe
region is significant, the gas can cross the co-orbital region over a
wide azimuthal range and the total torque is only weakly dependent
on $Z$. This results in a linear grow of the total torque $|\Gamma|$ with
$|Z|$ in the slow migration limit and a saturation of $|\Gamma|$
around $Z \approx -1.5$. In the fast migration $|\Gamma|$ decreases
slowly with $|Z|$.

The torque also depends on the rate of mass accumulation in the
planet's vicinity, which is related to the non-dimensional migration
rate (Fig.~\ref{fim3_mp}), and on the assumed scale height
for the circumplanetary disc. For $h_\rmn{p}=0.4$ the effective planet
mass grows slowly in the fast migration limit, since in this case the
horseshoe region contracts into a single tadpole-like region, and most
of the fluid crossing the co-orbital region does not enter the Roche
lobe. The gas orbits in the circumplanetary disc are found to be
strongly asymmetric, and the gas flow into the Roche lobe is
limited. The asymmetry of the horseshoe region and the circumplanetary
disc diminishes with decreasing $|Z|$, and the mass accumulation
rapidly increases for $ Z \approx -1.7$. This causes the torque generated
inside the Hill sphere to drop significantly and shortly after the
planet enters the slow migration limit, where the co-orbital flow is
limited to the a narrow stream in the planet's neighbourhood, and the
circumplanetary disc becomes more symmetric. This allows the planet to
capture a larger amount of the mass, but this process is limited by
the fact that the mass flux crossing the planet orbit decreases
quickly with $|Z|$, thus ending the mass accumulation phase close to
$Z \approx -0.6$. The fast mass accumulation builds up a strong
pressure gradient in the planet's proximity, which is only supported
by the gas inflow into the circumplanetary disc. After the mass
accumulation phase ends gas starts to flow away from the circumplanetary
disc, and short phase of outward migration can occur.

The growth of the effective planet mass is strongly dependent on the
assumed temperature profile in the circumplanetary disc, decreasing
for larger circumplanetary disc aspect ratios. For $h_\rmn{p} \ge 0.5$
we found no substantial mass accumulation, and gas is lost from the
the planet's environment during the entire slow migration limit. 

The planet mass plays an important role, even though the planet's
orbital evolution is only weakly dependent on it in the fast migration
limit. Since $Z \sim \widetilde M_\rmn{P}^{-2/3}$, the planet with
bigger mass (or lower $h_\rmn{p}$) makes the transition to the slow
migration limit sooner, and settle at larger radii than planets with 
lower mass (or larger $h_\rmn{p}$).

The position where the planet stops rapid type III migration thus
depends on the planet's mass, but also on the disc structure. The mass
of the co-orbital region defines the region of the disc where the
rapid migration is allowed. To start and maintain type III migration
the mass of the co-orbital region has to be comparable to the planet
mass. This means that an increase of the disc mass gives a similar
effect as a decrease of the planet's mass.

The planet's orbital evolution is determined by the changes of the
co-orbital mass deficit $M_\rmn{\Delta}$, expressing the mass and the
asymmetry of the co-orbital region. The mass is the product of the density
profile and the volume of the co-orbital region. For discs with 
$\alpha_\rmn{\Sigma} > -2$ it therefore decreases with radius, and
inward rapid migration always slows down. This is the stopping mechanism
for most of the simulations presented in this paper. Type III
migration can also be stopped at significant density jumps in the
disc (such as a disc edge). In this case the asymmetry of the
horseshoe region is lost due to the interaction with the low density
region and the planet rapidly makes the transition to the slow
migration regime $|Z| \ll 1$.

In all our simulations the eccentricity did not exceed $0.04$ and was
damped during the rapid migration phase. We found the damping
time-scale to be independent of the planet mass, but inversely
proportional to the total disc mass. The measured damping time-scales
are few times longer than the one predicted by the analytical
formulation.

The results in this paper show that type III migration can operate
effectively, but does require the mass in the co-orbital region to be
larger than the planet mass in order to get started, making it easier
to start type III migration for lower mass planets, and for positions
further out in the disk. The mass of the co-orbital region also sets a
limit on how far the planet can migrate inward. The minimum orbital
radius it can attain is again set by the position where the co-orbital
mass exceeds twice the planet mass. It can however stop at larger radii than
that, depending on the details of the density structure of the
co-orbital region. For all cases the use of the dimensionless
migration rate $Z$ is useful to characterise the migration behaviour
and the structure of the co-orbital region. Calculating $Z$ can for
example be used to check whether type I or type II migration will lead
to type III migration.

The simulations presented in this paper were set up to study inward
migration in the type III regime. In a real disc-planet interaction,
type III migration may also lead to outward migration, this depending
on the details of the previous migration pattern. This outward
migration mode will be studied in detail in a subsequent paper.


\section*{Acknowledgements}

The software used in this work was in part developed by the
DOE-supported ASC / Alliance Center for Astrophysical Thermonuclear
Flashes at the University of Chicago. We thank F.\ Masset for
interesting and useful comments. Calculations reported in this paper
were performed at High Performance Computing Centre North (HPC2N) and
National Supercomputing Centre, Link\"oping, and on the Antares
cluster funded at Stockholm Observatory purchased using funding
provided to PA by Vetenskapsradet, Sweden. The authors acknowledge
support of the European Community's Human Potential Programme under
contract HPRN-CT-2002-00308 (PLANETS), as well as the NSERC, Canada,
Discovery grant (2005-2008).

%
%
\bibliographystyle{mn2e}
\bibliography{articles_astro}

\label{lastpage}

\end{document}